\documentclass[12pt,draftcls,onecolumn]{IEEEtran}
\usepackage{algorithm}
\usepackage{algorithmic}
\renewcommand{\algorithmicrequire}{\textbf{Input:}}
\renewcommand{\algorithmicensure}{\textbf{Output:}}

\usepackage[normalem]{ulem}
\usepackage[usenames,dvipsnames]{color}
\usepackage{amsfonts}
\newcommand{\rd}{\textcolor{Red}}

\usepackage{cite}

 \ifCLASSINFOpdf
   \usepackage[pdftex]{graphicx}
 \else
   \usepackage[dvips]{graphicx}
 \fi

\usepackage[cmex10]{amsmath}

\usepackage{subfig}

\begin{document}
\title{
Waveform Inversion with Source Encoding for Breast Sound Speed Reconstruction
in Ultrasound Computed Tomography 
}

\author{Kun~Wang,~\IEEEmembership{Member,~IEEE,}
        Thomas~Matthews,~\IEEEmembership{Student Member,~IEEE,}
        Fatima~Anis, 
        Cuiping~Li, 
        Neb~Duric,\\
        and 
        Mark~A.~Anastasio,~\IEEEmembership{Senior Member,~IEEE}
     % ~\IEEEmembership{Life~Fellow,~IEEE}% <-this % stops a space
\thanks{K.~Wang, T.P.~Matthews, F.~Anis, and M.A.~Anastasio are with the Department
of Biomedical Engineering, 
Washington University in St. Louis, 
St. Louis, MO 63130,
e-mail: anastasio@wustl.edu}% <-this % stops a space
\thanks{C.~Li and N.~Duric are with Delphinus Medical Technologies,
Plymouth, MI 48170} 
\thanks{N.~Duric is also with Karmanos Cancer Institute, Wayne State University, 
4100 John R.~Street, 5 HWCRC, Detroit, MI 48201}
% 6550 Mapleridge Street, Suite 124, Houston, TX 77081-4629.}% <-this % stops a space
% \thanks{Manuscript received April 19, 2005; revised December 27, 2012.}
}

\maketitle

\begin{abstract}

Ultrasound computed tomography (USCT) holds great promise for improving the detection and management of breast cancer. 
Because they are based on the acoustic wave equation,
 waveform inversion-based reconstruction methods
 can produce images that possess improved spatial resolution properties over those produced by ray-based methods. 
However, waveform inversion methods are computationally demanding and have not
been applied widely in USCT breast imaging. 
In this work,  source encoding concepts are employed to develop an accelerated USCT reconstruction 
method that circumvents the large computational burden of conventional waveform inversion methods.
This method, referred to as the waveform inversion with source encoding (WISE) method,
encodes the measurement data using a random encoding vector and determines an estimate of the sound speed
distribution by
 solving a stochastic optimization problem by use of a stochastic gradient descent algorithm. 
% for USCT imaging system with a typical circular measurement geometry. 
% Instead of performing a waveform inversion sequentially for every acoustic pulse transmission, the WISE method combines all pulse transmissions into a single encoded pulse and applies waveform inversion to the encoded pulse transmission. 
% Using specially designed random codes, the reconstructed images can be automatically decoded during the iteration through a stochastic optimization algorithm. 
% In order to apply the WISE method to breast imaging applications with a circular measurement geometry, a heuristic data-filling strategy has been proposed. 
Both computer-simulation and experimental phantom studies are conducted to demonstrate the use of the WISE method.
The results suggest that the WISE method maintains the high spatial resolution of  waveform inversion 
methods while significantly reducing the computational burden. 
% This study demonstrates the feasibility of source encoding accelerated waveform inversion algorithm for USCT breast imaging with certain practical considerations.

\end{abstract}


% \fi
\IEEEpeerreviewmaketitle

\section{Introduction}

After decades of research \cite{glover1979characterization,Carson81:ClinUSCT,Schreiman84:USCT,Andre97:ClinDiffr}, advancements in hardware and computing technologies are now facilitating the clinical translation of
 ultrasound computed tomography (USCT) for breast imaging applications \cite{Carson81:ClinUSCT,Johnson99:Patent,Duric2007:MedPhys,Ruiter11:3D,Neb12:BookChapter}. 
USCT holds great potential for improving the detection and management of breast cancer
  since it provides novel acoustic tissue contrasts, is radiation- and breast-compression-free, and
is relatively inexpensive. \cite{Ruiter12:NewEra,Duric13:ClinReflect}. 
Several studies have reported the feasibility of USCT for characterizing breast tissues \cite{Carson81:ClinUSCT,Andre97:ClinDiffr,Johnson99:Patent,Duric2007:MedPhys,Li09:ClinSOS,Duric13:ClinReflect}.  
Although some USCT systems are
 capable of generating three images that depict the breast's acoustic reflectivity, acoustic attenuation, and sound speed 
distributions, 
this study will focus on the reconstruction of the sound speed distribution.

A variety of USCT imaging systems have been developed for breast sound speed imaging \cite{Johnson99:Patent,Wiskin12:WaveRecon,Duric13:ClinReflect,Ruiter11:3D,manohar,AAO12:LUS,Jun13:USCT}. 
In a typical USCT experiment,
 acoustic pulses that are generated by  different transducers
are employed, in turn, to insonify the breast.
The resulting  wavefield data are measured by an array of ultrasonic transducers that are
 located outside of the breast. 
Here and throughout the manuscript, a transducer that produces an acoustic pulse will be referred to as an
 emitter; the transducers that receive the resulting wavefield data  will be referred to as receivers.
%The measured data contain information about the acoustic properties of the breast.
From the collection of recorded wavefield data,
an image reconstruction method is utilized to 
 estimate the sound speed distribution within the breast \cite{Johnson99:Patent,Duric13:ClinReflect,Ruiter11:3D}.

The majority of USCT image reconstruction methods for breast imaging investigated to date have been
 based on approximations to the acoustic wave equation
 \cite{KakBook,Oelze09:DBIM,Oelze10:DBIM,Chew10,Huthwaite12:Combine,Simonetti06:SubWavelength,Los12:SOSPE,Duric2007:SOSWave,Wiskin12:WaveRecon,Huthwaite12:Validate}. 
A relatively popular class of methods is based on geometrical acoustics, and are commonly
referred to as `ray-based' methods.
These methods involve two steps. 
First,  time-of-flight (TOF) data corresponding to each emitter-receiver pair
are estimated \cite{Duric09:TOFPicker}.
Under a geometrical acoustics approximation, the TOF data are related to the sound speed distribution
via an integral geometry, or ray-based, imaging model \cite{KakBook,Roy10:BentRay}.
Second, by use of the measured TOF data and the ray-based imaging model,
  a reconstruction algorithm is employed to
 estimate the sound speed distribution.
Although ray-based methods can be computationally efficient,  the spatial resolution of the images
they produce is limited due to the fact that diffraction effects are not modeled
\cite{Bates1991185,Duric2007:SOSWave}.  This is undesirable for breast imaging applications,
in which the ability to resolve fine features, e.g., tumor spiculations, is important for distinguishing healthy from diseased tissues.

%Both straight-ray \cite{KakBook} and bent-ray-based \cite{Roy10:BentRay} imaging models have been investigated in the literature. 
%The ray-based image reconstruction algorithms ignore high-order diffractions, which may limit the spatial resolution of the reconstructed \rd{sound speed} images \cite{Bates1991185,Duric2007:SOSWave}. 
%Also, because the measurement noise is non-linearly mapped to the TOF data, it becomes more difficult to exploit the noise statistics to optimize the image variance \cite{Wernickbookchap21}.
%Recently, a hybrid diffraction tomography method has been employed to improve the spatial resolution for cases in which scattering is relatively weak (satisfying the Born approximation) \cite{Huthwaite12:Combine,Huthwaite12:Validate}. 
%% While its numerical properties remain unclear, 
%As an analytical image reconstruction method, the hybrid method is subject to the same difficulties in mitigating image noise.

USCT reconstruction methods based on the acoustic wave equation, also known as
full-wave inverse scattering or waveform inversion methods,
have also been explored for a variety of applications including medical imaging \cite{Duric2007:SOSWave,Roy2010:SOSWave,Los12:SOSPE,Wiskin12:WaveRecon} and geophysics \cite{Krebs:SE,Herrmann2012:SE,Herrmann13:GaussCode}.
Because they account for higher-order diffraction effects,
waveform inversion methods can produce images that possess higher spatial resolution than those produced
by ray-based methods \cite{Duric2007:SOSWave,Roy2010:SOSWave}. 
However, conventional waveform inversion methods are iterative in nature
and  require the wave equation to be solved numerically
a large number of times at each iteration.
Consequently, such methods can be extremely computationally burdensome.
For special geometries \cite{Wiskin12:WaveRecon,Wiskin13:3D},
efficient numerical wave equation solvers have been reported. %that mitigate this.
However, apart from special cases, the large computational burden of waveform inversion methods
has hindered their widespread application.

A natural way to reduce the computational complexity of the reconstruction problem is to reformulate
it in a way that permits a reduction in the number of times the wave equation needs to be solved.
In the geophysics literature, source encoding methods have been proposed to achieve this \cite{Krebs:SE,Herrmann2012:SE,Herrmann13:GaussCode}.
When source encoding is employed, at each iteration of a prescribed reconstruction algorithm,
   all of the acoustic pulses produced by the emitters  are combined (or `encoded') 
by use of a random encoding vector. The measured wavefield data are combined in the same  way.
% are combined in the same way.  
As a result, the wave equation may need to be
solved as few as twice at each algorithm iteration.
In conventional waveform inversion methods, this number would be equal
to twice the number of emitters employed.   Although conventional waveform inversion
methods may require fewer algorithm iterations to obtain a specified image
accuracy compared to source encoded methods, as demonstrated later, the latter
can greatly reduce the overall number of times the wave equation needs to be solved. 

\iffalse
To determine a \rd{sound speed} estimate from the encoded sources and measurement data, 
the  reconstruction problem is formulated as a stochastic optimization problem.
Inspired by these studies, a phase-encoded waveform inversion method has been reported
 with noise-free computer-simulation studies for a linear detection geometry \cite{Los12:SOSPE}.
However, a systematic investigation of source encoding methods for breast \rd{sound speed} reconstruction 
that employs realistic computer-simulated data and experimental data
remains largely unexplored.
\fi

%For systems with a spherical incident wave, a source encoding method has been proposed in the geophysics literature \cite{Krebs:SE,Herrmann2012:SE,Herrmann13:GaussCode}. 

\if 0
The source encoding technique requires a fixed measurement geometry for all emitters.
This condition can be easily satisfied by a system with a linear detection geometry. 
A phase-encoded waveform inversion method has been reported in the literature with noise-free computer-simulation studies for a linear detection geometry \cite{Los12:SOSPE}. 
However, when a circular measurement geometry is employed, the acoustic source will contaminate its neighboring receivers due to electrical/acoustical crosstalk, resulting in unreliable measurements near the emitter \cite{Duric2007:SOSWave}. 
Because of these challenges, source-encoding accelerated waveform inversion remains largely unexplored for medical imaging applications. 
\fi

In this study,  a waveform inversion with source encoding (WISE) method for USCT sound speed reconstruction is developed and investigated
for breast imaging with a circular transducer array.
The WISE method
determines an estimate of the sound speed
distribution by
 solving a stochastic optimization problem by use of a stochastic gradient descent algorithm \cite{Herrmann2012:SE,ArXiv:Haber14}.
Unlike previously studied waveform inversion methods that were based on the Helmholtz equation \cite{Duric2007:SOSWave,Los12:SOSPE},
the WISE method is formulated by use of the time-domain acoustic wave equation \cite{MortonBook:2005,Mast01:2ndKSpace,Tabei02:1stKSpace}
and utilizes broad-band measurements.
The wave equation is solved by use of a computationally efficient k-space method that is accelerated by use of graphics processing units (GPUs). 
In order to mitigate the interference of the emitter on its neighboring receivers, a heuristic data replacement strategy is proposed. 
The method is validated in computer-simulation studies that include
modeling errors and other physical factors.
The practical applicability of the method is further demonstrated
in studies involving  experimental breast phantom data.

The remainder of the paper is organized as follows. 
In Section II, USCT imaging models in their continuous and discrete forms are reviewed.
A conventional waveform inversion method and the WISE method for sound speed reconstruction
are formulated in Section III.
The computer-simulation studies and corresponding numerical results are presented in Sections IV and V, respectively. 
In Section VI,  the WISE method is further validated in experimental breast phantom studies.
Finally, the paper concludes with a discussion in Section VII.

\section{Background: USCT imaging models}
\label{Sect:BckGround}

%Although a practical digital imaging system should be described as a continuous-to-discrete (C-D) map (See Chapter 7 in \cite{BarrettBook}), in this section, for simplicity, we describe the USCT imaging system in its continuous form.  
%In this way, the waveform inversion method possesses a unified formulation that is not tied to a particular digital data acquisition system or numerical discretization strategy. 
%The effects of sampling are properly treated subsequently in Section \ref{Sect:WISE}. 

In this section, imaging models that provide the basis for image reconstruction in waveform inversion-based USCT are
reviewed.

\subsection{USCT imaging model in its continuous form}

Although a digital imaging system is properly described as a continuous-to-discrete (C-D) mapping
 (See Chapter 7 in \cite{BarrettBook}), for simplicity, a USCT imaging system
is initially described  in its continuous form below.

In USCT breast imaging, a sequence of acoustic pulses is transmitted through the breast. 
We denote each acoustic pulse by $s_m(\mathbf r, t)\in\mathbb L^2(\mathbb R^3\times[0,\infty))$, where 
each pulse is indexed by an integer $m$ for $m=0,1,\cdots,M-1$  with $M$ denoting the total number of acoustic pulses. 
Although it is spatially localized at the
emitter location, each acoustic pulse can be expressed as a function of space and time.
% $s_m(\mathbf r, t)$ has a spatial and temporal dependence with $\mathbf r$ and $t$ denoting the spatial and the temporal coordinate, respectively.
When the $m$-th pulse propagates through the breast, it generates a pressure wavefield distribution denoted by $p_m(\mathbf r, t)\in\mathbb L^2(\mathbb R^3\times[0,\infty))$.
If acoustic absorption and mass density variations are negligible, $p_m(\mathbf r, t)$ in an unbounded medium satisfies the acoustic wave equation \cite{KinslerBook}: 
\begin{equation}\label{eqn:WaveEqn}
  \nabla^2 p_m(\mathbf r, t) 
 - \frac{1}{c^2(\mathbf r)} \frac{\partial^2}{\partial t^2}p_m(\mathbf r, t) 
 = -4\pi s_m(\mathbf r, t), % \delta(\mathbf r-\mathbf r_m^{\rm e}), 
\end{equation}
where $c(\mathbf r)$ is the sought-after sound speed distribution. 
% For an unbounded medium, $p_m(\mathbf r, t)$ and $s_m(\mathbf r, t)$ are related by a linear mapping as
Equation \eqref{eqn:WaveEqn} can be expressed in operator form as 
\begin{equation}\label{eqn:CImagingModel1}
  p_m(\mathbf r, t) = \mathcal H^{\rm c} s_m(\mathbf r, t), 
\end{equation}
where the linear operator $\mathcal H^{\rm c}:\mathbb L^2(\mathbb R^3\times[0,\infty))\mapsto \mathbb L^2(\mathbb R^3\times[0,\infty))$ denotes the action of the wave equation and is independent of the index of $m$.
The superscript `c' indicates the dependence of  $\mathcal H^{\rm c}$ on $c(\mathbf r)$. 
%This way of mathematically describing the problem will be shown to be convenient when
%This notation will be convenient when developing the WISE method in Section III.

Consider that $p_m(\mathbf r, t)$ is recorded outside of the object for $\mathbf r\in\Omega_m$ and $t\in[0,T]$, where $\Omega_m\subset\mathbb R^3$ denotes a continuous measurement aperture.
In this case, when discrete sampling effects are neglected, the imaging model can be described as a continuous-to-continuous (C-C) mapping as: 
\begin{equation}\label{eqn:CImagingModel2}
  g_m(\mathbf r, t) = \mathcal M_m \mathcal H^{\rm c} s_m(\mathbf r, t),
  \quad\text{for}\quad m=0,1,\cdots,M-1,
\end{equation}
where $g_m(\mathbf r,t)\in\mathbb L^2(\Omega_m\times[0,T])$ denotes the measured data function 
and the operator $\mathcal M_m$ is the restriction of $\mathcal H^{\rm c}$ to $\Omega_m \times [0,T]$.
The $m$-dependent operator $\mathcal M_m$ allows Eqn.~\eqref{eqn:CImagingModel2} to describe USCT imaging systems in which the measurement aperture varies with emitter location. 
Here and throughout the manuscript, we will refer to the process of firing one acoustic pulse and acquiring
 the corresponding wavefield data as one data acquisition indexed by $m$. 
%Also, $c(\mathbf r)$ will be referred to as the object function. 
%Note that in Eqn.~\eqref{eqn:CImagingModel2}, the explicit form of the imaging operator $\mathcal H^{\rm c}$ is determined by the sought-after object function $c(\mathbf r)$. 
% Unlike many linear imaging models \cite{BarrettBook}, in which an object function is mapped to a data function, our model (see Eqn.~\eqref{eqn:CImagingModel2}) formulates the object function as a distributed parameter that determines the linear mapping from a known function to the data function. 
The  USCT reconstruction problem in its continuous form
 is to estimate the sound speed distribution $c(\mathbf r)$  
by use of Eqn. (\ref{eqn:CImagingModel2}) and the
data functions $\{g_m(\mathbf r, t)\}_{m=0}^{M-1}$.

\subsection{USCT imaging model in its discrete forms}

A digital imaging system is accurately described by a continuous-to-discrete (C-D) imaging model, 
which is typically approximated in practice by a discrete-to-discrete (D-D) imaging
 model to facilitate the application of iterative image reconstruction algorithms. 
A C-D description of the USCT imaging system is provided in Appendix \ref{Sect:C-DModel}. 
Below, a D-D imaging model for waveform-based  USCT is presented. 
This imaging model will be employed subsequently in the development of the WISE method in Section \ref{Sect:WISE}.

% In order to apply iterative, also known as optimization-based, image reconstruction algorithms,
% a D-D imaging model is typically required \cite{BarrettBook}.
Construction of a D-D imaging model requires the introduction of a finite-dimensional approximate representations of
the functions $c(\mathbf r)$ and $s_m(\mathbf r,t)$, which will be denoted by the vectors $\mathbf c  \in\mathbb R^{N}$
and $\mathbf s_m \in\mathbb R^{NL}$.
Here, $N$ and $L$ denote the number of spatial and temporal
 samples, respectively, employed by the numerical wave equation solver.
%Unlike for many integral transform-based imaging models \cite{BarrettBook,Wernickbookchap21}, construction of a D-D waveform-based USCT imaging model is largely restricted by the numerical method employed to solve Eqns.~\eqref{eqn:WaveEqn} and \eqref{eqn:AdjWaveEqn}.
% referred to as a numerical wave equation solver (\rd{numerical solver}).
In waveform-based USCT, the way in which $c(\mathbf r)$ and $s_m(\mathbf r, t)$ are discretized to form
$\mathbf c$ and $\mathbf s_m$  is dictated by the 
numerical method employed to solve the acoustic wave equation.
% referred to as a \rd {numerical solver}.
In this study, we employ a pseudospectral  k-space method \cite{MortonBook:2005,Mast01:2ndKSpace,Tabei02:1stKSpace}.
%The pseudospectral method employs a uniform Cartesian grid in space.
Accordingly,  $c(\mathbf r)$ and  $s_m(\mathbf r, t)$ are sampled on Cartesian grid points as
\begin{equation}
  [\mathbf c]_n =  c(\mathbf r_n), \quad\text{and}\quad
  [\mathbf s_m]_{nL+l} =  s_m(\mathbf r_n, l\Delta^t),
  \quad {\rm for}\quad
  \substack{ n=0,1,\cdots,N-1\\
             l=0,1,\cdots,L-1
      },
\end{equation}
where % $N$ denotes the total number of grid points 
$\Delta^t$ denotes the temporal sampling interval
and
$\mathbf r_n$  denotes the location of the $n$-th point.
% and $\Delta^{\rm s}$ is the grid spacing of the Cartesian grid.

For a given $\mathbf c$ and $\mathbf s_m$, the pseudospectral k-space method can be described in operator form as
\begin{equation}\label{eqn:WaveDis}
  \mathbf p^{\rm a}_m = \mathbf H^{\rm c} \mathbf s_m,
\end{equation}
where the matrix $\mathbf H^{\rm c}$ is of dimension $NL\times NL$ and represents a discrete approximation of the wave operator $\mathcal H^{\rm c}$ defined in Eqn.~\eqref{eqn:CImagingModel1},
and the vector $\mathbf p^{\rm a}_m$ represents the estimated pressure data at the grid
point locations and
 has the same dimension as $\mathbf s_m$.
The superscript `a' indicates that these values are approximate, i.e.,
 $[\mathbf p^{\rm a}_m]_{nL+l} \approx  p_m(\mathbf r_n, l\Delta^t)$.
%The matrix $\mathbf H^{\rm c}$ is of dimension $NL\times NL$, and represents a discrete approximation of the wave operator $\mathcal H^{\rm c}$ defined in Eqn.~\eqref{eqn:CImagingModel1}.
We refer the readers to \cite{MortonBook:2005,Mast01:2ndKSpace,Tabei02:1stKSpace} for
 additional details regarding the pseudospectral k-space method.

Because the pseudospectral k-space method yields sampled values
of the pressure data on a Cartesian grid,
a sampling matrix $\mathbf M_m$ is introduced to model the USCT data acquisition process as
% in order to resemble a practical data acquisition, a matrix $\mathbf M_m$ is introduced to generate the data vector $\mathbf g_m$ by sampling/interpolating the output of the numerical wave equation solver $\mathbf p_m^{\rm a}$ as
\begin{equation}\label{eqn:DDFwd}
  \mathbf g^{\rm a}_m = \mathbf M_m \mathbf p^{\rm a}_m
 \equiv \mathbf M_m \mathbf H^{\rm c} \mathbf s_m.
\end{equation}
Here, the $N^{\rm rec}L\times NL$ sampling matrix $\mathbf M_m$ extracts
 the pressure data corresponding to the receiver locations
on the measurement aperture $\Omega_m$,
with $N^{\rm rec}$ denoting the number of receivers.
The vector $\mathbf g^{\rm a}_m$ denotes the predicted data that approximates the true measurements. % vector $\mathbf g_m$ in Eqn.~\eqref{eqn:CDModel}.
In principle, $\mathbf M_m$ can be constructed to incorporate transducer characteristics, such as finite aperture size and temporal delays.
For simplicity, we assume that the transducers are point-like in this study.
When the receiver and grid point locations do not coincide, an interpolation method is required.
As an example, when a nearest-neighbor interpolation method is employed, the elements of $\mathbf M_m$ are defined as
\begin{equation}\label{eqn:Mm}
  [\mathbf M_m]_{n^{\rm rec}L+l,nL+l} =
 \left\{\begin{array}{ll} 1, &
   {\rm for}\quad n=\mathcal I_m (n^{\rm rec}),\\
   0, & \text{otherwise},
  \end{array}
     \right.
\end{equation}
where $[\mathbf M_m]_{n^{\rm rec}L+l,nL+l}$ denotes the element of $\mathbf M_m$ at the $(n^{\rm rec}L+l)$-th row and the $(nL+l)$-th column,
and $\mathcal I_m(n^{\rm rec})$ denotes the index of the grid point that is closest to
% the $n^{\rm rec}$-th transducer at the $m$-th data acquisition, i.e.,
$\mathbf r(m, n^{\rm rec})$.
Here, $\mathbf r(m, n^{\rm rec})$ denotes the location of the $n^{\rm rec}$-th receiver in the $m$-th data acquisition.
Equation~\eqref{eqn:DDFwd} represents the D-D imaging model that will be employed in the remainder of this study.
%Note that because of the dependence of $\mathbf M_m$ on $m$,  a varying detection geometry among data acquisitions
%can be described by use of this model.

\section {Waveform inversion with source encoding for USCT}
\label{Sect:WISE}

\subsection{Sequential waveform inversion in its discrete form}
\label{SubSect:SeqWaveInv}

A conventional waveform inversion method that does not utilize source encoding
will be employed as a reference for  the developed WISE method and is briefly
described below.
Like other conventional approaches, this  method
 sequentially processes the data acquisitions  $\mathbf g_m$
 for  $m=0, 1,\cdots, M-1$ at each iteration of the associated algorithm.
As such, we will  refer to the conventional method as a sequential waveform inversion method.

A sequential waveform inversion method can be formulated as a non-linear numerical  optimization problem:
\begin{equation}\label{eqn:DCostFunc}
  \hat{\mathbf c} = \arg\min_{\mathbf c} \{\mathcal F(\mathbf c) + \beta \mathcal R(\mathbf c)\}, 
\end{equation}
where $\mathcal F(\mathbf c)$, $\mathcal R(\mathbf c)$, and $\beta$ denote the data fidelity term, the penalty term, and the regularization parameter, respectively. 
The data fidelity term $\mathcal F(\mathbf c)$ is defined as  a sum
of squared $\ell^2$-norms of the data residuals corresponding
to all data acquisitions as:
\begin{equation}\label{eqn:DDataMisfit}
  \mathcal F(\mathbf c) = \frac{1}{2}
      \sum_{m=0}^{M-1} \Vert \underline{\mathbf g_m} - 
     \mathbf M_m \mathbf H^{\rm c} \mathbf s_m \Vert^2, 
\end{equation}
where $\underline{\mathbf g_m}\in\mathbb R^{N^{\rm rec}L}$ denotes the measured data vector at the $m$-th data acquisition. 
The choice of the penalty term will be addressed in Section \ref{Sect:SimuDesc}.

\iffalse 
As stated in Section \ref{Sect:BckGround}, in order to solve the optimization problem defined in Eqn.~\eqref{eqn:DCostFunc}, it is crucial to calculate the gradient of $\mathcal F(\mathbf c)$ with respect to $\mathbf c$, which will be denoted by $\mathbf J$. 
To our knowledge, approximating $\mathbf J$ by a discretized version of $\nabla_{\rm c}\mathcal F^{\rm CC}$ serves as the most computational efficient strategy.
This strategy will be employed in this study as 
% In this study, we approximate the gradient using a numerical approximation of $\nabla_{\rm c}\mathcal F^{\rm CC}$ in Eqn.~\eqref{eqn:Frecht} given as 
\fi

The gradient of $\mathcal F(\mathbf c)$ with respect to $\mathbf c$, denoted by $\mathbf J$, will be computed by
discretizing an expression for the  Fr\'echet derivative that is derived assuming
a continuous form of Eqn.\ (\ref{eqn:DDataMisfit}). The Fr\'echet derivative  is described in Appendix \ref{Sect:Frechet}.
Namely, the gradient is approximated as
\begin{equation}\label{eqn:DGradient}
  [\mathbf J]_n    \equiv  \sum_{m=0}^{M-1} [\mathbf J_m]_n
      \approx
     \frac{1}{ [\mathbf c]_n^3} 
      \sum_{m=0}^{M-1}
      \sum_{l=1}^{L-2} 
       [\mathbf q^{\rm a}_m]_{nL+(L-l)} 
       \frac{  [\mathbf p^{\rm a}_m]_{nL+l-1}
              -2[\mathbf p^{\rm a}_m]_{nL+l}
              +[\mathbf p^{\rm a}_m]_{nL+l+1} }
           {\Delta^{\rm t}} ,
\end{equation}
where $\mathbf J_m$  denotes the gradient of $\frac{1}{2}\Vert\underline{\mathbf g_m}-\mathbf M_m\mathbf H^{\rm c}\mathbf s_m\Vert^2$ with respect to $\mathbf c$ 
and the vector  $\mathbf q^{\rm a}_m$ contains samples that approximate adjoint wavefield $q_m(\mathbf r, t)$
that satisfies Eqn.~\eqref{eqn:AdjWaveEqn} in Appendix \ref{Sect:Frechet}.
%, i.e.,  $[\mathbf q^{\rm a}_m]_{nL+l}\approx q_m(\mathbf r_n, t_l)$.
By use of the  pseudospectral k-space method,  $\mathbf q^{\rm a}_m$ can be calculated as
\begin{equation}\label{eqn:DAdj}
  \mathbf q^{\rm a}_m = \frac{1}{4\pi}\mathbf H^{\rm c} \boldsymbol \tau_m,
\end{equation}
where 
\begin{equation}\label{eqn:AdjointSource}
[\boldsymbol\tau_m]_{nL+l} = \left\{\begin{array}{ll}
[\mathbf g^{\rm a}_m - \underline{\mathbf g_m}]_{\mathcal I_m^{\rm -1}(n)L+(L-l)},& {\rm if}\,\,
 n\in\mathbb N_m, \\ 
0, & {\rm otherwise} \end{array}
\right. .
\end{equation}
Here, $\mathbb N_m = \{n: \mathcal I_m(n^{\rm rec}), n^{\rm rec}=0,1,\cdots,N^{\rm rec}-1\}$, 
and $\mathcal I_m^{-1}$ denotes the inverse mapping of $\mathcal I_m$.

Given the explicit form of $\mathbf J$ in Eqn.~\eqref{eqn:DGradient}, a variety of optimization algorithms can be employed to solve Eqn.~\eqref{eqn:DCostFunc} \cite{Nash96Book}. 
%When a gradient descent algorithm is employed,
% the sequential waveform inversion method is given by
 Algorithm 1 describes a gradient descent-based sequential waveform inversion method. 
Note that at every algorithmic iteration, the sequential waveform inversion method updates the sound speed estimate only once using the gradient $\mathbf J$ accumulated over all $\mathbf J_m$ for $m=0, 1, \cdots, M-1$.
This is unlike the Kaczmarz method---also known as the algebraic reconstruction technique \cite{Chew10,KakBook,Hesse2013:USCT}---that updates the sound speed estimate multiple times in one algorithmic iteration.
 In Line-10 of Algorithm 1, $\mathbf J^{\rm R}$ denotes the gradient of $\mathcal R(\mathbf c)$ with respect to $\mathbf c$. 
\begin{algorithm}[H]
\caption{\label{Alg:DGradDesc}
Gradient descent-based sequential waveform inversion. 
}
\algsetup{indent=2em}
\begin{algorithmic}[1]
\REQUIRE $\{\underline{\mathbf g_m}\}$, $\{\mathbf s_m\}$, $\mathbf c ^{(0)}$
\ENSURE $\hat{\mathbf c}$
  \STATE {$k \gets 0$} \COMMENT{$k$ is the number of algorithm iteration.}
  \WHILE {stopping criterion is not satisfied}
  \STATE{$k \gets k+1$}
  \STATE{$\mathbf J \gets \mathbf 0$}
  \FOR{$m:=0$ \TO $M-1$} 
  \STATE{$\mathbf p^{\rm a}_m \gets \mathbf H^{\rm c} \mathbf s_m$}
  \COMMENT{$m$ is the index of the emitter.} 
  \STATE{$\mathbf q^{\rm a}_m \gets \mathbf H^{\rm c} \boldsymbol\tau_m$}
  \COMMENT{$\boldsymbol \tau_m$ is calculated via Eqn.~\eqref{eqn:AdjointSource}.}
  \STATE{$\mathbf J \gets \mathbf J + \mathbf J_m$}
  \COMMENT{$\mathbf J_m$ is calculated via Eqn.~\eqref{eqn:DGradient}.}
  \ENDFOR
  \STATE {$\mathbf J \gets \mathbf J + \beta\mathbf J^{\rm R}$}
  \STATE {Determine step size $\lambda$ via a line search}
  \STATE {$\mathbf c^{(k)} \gets \mathbf c^{(k-1)} - \lambda \mathbf J$}
  \ENDWHILE
  \STATE {$\hat{\mathbf c} = \mathbf c^{(k)}$}
\end{algorithmic}
\end{algorithm}

In Algorithm \ref{Alg:DGradDesc}, $\mathbf H^{\rm c}$ is the most computationally burdensome operator, representing one run of the wave equation solver.
Note that it  appears in Lines-6, -7, and -11.
Because Lines-6 and -7 have to be executed $M$ times to process all of the data acquisitions, the
 wave equation solver has to be executed
  at least $(2M+1)$ times at each algorithm iteration.   
The line search in Line-11 searches for a step size along the direction of $-\mathbf J$ so that the cost function is reduced by use of a classic trial-and-error approach \cite{Nash96Book}. 
Note that, in general, the line search will require more than one application of $\mathbf H^{\rm c}$, so 
 $(2M+1)$ represents a lower bound on the total number of wave equation solver runs per iteration.

\subsection{Stochastic optimization-based waveform inversion with source encoding (WISE)}
\label{SubSect:WISE}

%In Algorithm \ref{Alg:DGradDesc}, $\mathbf H^{\rm c}$ is the most computationally burdensome operator, representing one run of the \rd{numerical solver}. 
%Note that it  appears in Lines-6, -7, and -11.
%Because Lines-6 and -7 have to be executed $M$ to process all of the data acquisitions, the \rd{numerical solver} has to be executed  $(2M+1)$ times at
%each algorithm iteration, which presents a computational challenge in practice. 
\iffalse
In order to alleviate the large computational burden presented by sequential
waveform inversion methods (e.g., Algorithm 1), a source encoding method, originated in the geophysics literature \cite{Krebs:SE,Romero00:PE}, has been investigated through breast USCT computer-simulation studies \cite{Los12:SOSPE}.  
\fi 
In order to alleviate the large computational burden  presented by sequential
waveform inversion methods (e.g., Algorithm 1), a source encoding method has been proposed \cite{Krebs:SE,Romero00:PE,Los12:SOSPE}.  
This method has been formulated as a stochastic optimization problem and solved by various stochastic gradient-based algorithms \cite{Herrmann2012:SE,Herrmann13:GaussCode}. 
In this section, we adapt the stochastic optimization-based formulation in \cite{Herrmann2012:SE}
 to find the solution of Eqn.~\eqref{eqn:DCostFunc}. 
%The method will be referred to as a waveform inversion with source encoding (WISE) method. 
\begin{algorithm}[b!]
\caption{\label{Alg:WISE}
Waveform inversion with source encoding (WISE) algorithm.
}
\algsetup{indent=2em}
\begin{algorithmic}[1]
\REQUIRE $\{\underline{\mathbf g_m}\}$, $\{\mathbf s_m\}$, $\mathbf c ^{(0)}$
\ENSURE $\hat{\mathbf c}$
  \STATE{$k \gets 0$} \COMMENT{$k$ is the number of algorithm iteration}
  \WHILE {stopping criterion is not satisfied}
  \STATE{$k \gets k+1$}
  \STATE{Draw elements of $\mathbf w$ from independent and identical Rademacher distribution.}
  \STATE {$\mathbf p^{\rm w} \gets \mathbf H^c \mathbf s^{\rm w}$}
  \COMMENT{$\mathbf s^{\rm w}$ is calculated via Eqn.~\eqref{eqn:Coding}. }
  \STATE {$\mathbf q^{\rm w} \gets \mathbf H^c \boldsymbol \tau^{\rm w}$}
  \COMMENT{$\boldsymbol \tau^{\rm w}$ is calculated via Eqn.~\eqref{eqn:tauw}.}
  \STATE {$\mathbf J\gets \mathbf J^{\rm w}+\beta\mathbf J^{\rm R}$}
  \COMMENT{$\mathbf J^{\rm w}$ is calculated via Eqn.~\eqref{eqn:CalJ}}
  \STATE {Determine step size $\lambda$ by use of line search }
  \STATE {$\mathbf c^{(k)} \gets \mathbf c^{(k-1)} - \lambda \mathbf J$}
  % \STATE {$\boldsymbol \eta \gets \mathbf c^{(k-1)} - \lambda \mathbf J^{\rm w}$}
  % \STATE {$\mathbf c^{(k)} \gets \frac{1}{k+1}(\boldsymbol\eta  + k\mathbf c^{(k-1)})$}
  \ENDWHILE
  \STATE {$\hat{\mathbf c} = \mathbf c^{(k)}$}
\end{algorithmic}
\end{algorithm}

The WISE method seeks to minimize the same  cost function as the
sequential waveform inversion method, namely,  Eqn.~\eqref{eqn:DCostFunc}.
However, to accomplish this,  the data fidelity term in Eqn.~\eqref{eqn:DDataMisfit} is reformulated as the expectation of a random quantity as \cite{Krebs:SE,Romero00:PE,Herrmann13:GaussCode,Herrmann2012:SE,ArXiv:Haber14,ArXiv:Ascher14}
\begin{equation}\label{eqn:StoCost}
  \mathcal F_s(\mathbf c) = \mathbf E_{\mathbf w} \big\{ \frac{1}{2}\Vert 
    \underline{\mathbf g}^{\rm w} 
  - \mathbf M
      \mathbf H^{\rm c} \mathbf s^{\rm w}
      \Vert^2  \big\},
\end{equation}
where $\mathbf E_{\mathbf w}$ denotes the expectation operator with respect to the
 random source encoding vector $\mathbf w\in\mathbb R^{M}$,  
$\mathbf M \equiv \mathbf M_m$ is the sampling matrix that is assumed to be identical for $m=0, 1, \cdots, M-1$, 
and $\underline{\mathbf g}^{\rm w}$ and  $\mathbf s^{\rm w}$ denote the $\mathbf w$-encoded data  and source vectors, defined as 
\begin{equation}\label{eqn:Coding}
  \underline{\mathbf g}^{\rm w} = 
    \sum_{m=0}^{M-1} [{\mathbf w}]_m \underline {\mathbf g_m}, 
 \quad \rm {and} \quad 
  {\mathbf s}^{\rm w} = 
  \sum_{m=0}^{M-1} [{\mathbf w}]_m {\mathbf s}_m, 
\end{equation}
 respectively. 
It has been demonstrated that  Eqns.~\eqref{eqn:DDataMisfit} and \eqref{eqn:StoCost}
are mathematically equivalent when
 $\mathbf w$ possesses a zero mean and an identity covariance matrix \cite{Herrmann2012:SE,ArXiv:Haber14,ArXiv:Ascher14}.
%A stochastic gradient descent (SGD) algorithm can be employed to solve the stochastic optimization problem
% employed to solve Eqn.~\eqref{eqn:DCostFunc} with $\mathcal F(\mathbf c)$ given in Eqn.~\eqref{eqn:StoCost}, rewritten as  \cite{Zhang04:SGD,Nemirovski09:SA,Herrmann2012:SE}
In this case, the optimization problem whose solution specifies the sound speed estimate can be re-expressed in a 
stochastic framework as 
\begin{equation}\label{eqn:WISE}
  \hat {\mathbf c} = \arg\min_{\mathbf c}
  \mathbf E_{\mathbf w} \big\{ \frac{1}{2}\Vert
    \underline{\mathbf g}^{\rm w}
  - \mathbf M
      \mathbf H^{\rm c} \mathbf s^{\rm w}
      \Vert^2  \big\}
   + \beta \mathcal R(\mathbf c),
\end{equation}
which we refer to as the  waveform inversion with source encoding (WISE) method.  An implementation of the WISE method
that utilizes the stochastic gradient descent algorithm  is summarized in Algorithm \ref{Alg:WISE}.

In Algorithm \ref{Alg:WISE}, the wave equation solver needs to be run
 one time in each of  Lines-5 and 6.
In the line search to determine the step size in Line 8,
 the wave equation solver needs to be run at least one time, but in general will
require a small number of additional runs, just as in Algorithm \ref{Alg:DGradDesc}. 
  Accordingly, the lower bound
on the number of required wave equation solver runs per iteration is 3, as opposed to $(2M+1)$ 
for the conventional sequential waveform inversion method described
by Algorithm 1.
As demonstrated in geophysics applications \cite{Krebs:SE,Romero00:PE,Herrmann13:GaussCode} and the breast imaging studies below,
the WISE method provides a substantial reduction in reconstruction times over use of the standard
sequential waveform inversion method.
In Line-7, $\mathbf J^{\rm w}$ can be calculated analogously to Eqn.~\eqref{eqn:DGradient} as 
\begin{equation}\label{eqn:CalJ}
  [\mathbf J^{\rm w}]_n    \approx
     \frac{1}{ [\mathbf c]_n^3} 
      \sum_{l=1}^{L-2} 
       [\mathbf q^{\rm w}]_{nL+(L-l)} 
       \frac{  [\mathbf p^{\rm w}]_{nL+l-1}
              -2[\mathbf p^{\rm w}]_{nL+l}
              +[\mathbf p^{\rm w}]_{nL+l+1} }
           {\Delta^{\rm t}} ,
\end{equation}
where $\mathbf p^{\rm w}=\mathbf H^{\rm c}\mathbf s^{\rm w}$ and 
$\mathbf q^{\rm w}=\mathbf H^{\rm c}\boldsymbol \tau^{\rm w}$ with 
$\boldsymbol \tau^{\rm w} \in \mathbb R^{NL}$ calculated by 
 \begin{equation}\label{eqn:tauw}
[\boldsymbol\tau^{\rm w}]_{nL+l} = \left\{\begin{array}{ll}
[\mathbf M\mathbf p^{\rm w} - \underline{\mathbf g}^{\rm w}]_{\mathcal I^{\rm -1}(n)L+(L-l)},& {\rm if}\,\,
 n\in\mathbb N, \\ 
0, & {\rm otherwise} \end{array}
\right. .
\end{equation}
Here, we drop the subscript $m$ of both $\mathcal I^{\rm -1}(n)$ and $\mathbb N$ because we assume $\mathbf M$ to be identical for all data acquisitions.
Various probability density functions have been proposed to describe the source encoding vector $\mathbf w$ \cite{Krebs:SE,Romero00:PE,Herrmann13:GaussCode}.
In this study, we employed a Rademacher distribution as suggested by \cite{Krebs:SE},
in which case each element of $\mathbf w$ had a $50\%$ chance of being either $+1$ or $-1$.

\iffalse
\subsection{Heuristic data replacement for breast imaging}
These cases can be readily described by the D-D imaging model introduced in Section III-A.
Consider a circular transducer array containing uniformly distributed $M$ elements as depicted in Fig.~\ref{fig:Geo}. 
Each element is indexed by $m$ for $m=0,1,\cdots, M-1$, and is sequentially driven to emit a short acoustic pulse. 
When the $m$-th element emits the acoustic pulse, only the opposite $N^{\rm rec}$ transducers reliably measure acoustic signals while other transducers are contaminated with the emitter's interference.
Collecting the reliable measurements results in a data vector  denoted by $\underline{\mathbf g_m}$ with dimensions of $N^{\rm rec}L$, which can be approximated by use of the D-D imaging model in Eqn.~\eqref{eqn:DDFwd}. 
It is obvious that $\mathbf M_m$ will be different for all $m$'s since $\mathcal I_m(n^{\rm rec})\neq\mathcal I_{m'}(n^{\rm rec})$ when $m\neq m'$ in Eqn.~\eqref{eqn:Mm}. 
\fi

\section{Description of computer-simulation studies}
\label{Sect:SimuDesc}
Two-dimensional computer-simulation studies were conducted to validate the WISE method for breast sound speed imaging and demonstrate its computational advantage over the standard sequential waveform inversion method. 
\subsection{Measurement geometry}

A circular measurement geometry was chosen to emulate a previously reported USCT breast imaging system \cite{Duric2007:SOSWave,Duric13:ClinReflect,Duric13:SoftVue}. 
As depicted in Fig.~\ref{fig:Geo}, $256$ ultrasonic transducers were uniformly distributed on a ring of radius $110$ mm. 
The generation of one USCT data set consisted of $M=256$ sequential data acquisitions.
In each data acquisition,  one emitter produced an acoustic pulse.
The acoustic pulse was numerically propagated through the breast phantom and the resulting wavefield data were recorded by all transducers in the array as described below.
Note that the location of the emitter in every data acquisition was different from those in other acquisitions, while the locations of receivers were identical for all acquisitions. 

\subsection{Numerical breast phantom}
\label{Subsect:Phantom}
\begin{figure}[h]
\centering
\subfloat[]{\includegraphics[height=6cm]{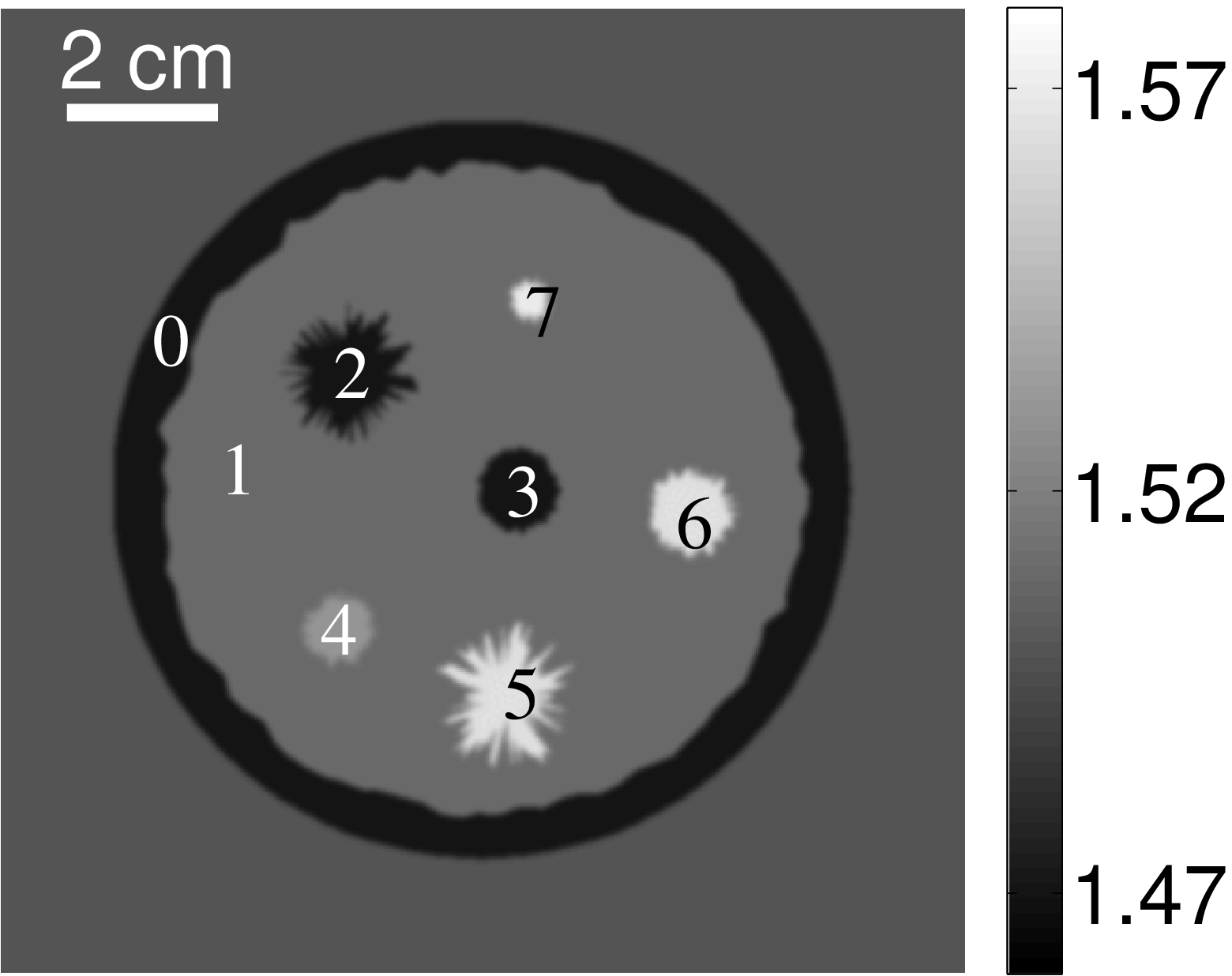}}
\hskip 0.5cm
\subfloat[]{\includegraphics[height=6cm]{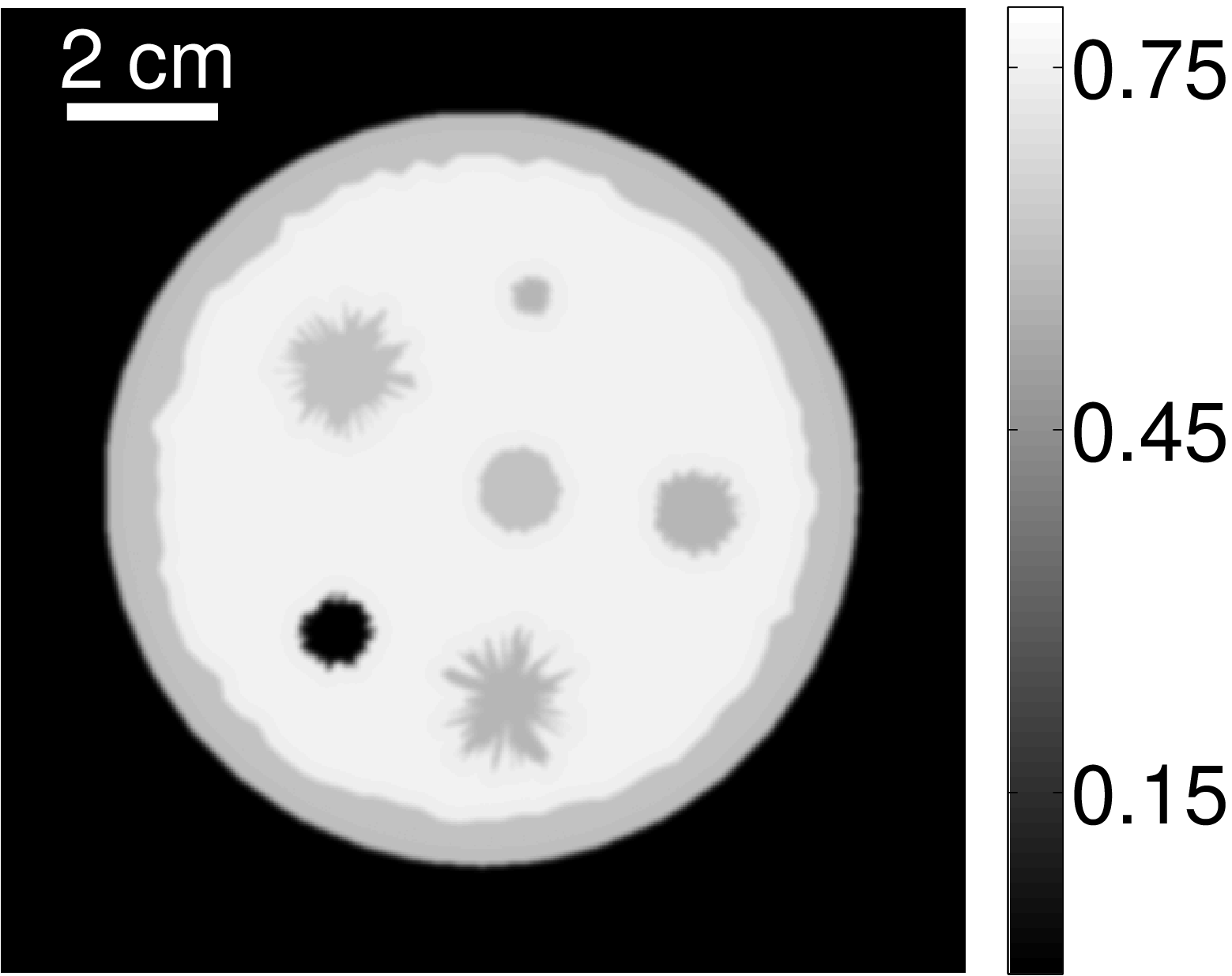}}
\caption{\label{fig:NumPhantom}
(a) Sound speed map [mm$\cdot\mu$s$^{-1}$] and (b) acoustic attenuation slope map [dB$\cdot($MHz$)^{-y}\cdot$cm$^{-1}$] of the numerical breast phantom. 
}
\end{figure}

A numerical breast phantom of diameter $98$ mm was employed. 
The phantom was composed of $8$ structures representing adipose tissues, parenchymal breast tissues, cysts, benign tumors, and malignant tumors, as shown in Fig.~\ref{fig:NumPhantom}. 
For simplicity, the acoustic attenuation of all tissues was described by a power law with a fixed exponent $y = 1.5$ \cite{Szabo04Book}. 
The corresponding sound speed and the attenuation slope values are listed in TABLE \ref{tab:NumPhantom} \cite{Szabo04Book,Glide07:MedPhy,Li08:SPIE}.
Both the sound speed and the attenuation slope distributions in Fig.~\ref{fig:NumPhantom} were sampled on a uniform Cartesian grid with spacing $\Delta^{\rm s} = 0.25$ mm. 
The finest structure (indexed by $7$ in Fig.~\ref{fig:NumPhantom}-(a)) was of diameter $3.75$ mm. 
% For simplicity, we let the background water $c_0=1.5$ mm/$\mu$s and no acoustic absorption, and let all tissues have a power law exponent $y = 1.5$ \cite{Szabo04Book}. 

\subsection{Simulation of the measurement data}
\label{Subsect:GenData}

\subsubsection{First-order numerical wave equation solver}
Acoustic wave propagation in acoustically absorbing media was modeled by three coupled first-order partial differential equations \cite{CoxIP10:Attenuation}: 
\begin{subequations}\label{eqn:AttWaveEqn}
\begin{align}
&\frac{\partial}{\partial t} \mathbf u(\mathbf r, t) = - \nabla p(\mathbf r, t) \\ 
&\frac{\partial}{\partial t} \mathbf \rho(\mathbf r, t) = - \nabla \cdot \mathbf u(\mathbf r, t) 
+ 4\pi\int_0^t d t' s(\mathbf r, t')\\
&p(\mathbf r, t) = c^2(\mathbf r)  
  \big[ 1 + \tau (\mathbf r) \frac{\partial}{\partial t} (-\nabla^2)^{y/2-1} 
    + \eta (\mathbf r) (-\nabla^2)^{(y+1)/2-1} \big]
        \rho(\mathbf r, t),
\end{align}
\end{subequations}
where $\mathbf u(\mathbf r, t)$, $p(\mathbf r, t)$, and $\rho(\mathbf r)$ denote the acoustic particle velocity, the acoustic pressure, and the acoustic density, respectively. 
The functions $\tau(\mathbf r)$ and $\eta(\mathbf r)$ describe acoustic absorption and dispersion during the wave propagation \cite{CoxIP10:Attenuation}: 
\begin{equation}
\tau(\mathbf r) = -2\alpha_0(\mathbf r) c_0(\mathbf r)^{y-1},\quad
\eta(\mathbf r) =  2\alpha_0(\mathbf r) c_0(\mathbf r)^y \tan(\pi y/2),
\end{equation}
where $\alpha_0(\mathbf r)$ and $y$ are the attenuation slope and the power law exponent, respectively. 
When the medium is assumed to be lossless, i.e., $\alpha_0(\mathbf r) = 0$, it can be shown that Eqn.~\eqref{eqn:AttWaveEqn} is equivalent to Eqn.~\eqref{eqn:WaveEqn}. 

Based on Eqn.~\eqref{eqn:AttWaveEqn}, a pseudospectral k-space method was employed to simulate acoustic pressure data \cite{Tabei02:1stKSpace,CoxIP10:Attenuation}. 
%The numerical scheme will be referred to as a first-order \rd{numerical solver}. 
This method was implemented by use of a first-order numerical scheme on GPU hardware.
%The first-order \rd{numerical solver} was implemented using graphic processing units (GPUs). 
The calculation domain was of size $512\times 512$ mm$^2$, sampled on a $2048\times 2048$ uniform Cartesian grid of spacing $\Delta^{\rm s} = 0.25$ mm. 
A nearest-neighbor interpolation was employed to place all transducers on the grid points.
On a platform consisting of dual quad-core CPUs with a $3.30$ GHz clock speed, $64$ gigabytes (GB) of random-accessing memory (RAM), and a single NVIDIA Tesla K20 GPU, the first-order pseudospectral k-space method required approximately $108$ seconds to complete one forward simulation.

\subsubsection{Acoustic excitation pulse}
\begin{figure}[h]
\centering
\subfloat[]{\includegraphics[height=5.6cm]{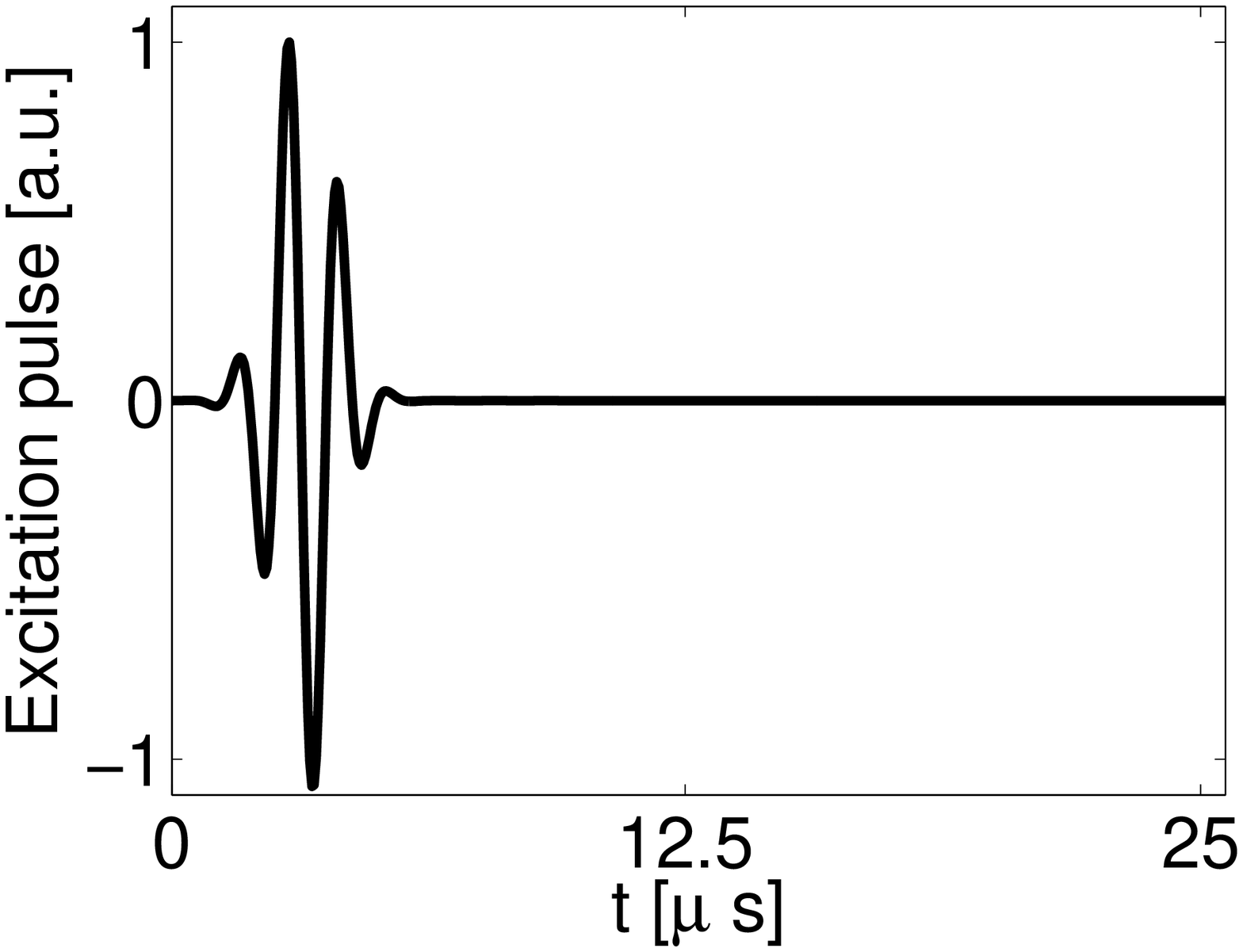}}
\subfloat[]{\includegraphics[height=5.6cm]{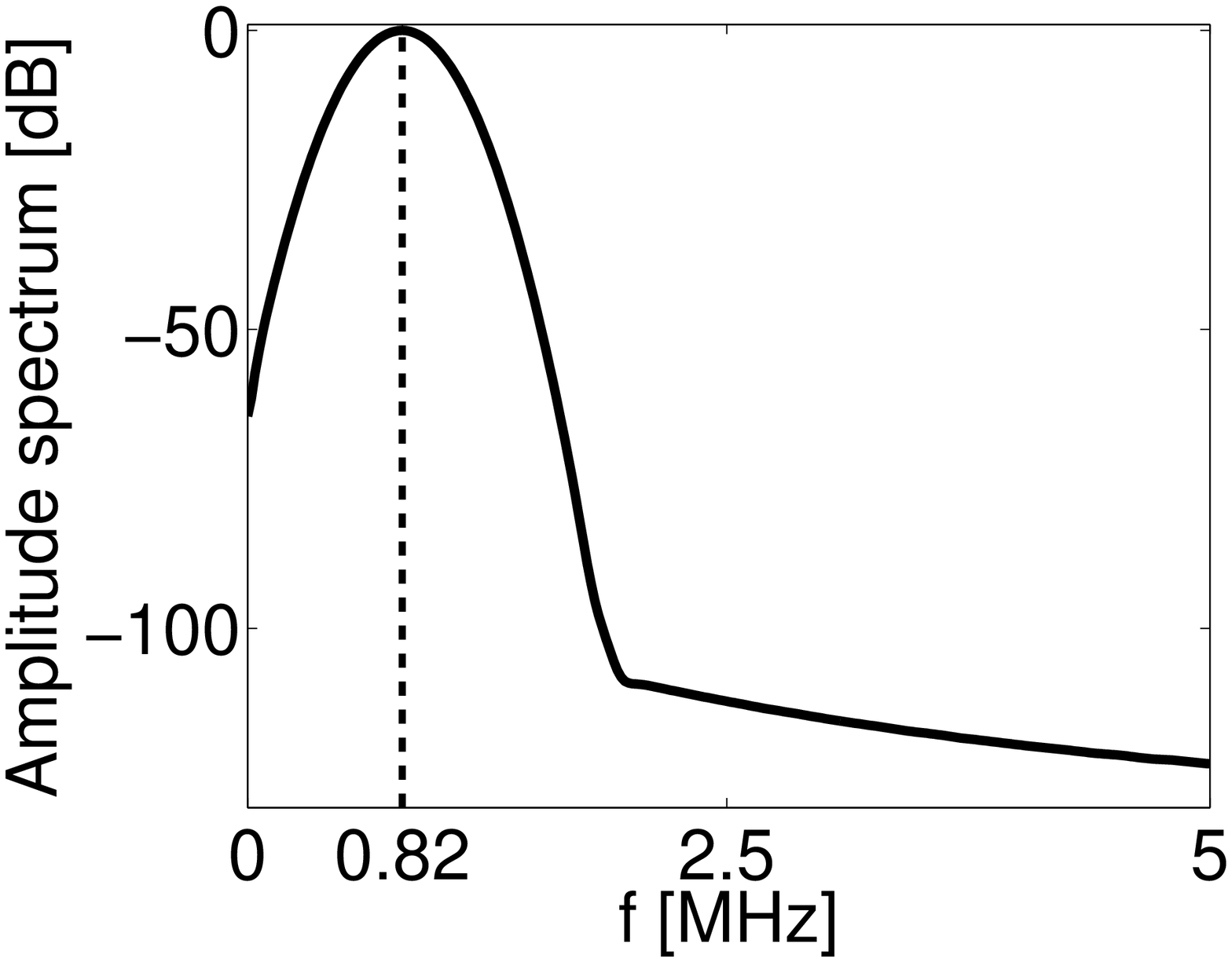}}
\caption{\label{fig:SimExctPulse} 
(a) Normalized temporal profile and 
(b) amplitude spectrum of the excitation pulse employed in
the computer-simulation studies. 
The dashed line in (b) marks the center frequency of excitation pulse at $0.82$ MHz.
}
\end{figure}

The excitation pulse employed in this study was assumed to be spatially localized at the emitter location while temporally it was a $f_{\rm c}=0.8$ MHz  sinusoidal function tapered by a Gaussian kernel with standard deviation $\sigma=0.5$ $\mu s$, i.e.,
\begin{equation}
  s_m(\mathbf r, t) = \left\{
  \begin{array}{ll}
  \exp\big(-\frac{(t-t_c)^2}{2\sigma^2}\big)\sin(2\pi f_c t), &\text{at the $m$-th emitter location} \\
       0, &\text{otherwise}, 
  \end{array}
  \right.
\end{equation}
where the constant time shift $t_{\rm c} = 3.2$ $\mu$s. 
The temporal profile and the amplitude frequency spectrum of the excitation pulse are plotted in  Fig.~\ref{fig:SimExctPulse}-(a) and -(b), respectively.
The excitation pulse contained approximately $3$ cycles.

\subsubsection{Generation of non-attenuated and attenuated noise-free data}

\begin{figure}[h]
\centering
\subfloat[]{\includegraphics[height=5.6cm]{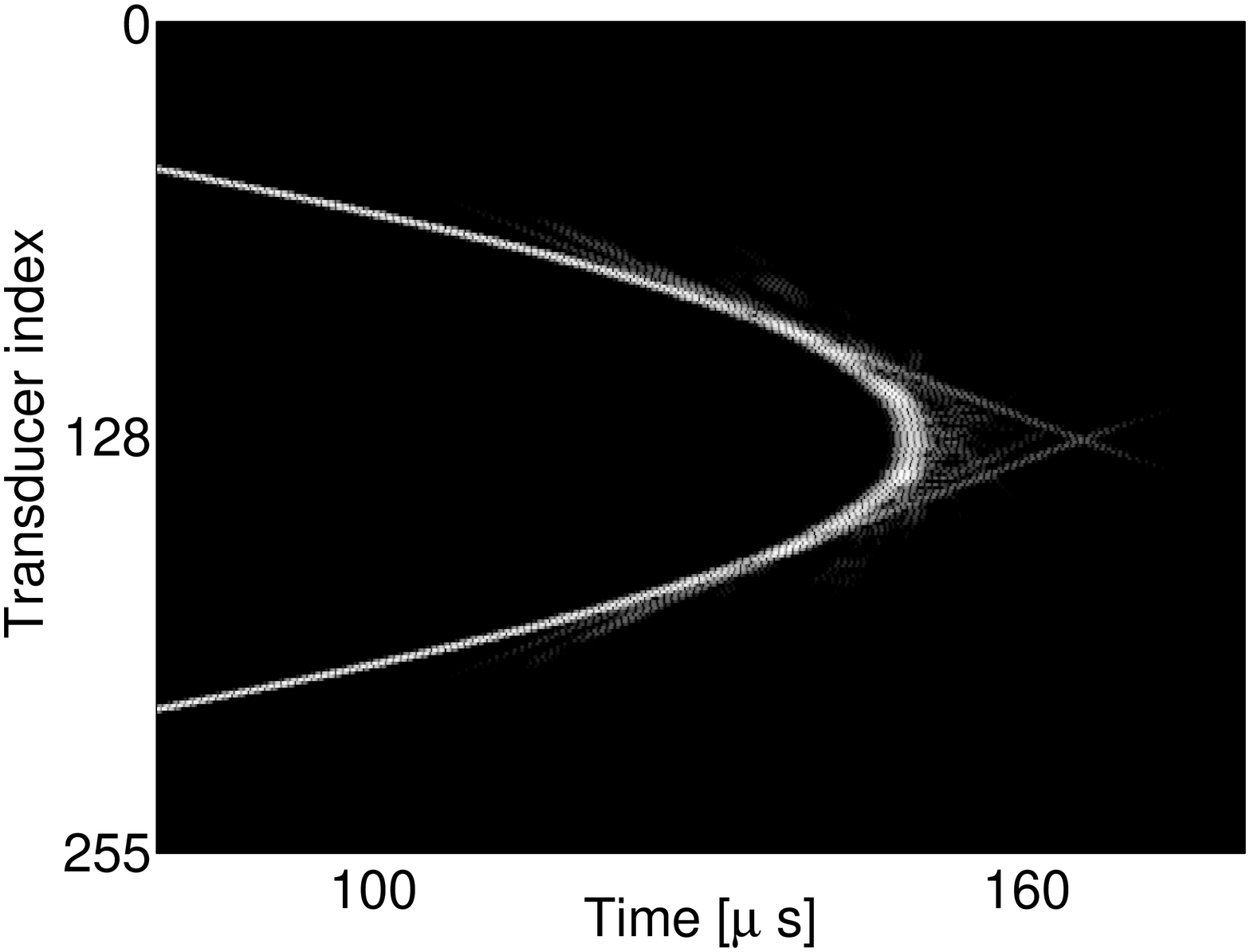}}
\subfloat[]{\includegraphics[height=5.6cm]{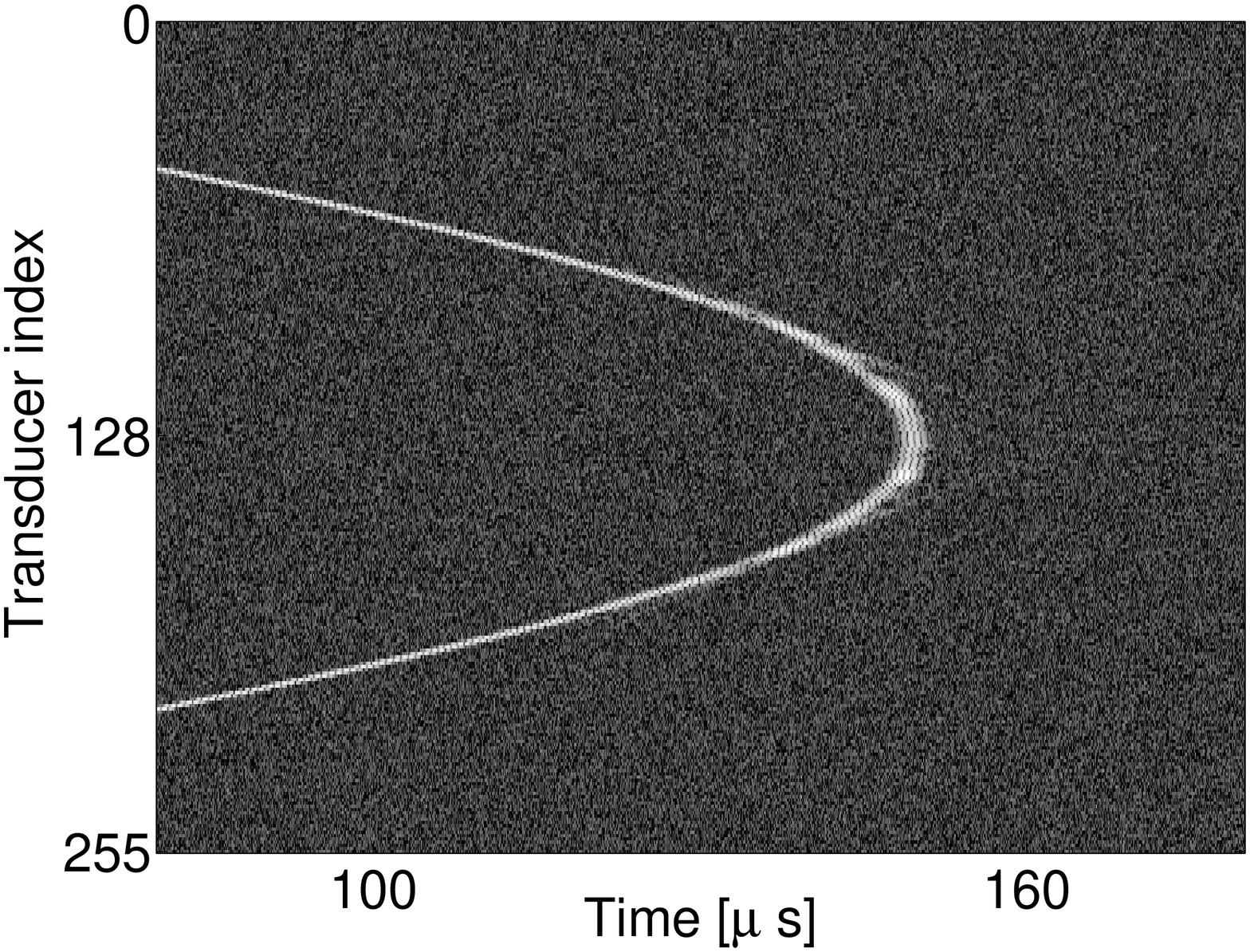}}\\
\subfloat[]{\includegraphics[height=5.6cm]{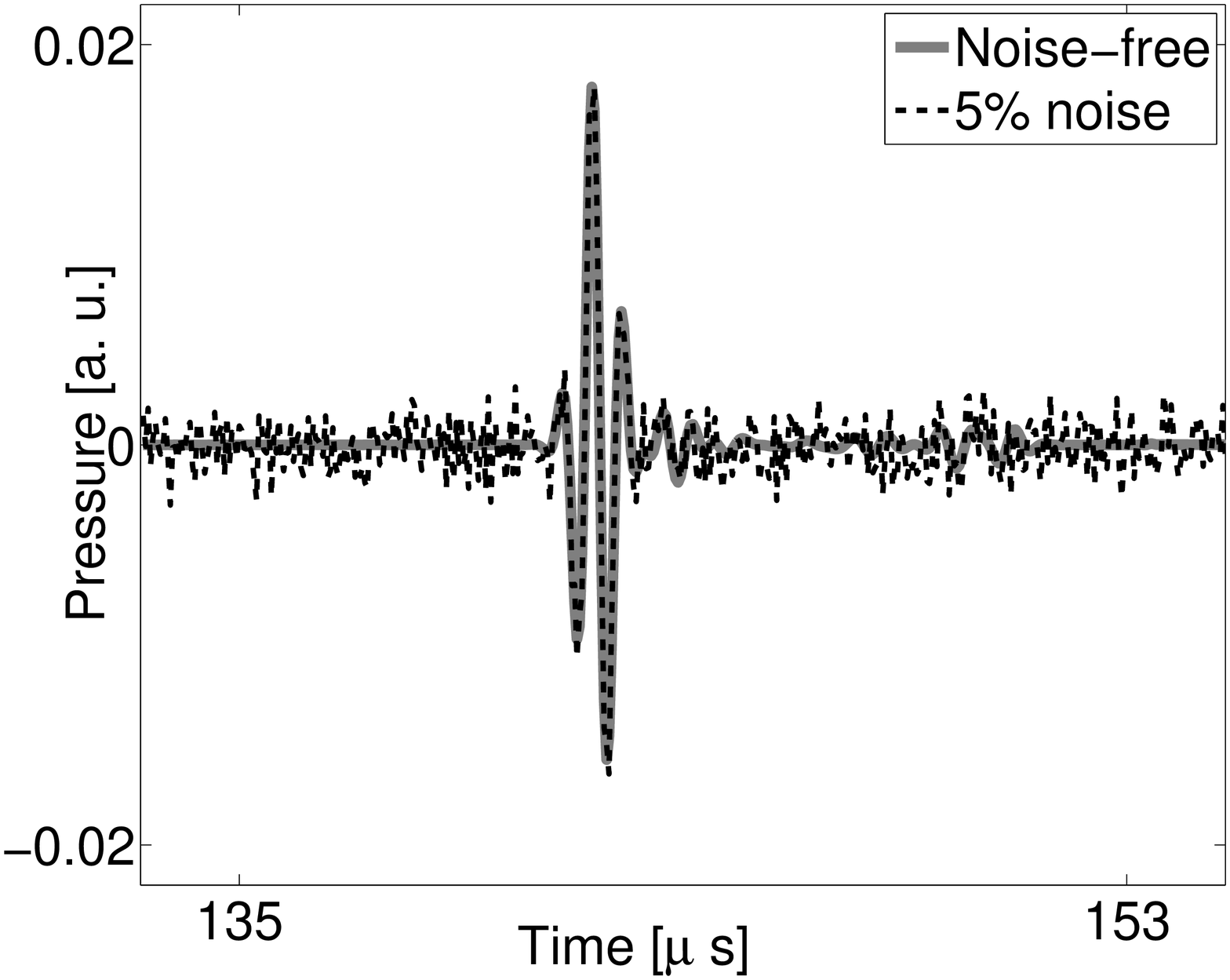}}
\caption{\label{fig:PreNoise} 
Computer-simulated (a) noise-free and (b) noisy data vectors at the $0$-th data acquisition. 
(c) Profiles of the pressure received by the $128$-th transducer. 
The grayscale window for (a) and (b) is $[-45,0]$ dB. 
}
\end{figure}

For every data acquisition (indexed by $m$), the first-order pseudospectral k-space 
method was run for $3600$ time steps with a time interval $\Delta^{\rm t} = 0.05$ $\mu$s (corresponding to a $20$ MHz sampling rate).  
Downsampling the recorded data by taking every other time sample resulted in a
 data vector $\underline{\mathbf g_m}$  (see Eqn.~\eqref{eqn:DDataMisfit}) that was effectively sampled at $10$ MHz and was of dimensions $ML$ with $M=256$ and $L=1800$. 
The data vector at the $0$-th data acquisition, $\underline{\mathbf g_0}$, is displayed as a 2D image in Fig.~\ref{fig:PreNoise}-(a). 
This undersampling procedure was introduced to avoid {\it inverse crime}  \cite{colton12:inversecrime} so that the data generation and the image reconstruction employed different numerical discretization schemes. 
% in the image reconstruction, which will be described in Section \ref{Subsect:Recon}. 
Repeating the calculation for $m=0,1,\cdots, 255$, we obtained a collection $\{\underline{\mathbf g_m}\}$
of data vectors that together represented one complete data set. 
Utilizing the absorption phantom described in Section \ref{Subsect:Phantom},  a complete attenuated
data set was computed.
An idealized, non-attenuated,  data set was also computed by setting $\alpha_0(\mathbf r) = 0$.

\subsubsection{Generation of incomplete data}
\label{Sect:incomplete}
\begin{figure}[h]
\centering
{\includegraphics[height=6cm]{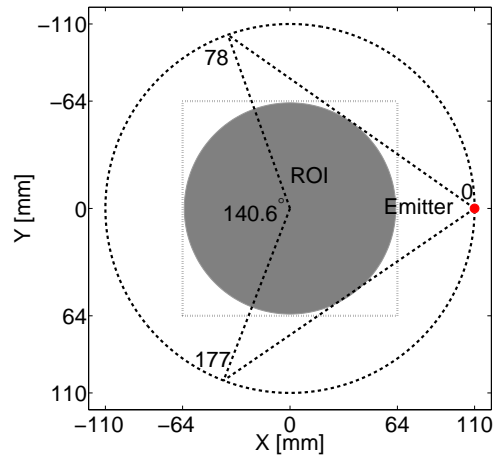}}
\caption{\label{fig:Geo}
Schematic of a USCT system with a circular transducer array whose elements are indexed from $0$ to $255$.  
It shows the first data acquisition, where element-0 (in red) is emitting an acoustic pulse, while all $256$ elements are receiving signals. 
The region-of-interest (ROI) is shaded in gray, and the dashed square box represents the physical dimensions ($128\times 128 $ mm$^2$) of all reconstructed images. 
}
\end{figure}

An incomplete data set in this study corresponds to one in which only $N^{\rm rec}$ receivers located on the opposite side of the emitter record the  pressure wavefield, with  $N^{\rm rec}<M$. 
Taking the $0$-th data acquisition as an example (see Fig.~\ref{fig:Geo}), only $N^{\rm rec}=100$ receivers, indexed from $78$ to $177$, record the wavefield, while other receivers record either unreliable or no measurements. 
Incomplete data sets formed in this way can emulate two practical scenarios: 
(1) Signals recorded by receivers near the emitter are unreliable and therefore discarded \cite{Duric2007:SOSWave};
%jIt has been reported in the literature \cite{Duric2007:SOSWave} that the signals received by transducers close to the emitter are unreliable (because of strong interference from the emitter) and are suggested to be discarded; 
and (2) An arc-shaped transducer array is employed that rotates with the emitter \cite{manohar,AAO12:LUS,TomoOUI:2014}.
% Several emerging laser-induced ultrasound USCT imaging systems employ an arc-shaped transducer array that rotates with the excitation source to record the transmitted wavefield \cite{manohar,TomoOUI:2014}. 

%In order to investigate the performance of the WISE method for these measurment configurations, we simulated an incomplete data set according to 
Specifically, incomplete data sets were generated as
\begin{equation}
  \big[\underline{\mathbf g^{\rm incpl}_m}\big]_{n^{\rm rec}L+l} = 
  \big[\underline{\mathbf g_m}\big]_{\mathcal J_m(n^{\rm rec})L+l},
  \quad{\rm for}\quad
  \substack{m=0,1,\cdots,M-1\\
            n^{\rm rec} = 0, 1, \cdots, N^{\rm rec}-1
           },
\end{equation}
where $\underline{\mathbf g^{\rm incpl}_m} $ is the incomplete $m$-th data acquisition, which is of dimensions $N^{\rm rec}L$, with $N^{\rm rec}<M$. 
The index map $\mathcal J_m:\{0,1,\cdots,N^{\rm rec}-1\}\mapsto\mathbb M_m^{\rm good}$ is defined as 
\begin{equation}
  \mathcal J_m(n^{\rm rec}) = \Big( m + n^{\rm rec} + \frac{M-N^{\rm rec}}{2}\Big)
  \mod M, 
\end{equation}
where $(m'\mod M)$ calculates the remainder of $m'$ divided by $M$,
and the index set $\mathbb M_m^{\rm good}$ collects indices of transducers that reliably record data at the $m$-th data acquisition
and is defined as
\begin{equation}
  \mathbb M_m^{\rm good} = \Big\{ k\!\!\!\!\mod M \big| k\in \big [m+(M-N^{\rm rec})/2,m+(M+N^{\rm rec})/2 \big)
          \Big\}.
\end{equation}
Here, for simplicity, we assume that $M$ and $N^{\rm rec}$ are even numbers. 
In this study, we empirically set $N^{\rm rec}=100$ so that the object can be fully covered by the fan region as shown in Fig.~\ref{fig:Geo}.

\subsubsection{Generation of noisy data}
An additive Gaussian white noise model was employed to simulate electronic measurement noise as 
\begin{equation}
  \underline{\tilde{\mathbf g}_m} = \underline{\mathbf g_m} + \tilde{\mathbf n}, 
\end{equation}
where $\underline{\tilde{\mathbf g}_m}$ and $\tilde{\mathbf n}$ are the noisy data vector and the Gaussian white noise vector, respectively. 
In this study, the maximum value of the pressure received by the $128$-th transducer at the $0$-th data acquisition with a homogeneous medium (water tank) was chosen as a reference signal amplitude.  
The noise standard deviation was set to be $5\%$ of this value. 
An example of a simulated noiseless and noisy data acquisition is shown  Fig.~\ref{fig:PreNoise}. 
% A typical noisy realization of $\underline{\tilde{\mathbf g}_m}$ is shown in Figs.~\ref{fig:PreNoise}-(b) and -(c).

\iffalse
\subsubsection{Generation of sparsely-sampled data}
In addition, we produced a half-data set to investigate the robustness of the WISE method to sparsely sampled data. 
The combined data set was downsampled by taking every other transducer, resulting a sequence of $\underline{\mathbf g^{\rm half}_m} = \underline{\mathbf g^{\rm comb}_{2m}}$ for $m=0,1,\cdots, 127$. 
\fi

\subsection{Image reconstruction}
\label{Subsect:Recon}

\subsubsection{Second-order pseudospectral k-space method}
In the reconstruction methods described below, the action of the
operator $\mathbf H^c$ (Eqn.\ (\ref{eqn:WaveDis})) was
computed by solving Eqn.~\eqref{eqn:WaveEqn} by use of a second-order pseudospectral k-space method.
%As mentioned in Section \ref{Subsect:GenData}, in order to avoid {\it inverse crime} \cite{colton12:inversecrime}, we implemented a \rd{numerical solver} \cite{Mast01:2ndKSpace} that was different from the first-order \rd{numerical solver} used to generate the data. 
%Though it was also based on the pseudo-spectral method, the new version of \rd{numerical solver} directly solved the second-order wave equation, i.e., Eqn.~\eqref{eqn:WaveEqn}, and, therefore, will be referred to as a second-order \rd{numerical solver}.  
This  was  implemented using GPUs. 
The calculation domain was of size $512\times 512$ mm$^2$, sampled on a $1024\times1024$ uniform Cartesian grid of spacing $\Delta^{\rm s} = 0.5$ mm for reconstruction. 
On a platform consisting of dual octa-core CPUs with a $2.00$ GHz clock speed, $125$ GB RAM, and a single NVIDIA Tesla K20C GPU, the second-order k-space method required approximately $7$ seconds to complete one forward simulation.
% sufficiently large to avoid boundary wrapping effects \cite{Mast01:2ndKSpace}.  

\subsubsection{Sequential waveform inversion}

To serve as a reference for the WISE method,
 we implemented the sequential waveform inversion method described in Algorithm 1.
\iffalse
Specifically, the method seeks the solution of Eqn.~\eqref{eqn:DCostFunc}
\begin{equation}\label{eqn:SWI}
   \hat {\mathbf c} = \arg\min_{\mathbf c} \frac{1}{2}
      \sum_{m=0}^{M-1} \Vert \underline{\mathbf g_m} - 
     \mathbf M_m \mathbf H^{\rm c} \mathbf s_m \Vert^2, 
\end{equation}
\fi
%where $\mathbf H^{\rm c}$ was implemented using the second-order \rd{numerical solver}. 
No penalty term was included ($\beta=0$) because, due to its extreme computational burden, we only investigated this method
 in preliminary studies  involving noise-free non-attenuated data. 
% Equation \eqref{eqn:SWI} was solved by use of Algorithm \ref{Alg:DGradDesc}. 
A uniform sound speed distribution was employed as the initial guess, which
corresponded to the known background value of $1.5$ mm/$\mu$s.
%We assumed that the background \rd{sound speed} was known and
 The object was contained in a square region-of-interest (ROI) of dimension $128\times128$ mm$^2$ (See Fig.~\ref{fig:Geo}), which was covered by $256\times 256$ pixels. 
%The reconstructed images will provide a reference to evaluate the accuracy of the images reconstructed by use of the WISE method. 
%However, the computation burden of this method is extremely large, which will be discussed in Section \ref{Sect:SimuResults}. 

\subsubsection{WISE method}

We implemented the WISE method by use of Algorithm 2.
%described in Section \ref{SubSect:WISE}.
% Specifically, the method seeks the solution of 
Two types of penalties were employed in this study: 
a quadratic penalty expressed as 
\begin{equation}
  \mathcal R^{\rm Q}(\mathbf c) = \sum_j \sum_i 
          ( [\mathbf c]_{jN_x+i} -[\mathbf c]_{jN_x+i-1}
           )^2
        +  ( [\mathbf c]_{jN_x+i} -[\mathbf c]_{(j-1)N_x+i}
           )^2
           ,
\end{equation}
where $N_x$ and $N_y$ denote the number of grid points along the `x' and `y' directions respectively, and a total variation (TV) penalty, defined as \cite{pmb08:tv,KunSPIETV:2011}
\begin{equation}
  \mathcal R^{\rm TV}(\mathbf c) = \sum_j \sum_i \sqrt{
          \epsilon +
          ( [\mathbf c]_{jN_x+i} -[\mathbf c]_{jN_x+i-1}
           )^2
        +  ( [\mathbf c]_{jN_x+i} -[\mathbf c]_{(j-1)N_x+i}
           )^2}
           ,
\end{equation}
where $\epsilon$ is a small number introduced to avoid dividing by $0$ in the gradient calculation.
In this study, we empirically selected $\epsilon=10^{-8}$. 
This value was fixed because we observed that it had a minor impact on the reconstructed images compared to the impact of $\beta$. 
The use of this parameter can be avoided when advanced optimization algorithms are employed \cite{Kun:PMB,Chao13:FullWave}.
As in the sequential waveform inversion case, it was assumed that the background sound speed was known and the object was contained in a square ROI of dimension $128\times128$ mm$^2$ (See Fig.~\ref{fig:Geo}), which corresponded to $256\times 256$ pixels. 
% The WISE method equiped with the quadratic and the TV penalties will be referred to as the \rd{WISE method with a quadratic penalty} and the \rd{WISE method with a TV penalty} methods respectively. 
The regularization parameters corresponding to the quadratic penalty and the TV penalty will be denoted by $\beta^{\rm Q}$ and $\beta^{\rm TV}$, respectively. 
Optimal regularization parameter values should ultimately be identified by use of task-based
measures of image quality \cite{BarrettBook}.
% depend on the specific medically relavent task that the reconstructed images are used for \cite{BarrettBook}. 
%Estimation of the optimal values requires a systematic investigation of the image statistics, which is out of the scope of this study.
In this preliminary study, we  investigated the impact of $\beta^{\rm Q}$ and $\beta^{\rm TV}$ on the reconstructed images by sweeping their values over a wide range.

\subsubsection{Reconstruction from incomplete data}
Because the WISE method requires  $\mathbf M_m$ to be identical for all $m$'s, image reconstruction from incomplete data remains challenging \cite{Herrmann2012:SE,ArXiv:Haber14,ArXiv:Ascher14}. 
In this study, two data completion strategies were investigated \cite{Herrmann2012:SE,ArXiv:Haber14,ArXiv:Ascher14} to synthesize a complete data set, from which the WISE method could be effectively applied. 
\iffalse
More specifically, the data fidelity term in Eqn.~\eqref{eqn:DDataMisfit} was approximated by 
\begin{equation}\label{eqn:CombDataMisfit}
   \mathcal F (\mathbf c) \approx 
           \frac{1}{2} \sum_{m=0}^{M-1}
         \Vert 
          \underline{\mathbf g_m^{\rm comb}}
        -\mathbf M \mathbf H^{\rm c}  \mathbf s_m 
          \Vert^2,
\end{equation}
\fi

% Denoting the synthesized complete data by $\underline{\mathbf g_m^{\rm comb}}\in\mathbb R^{ML}$, 
%Two data completion strategies were employed. 
% to complete the incomplete-data measurements described in Section \ref{SubSect:INCPL}, a data completion strategy was employed \cite{Herrmann2012:SE,ArXiv:Haber14,ArXiv:Ascher14}. 
One strategy was to fill the missing data with pressure corresponding to a homogeneous medium as 
\begin{equation}\label{eqn:NewData}
\lbrack \underline{\mathbf g^{\rm combH}_m} \rbrack_{m^{\rm rec}L+l} = 
 \left\{\begin{array}{ll}  
   [\underline{\mathbf g^{\rm incpl}_m}]_{\mathcal J^{-1}_m(m^{\rm rec})L+l}, & 
   {\rm if}\quad m^{\rm rec}\in \mathbb M^{\rm good}_m\\ 
   \lbrack\mathbf g^{\rm h}_m\rbrack_{m^{\rm rec}L+l}, & \text{otherwise},
  \end{array}
\right.
\end{equation}
for $m^{\rm rec}=0,1,\cdots,M-1$, where $\mathbf g_m^{\rm h}\in \mathbb R^{ML}$, $\underline{\mathbf g^{\rm incpl}_m} \in \mathbb R^{N^{\rm rec}L}$, and $\underline{\mathbf g^{\rm combH}_m}\in\mathbb R^{ML}$, 
denote the computer-simulated (with a homogeneous medium), the measured incomplete, and the combined complete data vectors at the $m$-th data acquisition, respectively. 
% Also, $\mathbf g_m^{\rm h}$ denotes the computer-simulated pressure data vector with a homogeneous medium. 
% acquisition when the object is absent, i.e., 
% \begin{equation}
%   \mathbf g_m^{\rm h} = \mathbf M \mathbf H^{\rm c_0} \mathbf s_m, 
% \end{equation}
% where $\mathbf H^{\rm c_0}$ denotes the numerical wave equation solver with a background \rd{sound speed} of value $c_0$, 
% corresponding to calibration measurements in practice \cite{Duric2007:MedPhys}. 
The mapping $\mathcal J^{-1}_m:\mathbb M_m^{\rm good} \mapsto \{0,1,\cdots,N^{\rm rec}-1\}$ denotes the inverse operator of $\mathcal J_m$ as 
\begin{equation}
  \mathcal J^{-1}_m (m^{\rm rec}) = 
   \left \{\begin{array}{ll}
     m^{\rm rec}-m -\frac{M-N^{\rm rec}}{2}, \quad {\rm if}\quad \frac{M-N^{\rm rec}}{2} \leq m^{\rm rec}-m < \frac{M+N^{\rm rec}}{2}\\
     m^{\rm rec}-m + \frac{M+N^{\rm rec}}{2}, \quad {\rm if}\quad \frac{-M-N^{\rm rec}}{2} \leq m^{\rm rec} -m < \frac{-M+N^{\rm rec}}{2} .
    \end{array}\right.
\end{equation}
This data completion strategy is based on the assumption that the
back-scatter from breast tissue in an appropriately sound speed-matched water bath is weak. 
This assumption suggests that the missing  measurements can be replaced by the corresponding
 pressure data that would have been produced in the absence of the object.

The second, more crude, data completion strategy was to simply fill the  missing data with zeros, i.e.,
\begin{equation}\label{eqn:NewData0}
\lbrack \underline{\mathbf g^{\rm comb0}_m} \rbrack_{m^{\rm rec}L+l} = 
 \left\{\begin{array}{ll}  
   [\underline{\mathbf g^{\rm incpl}_m}]_{\mathcal J^{-1}_m(m^{\rm rec})L+l}, & 
   {\rm if}\quad m^{\rm rec}\in \mathbb M^{\rm good}_m\\ 
   0, & \text{otherwise},
  \end{array}
\right.
\end{equation}
where $\underline{\mathbf g^{\rm comb0}_m}$ denotes the data completed with the second strategy. 

\subsubsection{Bent-ray image reconstruction} A bent-ray  method was also employed
to reconstruct images. Details regarding the time-of-flight estimation and algorithm
implementation are provided in Appendix \ref{Sect:BentRay}.

\section{Computer-simulation results}
\label{Sect:SimuResults}

\subsection{Images reconstructed from idealized data}

\begin{figure}[h]
\centering
\subfloat[]{\includegraphics[width=5.6cm]{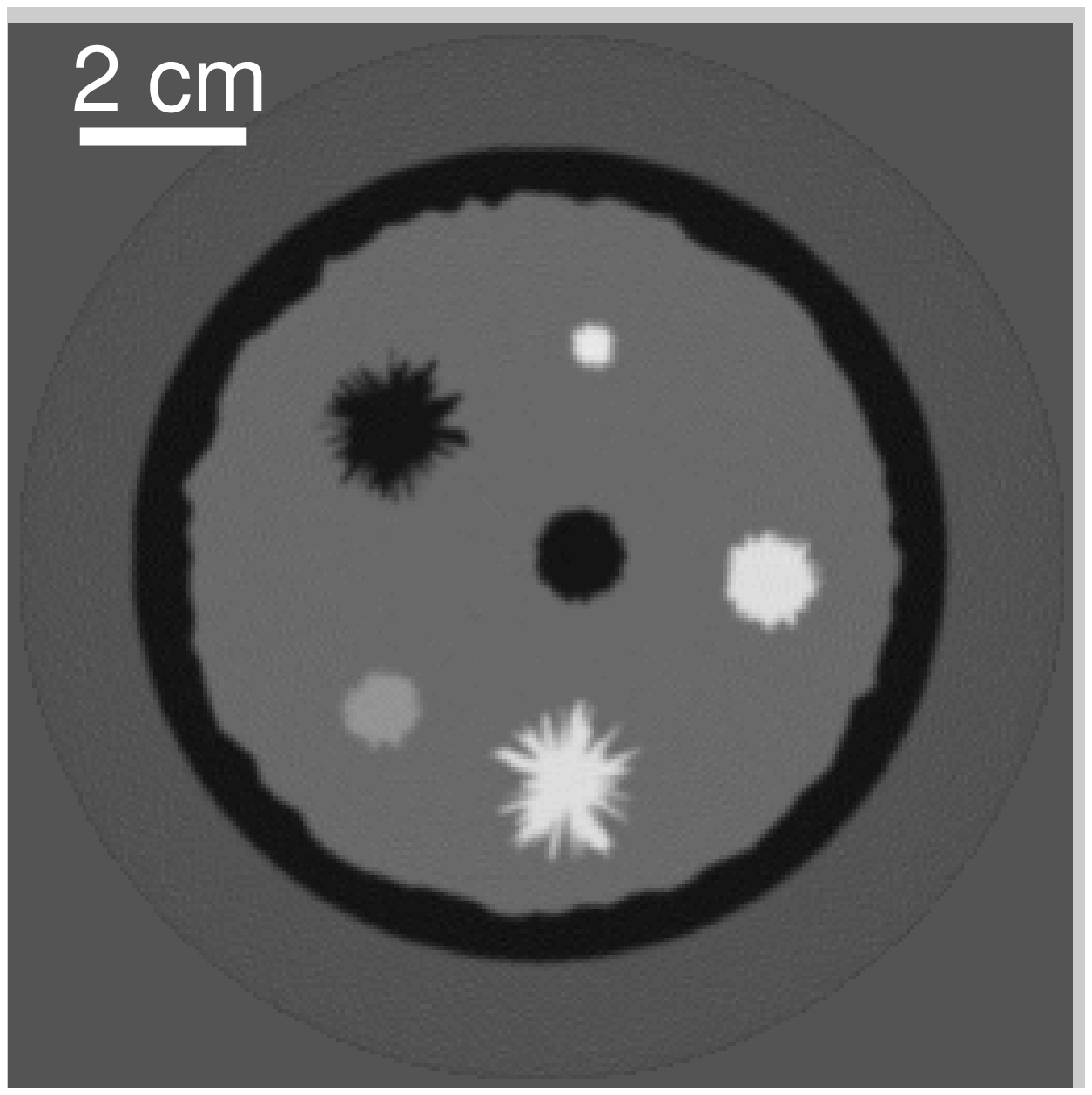}}
\subfloat[]{\includegraphics[width=5.6cm]{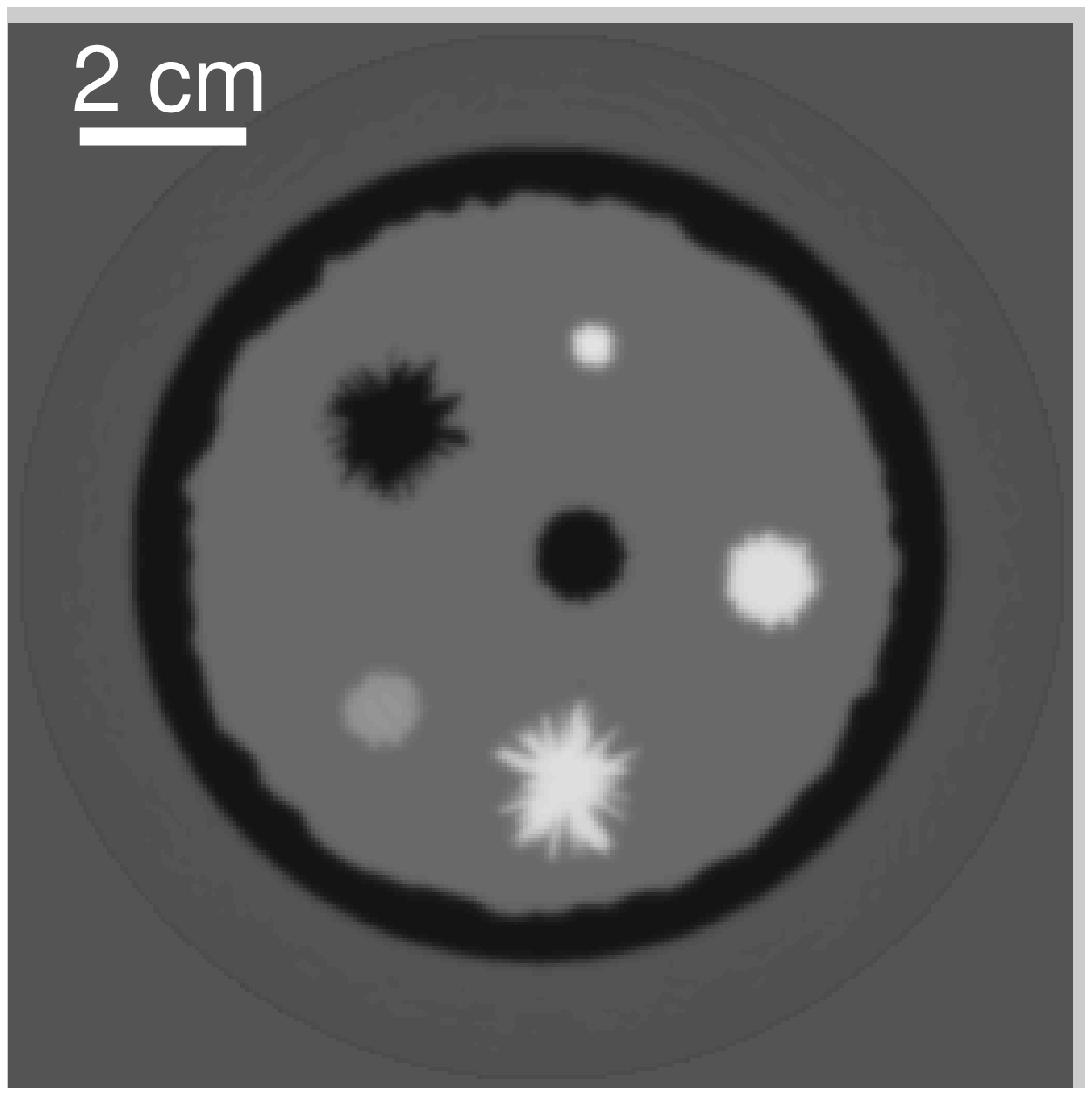}}\\
\subfloat[]{\includegraphics[width=5.6cm]{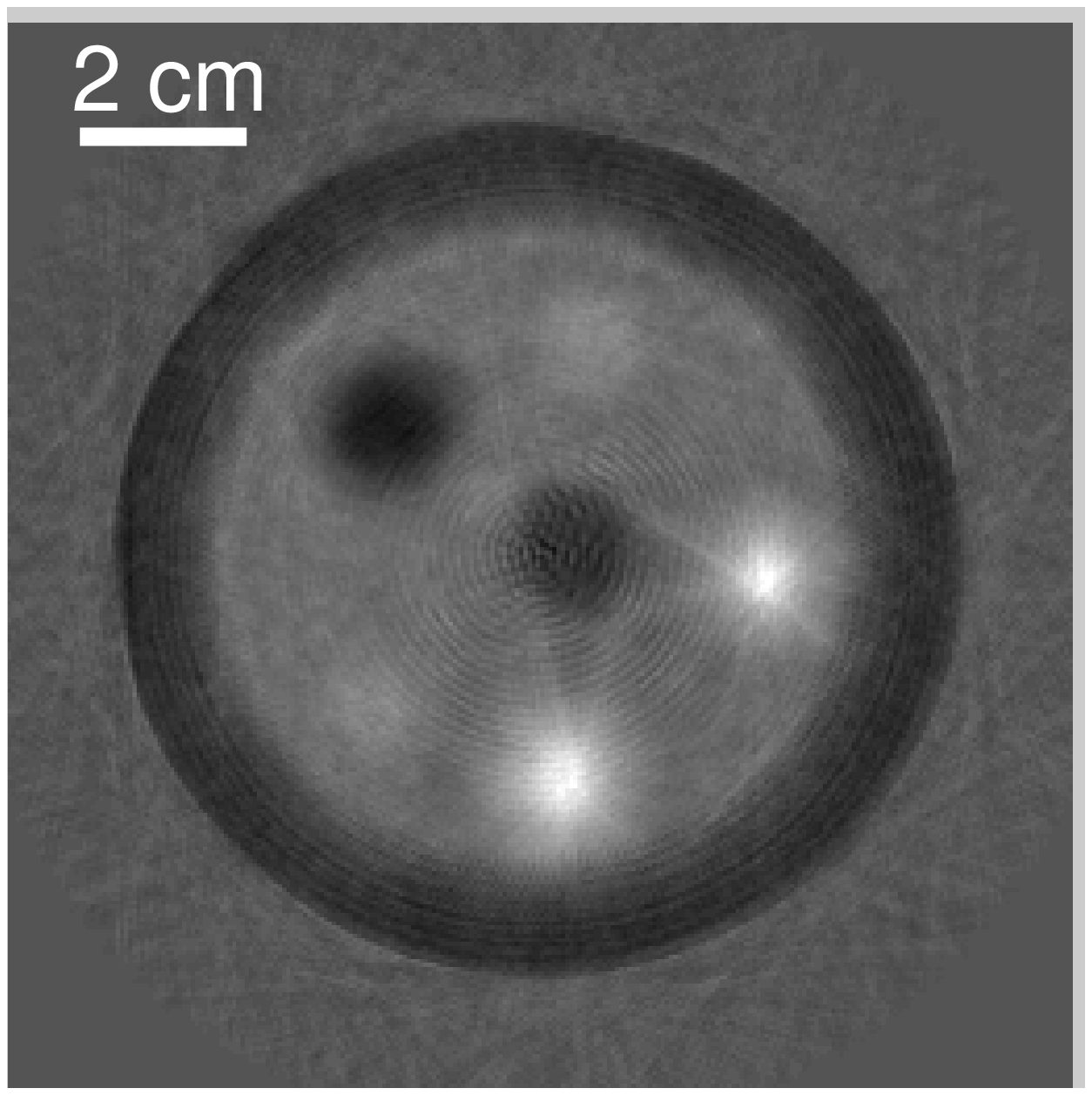}}
\subfloat[]{\includegraphics[width=5.6cm]{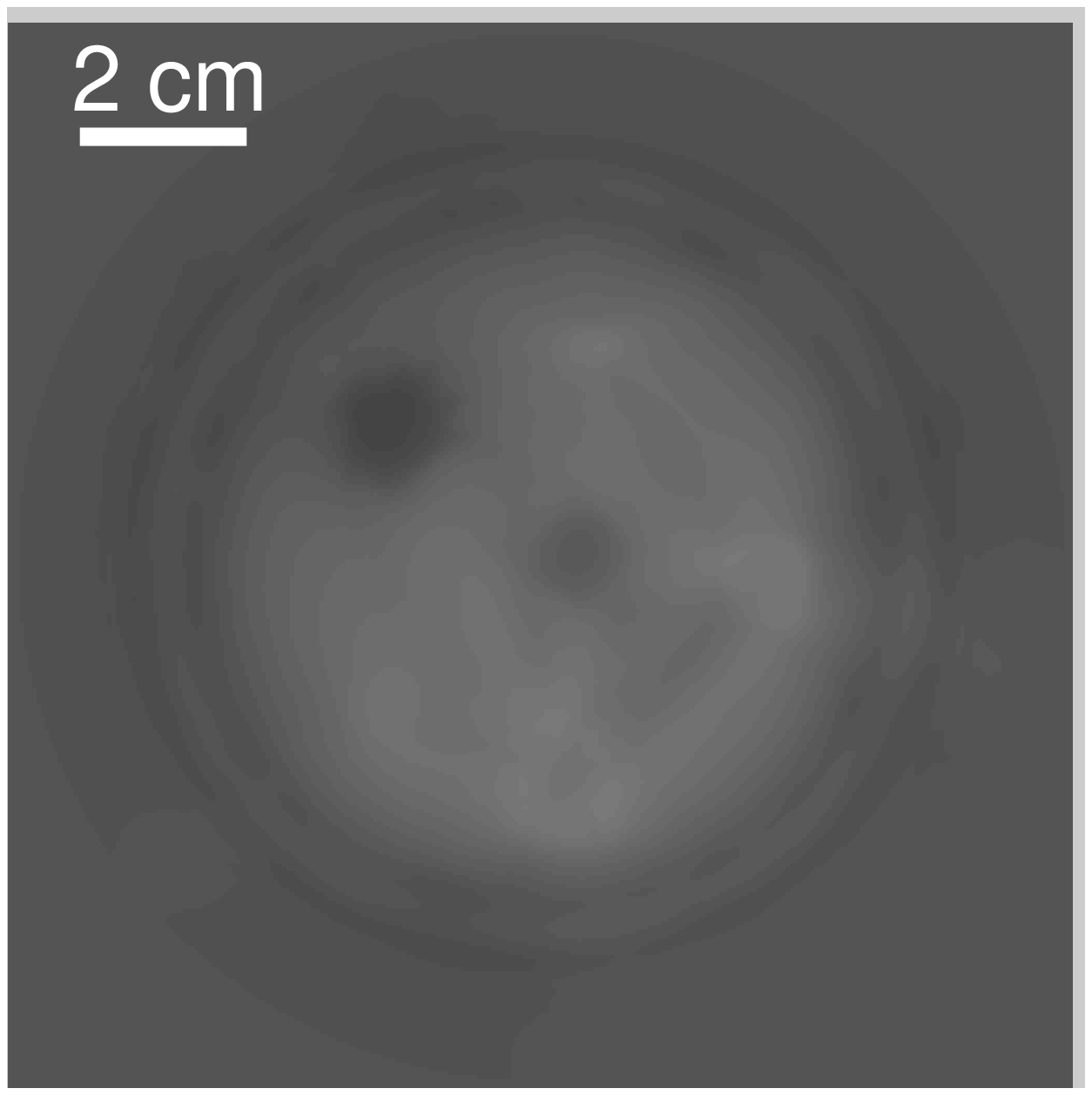}}
\caption{\label{fig:WISE} 
Images reconstructed by use of 
(a) the WISE method after the $199$-th iteration ($1,018$ runs of the wave equation solver), 
(b) the sequential waveform inversion algorithm after the $43$-rd iteration
($57,088$ runs of the wave equation solver),
(c) the bent-ray model-based sound speed reconstruction method, 
and 
(d) the sequential waveform inversion algorithm after the $1$-st iteration
($1,024$ runs of the wave equation solver)
from the noise-free non-attenuated data. 
The grayscale window is $[1.46,1.58]$ mm/$\mu$s.}
\end{figure}

The images reconstructed from the noise-free,  non-attenuated, data by use of the WISE method with $199$ iterations
and the sequential waveform inversion method with 43 iterations are  shown in Fig.~\ref{fig:WISE}-(a) and (b). 
% , the WISE method yields a highly-accurate image as shown in Figs.~\ref{fig:WISE}-(a) and \ref{fig:Prof_Ideal} for the profiles. 
%Using the sequential waveform inversion method, the \rd{sound speed} map can also be accurately reconstructed, as shown in Fig.~\ref{fig:WISE}-(b) after $43$ iterations.  
As expected \cite{Duric2007:SOSWave,Neb14:Wave}, both images are more accurate and possess higher
spatial resolution than the one reconstructed by use of the bent-ray  reconstruction algorithm displayed in Fig.~\ref{fig:WISE}-(c).
Profiles through the reconstructed images are displayed in Fig.~\ref{fig:Prof_Ideal}.
The images shown in Fig.~\ref{fig:WISE}-(a) and -(b) possess similar accuracies as measured by their root-mean-square errors (RMSEs),
namely, $1.08\times 10^{-3}$ for the former and $1.19\times 10^{-3}$ for the latter.
The RMSE was computed as the Euclidean distance between the reconstructed image and the sound speed phantom vector $\mathbf c$, averaged by the $256\times 256$ pixels of the ROI sketched in Fig.~\ref{fig:Geo}. 
However, the reconstruction of Fig.~\ref{fig:WISE}-(a) required only about $1.7\%$ of the computational time required to reconstruct Fig.~\ref{fig:WISE}-(b), namely, $1.4$ hours for the former and $81.4$ hours for the latter respectively. 
%The absolute Euclidean distance is defined as the $\Vert \hat{\mathbf c} - \mathbf c\Vert$, where $\mathbf c$ denotes the phantom \rd{sound speed} vector. 
This is because the WISE method required only $1018$ wave equation solver runs which is significantly less than the  $57088$ wave equation solver runs
required by the sequential waveform inversion method.
With a similar number of wave equation solver runs, (e.g., $1024$), one can complete only a single algorithm iteration by use of the sequential waveform inversion method.  
The corresponding image, shown in Fig.~\ref{fig:WISE}-(d), lacks quantitative accuracy as well as qualitative value for identifying features. 
The results suggest that the WISE method maintains the advantages of the sequential waveform inversion method while significantly reducing the computational time. 

\subsection{Convergence of the WISE method}

\begin{figure}[h]
\centering
{\includegraphics[width=7cm]{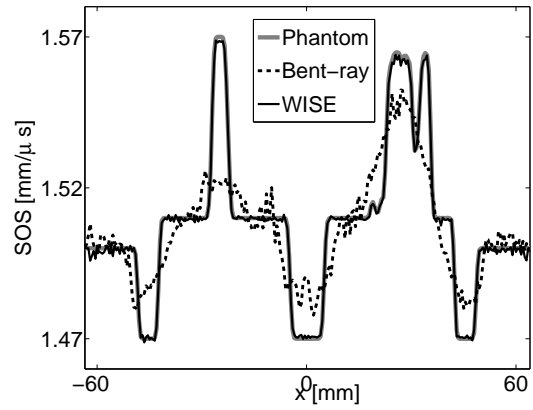}}
\caption{\label{fig:Prof_Ideal}
Profiles at $y = 6.5$ mm of the images reconstructed by use of the bent-ray TOF image reconstruction method and the WISE method from the noise-free non-attenuated data.
}
\end{figure}

\begin{figure}[h]
\centering
\subfloat[]{\includegraphics[width=5.6cm]{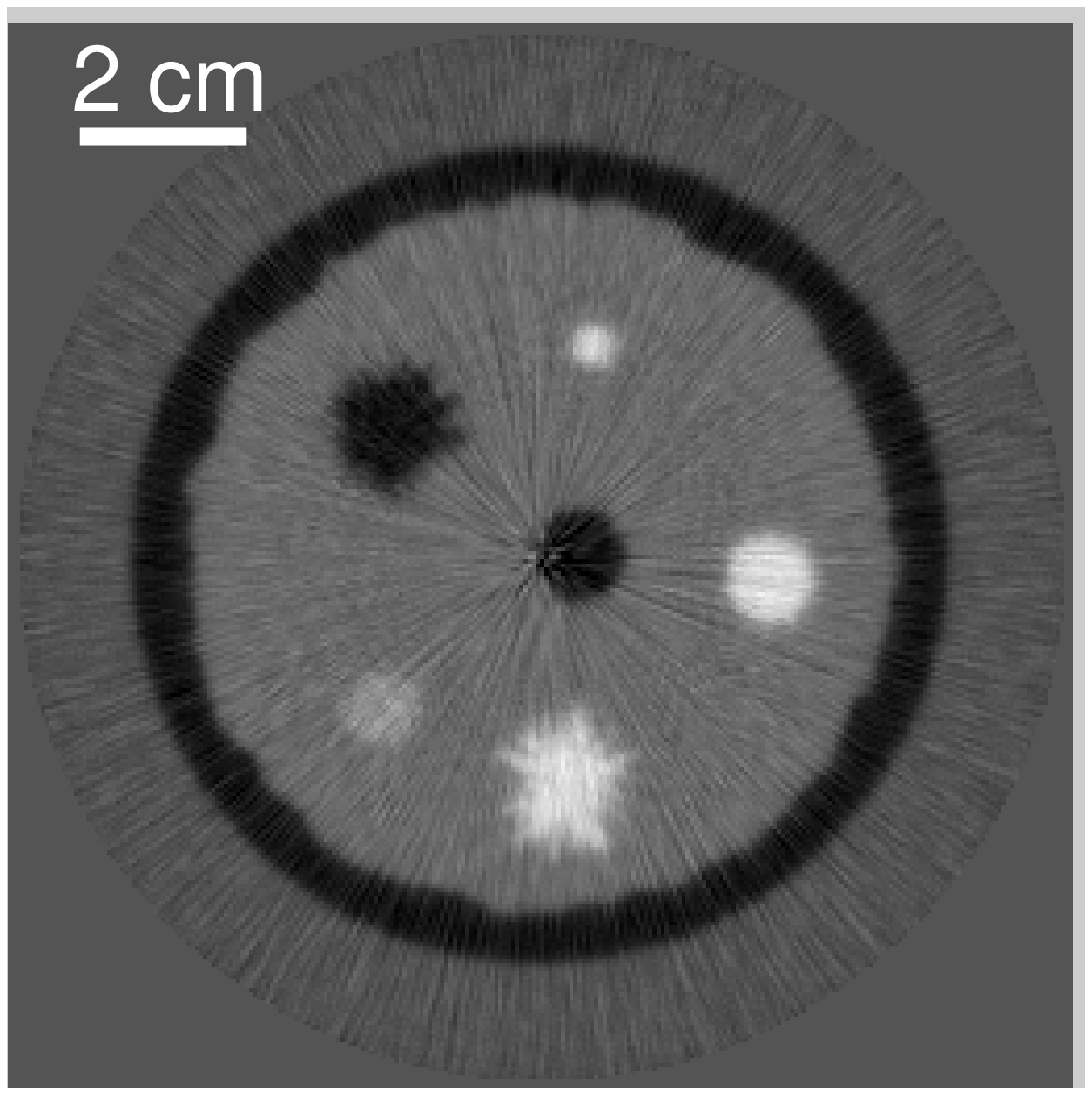}}
\subfloat[]{\includegraphics[width=5.6cm]{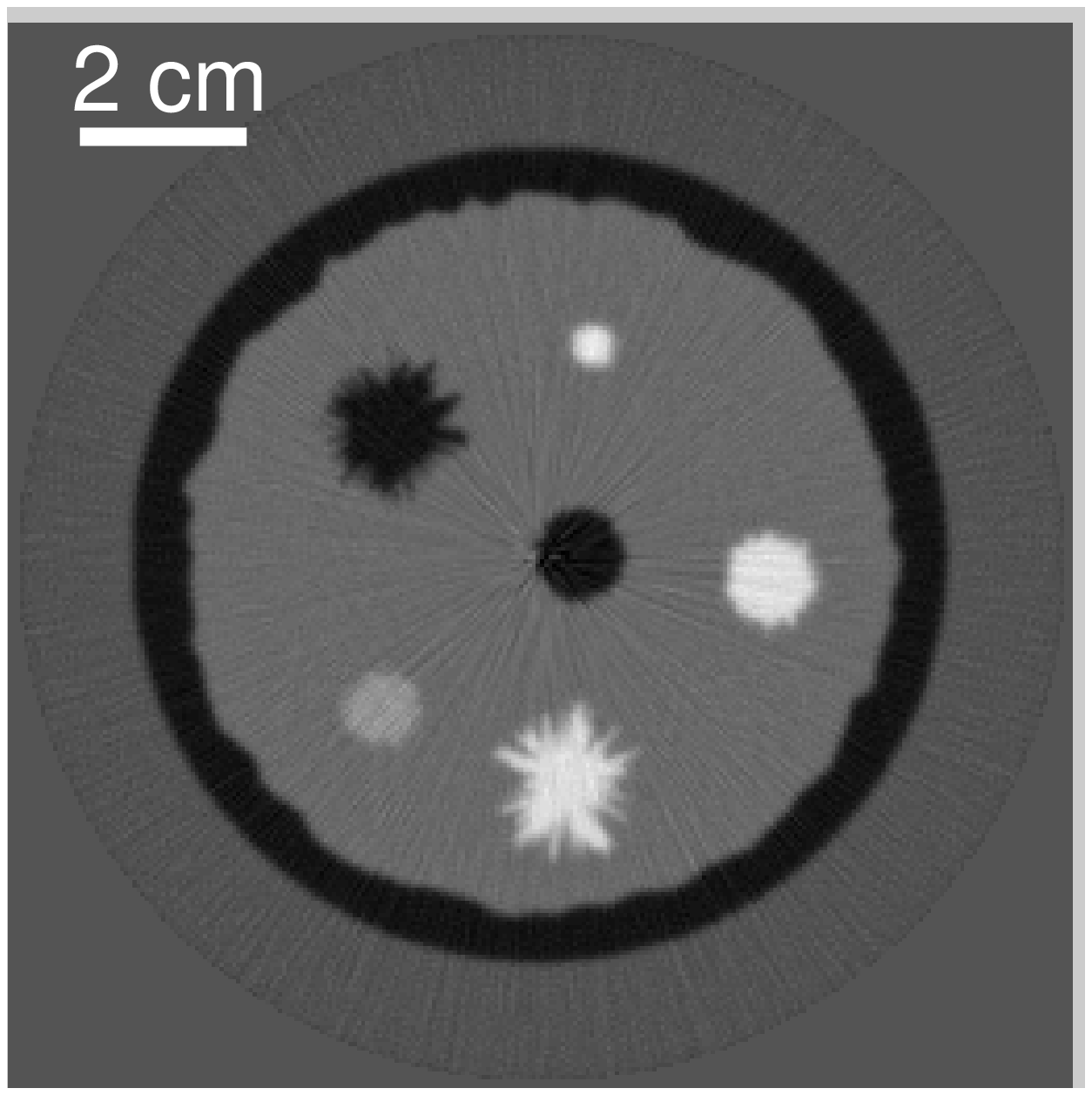}}\\
\subfloat[]{\includegraphics[width=5.6cm]{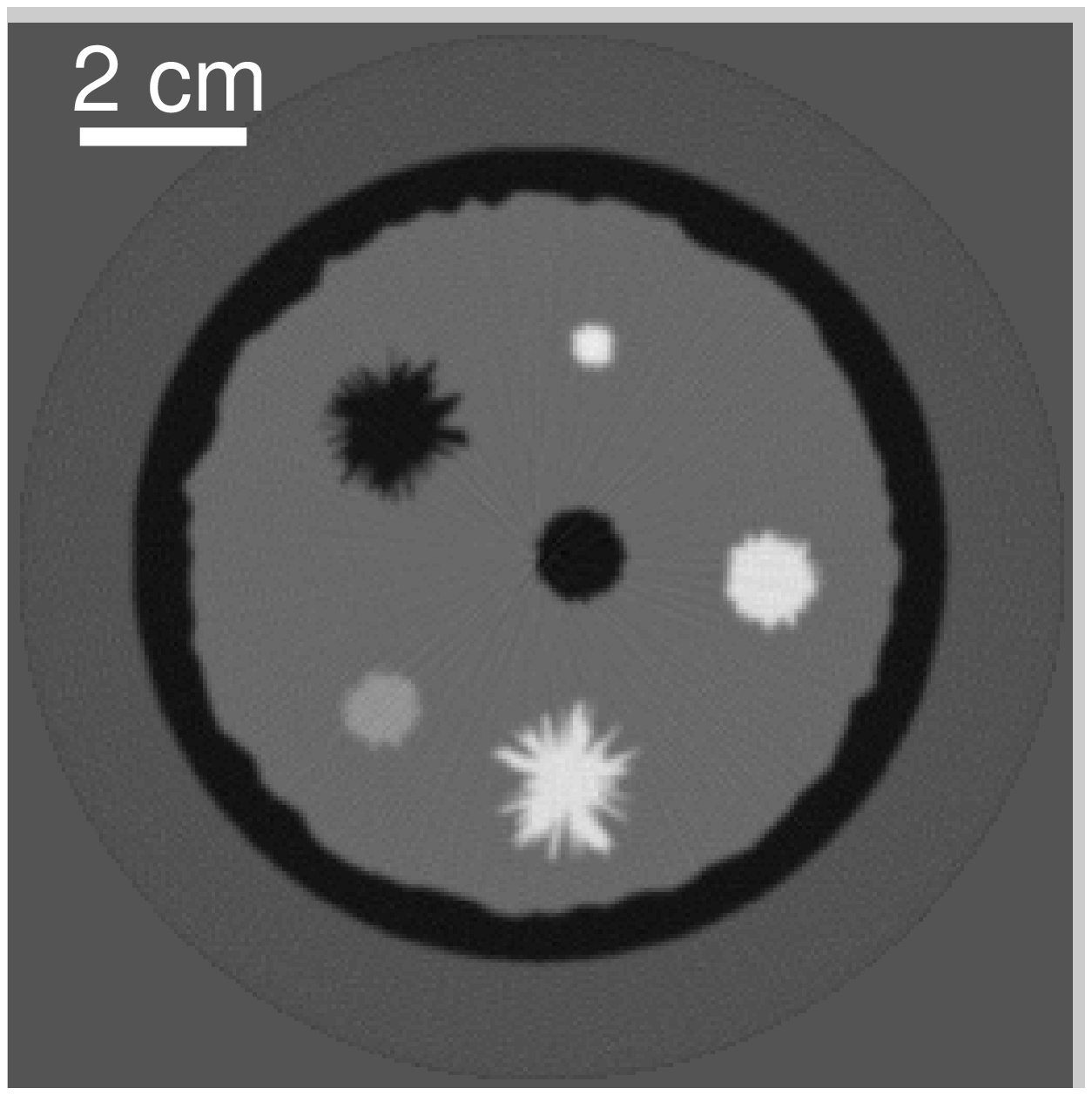}}
\subfloat[]{\includegraphics[width=5.6cm]{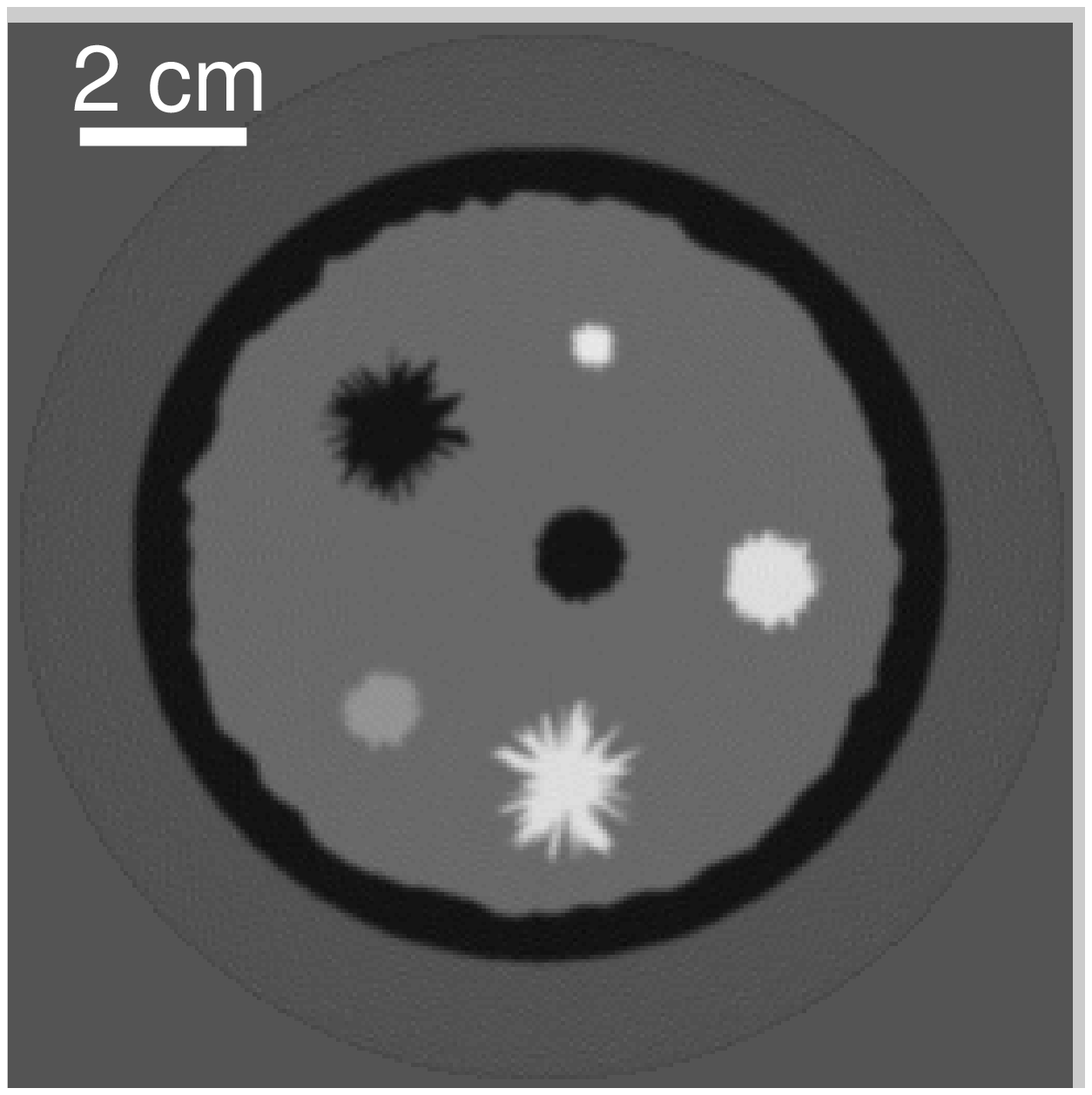}}
\caption{\label{fig:WISEConverge} 
Images reconstructed by use of the WISE method after (a) the $20$-th, 
(b) the $50$-th, (c) the $100$-th, and (d) the $250$-th iteration from the noise-free, non-attenuated
 data set.
The grayscale window is $[1.46,1.58]$ mm/$\mu$s.
}
\end{figure}

\begin{figure}[h]
\centering
{\includegraphics[width=7cm]{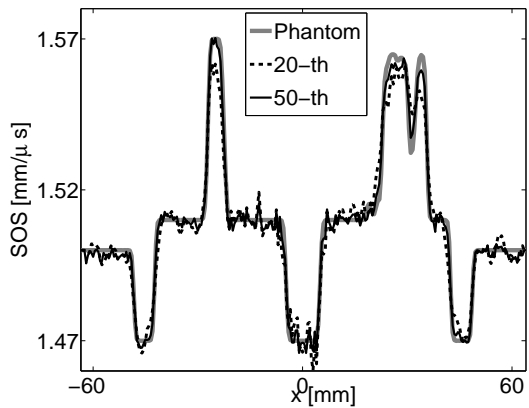}}
\caption{\label{fig:Prof_Conv}
Profiles of the images reconstructed by use of the WISE method from the noise-free non-attenuated data after different numbers of iterations.
}
\end{figure}

% As expected \cite{Romero00:PE,Herrmann13:GaussCode}, 
Images reconstructed from noise-free, non-attenuated, data by use of the WISE method contain
radial streak artifacts when the algorithm iteration number is less than $100$,
  as shown in Figs.~\ref{fig:WISEConverge}-(a-c). Profiles through these images are displayed in \ref{fig:Prof_Conv}.
The streaks artifacts are likely caused by  crosstalk introduced during the source encoding procedure \cite{Romero00:PE,Herrmann13:GaussCode}. 
However, these artifacts are effectively mitigated after more iterations as demonstrated by the image reconstructed after the $199$-th iteration in Fig.~\ref{fig:WISE}-(a) and its profile in Fig.~\ref{fig:Prof_Ideal}. 
%In this particular case, the WISE method appears to converge after $200$ iterations as suggested by Fig.~\ref{fig:Convergence}. 
%For example, the images reconstructed after 199  (Fig.~\ref{fig:WISE}-(a)) and 250 (Fig.~\ref{fig:WISEConverge}-(d))
%iterations are visually very similar.
%Little difference can be seen between the image after the $250$-th iteration (Fig.~\ref{fig:WISEConverge}-(d)) and the one after the $199$-th iteration (Fig.~\ref{fig:WISE}-(a)). 
The quantitative accuracy of the reconstructed images is improved with more iterations as shown in Fig.~\ref{fig:Prof_Conv}.
 
Figure \ref{fig:Convergence}-(a) reveals that the WISE method requires a larger number of algorithm iterations than does the sequential waveform inversion method to achieve the same RMSE.
%   Euclidean distance between the reconstructed image and original
% phantom, expressed as a percentage of $\Vert \mathbf c \Vert$. 
The RMSE of the images reconstructed by use of the WISE method
 appears to oscillate around $1.0\times 10^{-3}$ after the first $100$ iterations while the sequential waveform inversion method can achieve a lower RMSE.
However, as shown previously in Fig.~\ref{fig:WISE}-(a) and the corresponding profile
in Fig.\ \ref{fig:Prof_Ideal}, after additional iterations
the image reconstructed by use of the WISE method achieves a high accuracy.
% the image reconstructed by use of the WISE method is highly accurate after the $199$-th iteration. 
Moreover, to achieve the same accuracy as the sequential waveform inversion
method, the  WISE method requires a computation time that is reduced by
approximately two-orders of magnitude,
% is about two-orders of magnitude smaller than that for the sequential waveform inversion method
 as suggested by Fig.~\ref{fig:Convergence}-(b).
% For example, to achieve the accuracy with an Euclidean distance $0.08\%$ of $\Vert\mathbf c\Vert$, the WISE method took $58$ iterations while the sequential waveform inversion method took $58112$ numerical solver runs. 
%This is because of the significant computation reduction per iteration when the WISE method is employed. 
We also plotted the cost function value against the number of iterations in Fig.~\ref{fig:Convergence}-(c). 
Note that for the WISE method, the cost function value was approximated by the current realization of $\frac{1}{2}\Vert \underline{\mathbf g}^{\rm w} - \mathbf M \mathbf H^{\rm c} \mathbf s^{\rm w}\Vert^2$.
\iffalse
It is interesting to observe the oscillations in the curve of the relative Euclidean distance in Fig.~\ref{fig:Convergence}, where the relative Euclidean distance is defined as $\Vert \mathbf c^{(k)} - \mathbf c^{(k-1)}\Vert$, with $k$ indexing the number of iterations. 
The oscillations are likely due to the random source encoding technique.
% which introduced randomness between iterations. 
However, the magnitude of these oscillations are below the absolute Euclidean distance of the reconstructed images, suggesting a subtle impact on the reconstructed images. 
\fi
These plots suggest that, in this particular case, the WISE method appears to approximately converge after $200$ iterations.
For example, the images reconstructed after 199  (Fig.~\ref{fig:WISE}-(a)) and 250 (Fig.~\ref{fig:WISEConverge}-(d))
iterations are nearly identical.

\subsection{Images reconstructed from non-attenuated data containing noise}

\begin{figure}[h]
\centering
\subfloat[]{\includegraphics[width=5.6cm]{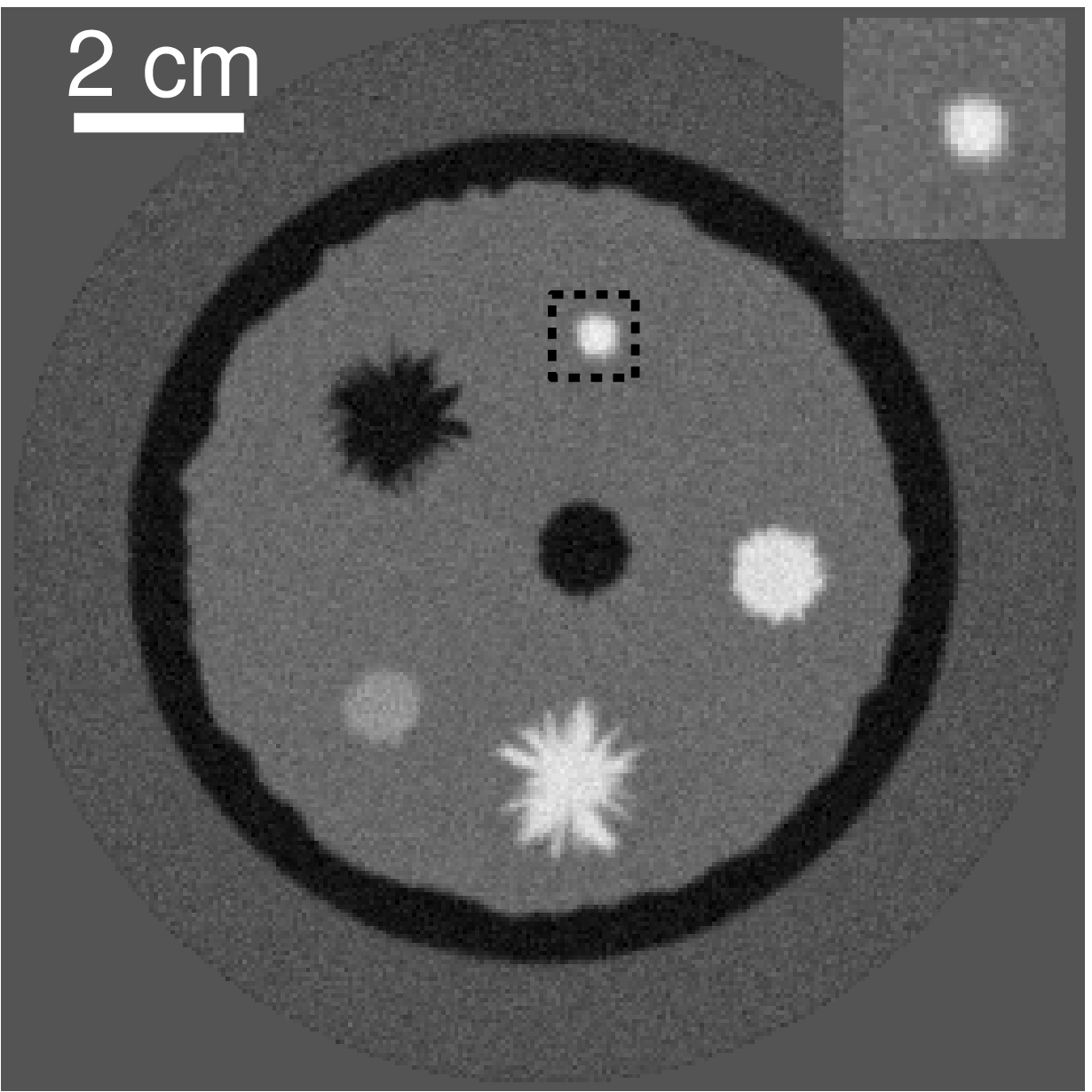}}
\subfloat[]{\includegraphics[width=5.6cm]{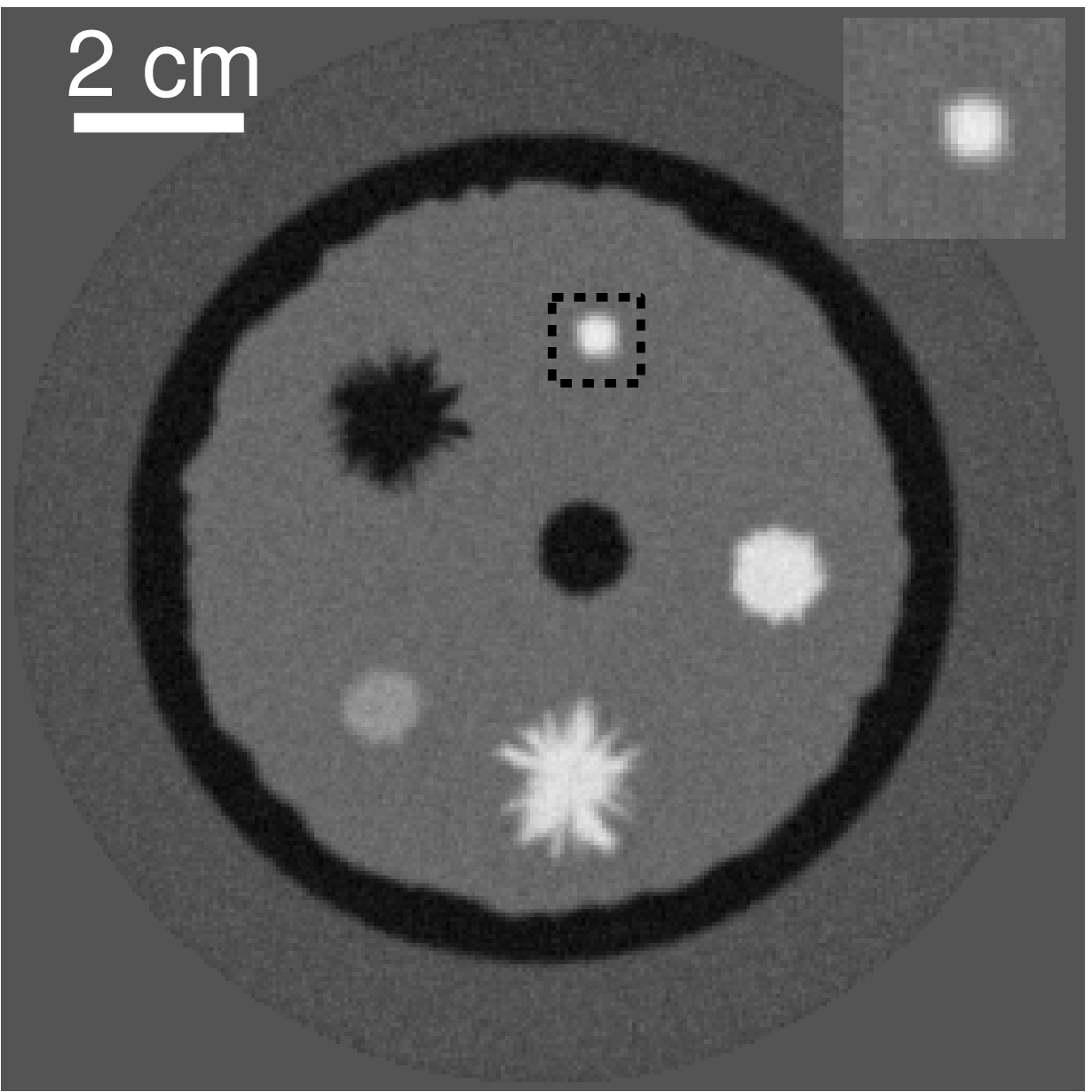}}\\
\subfloat[]{\includegraphics[width=5.6cm]{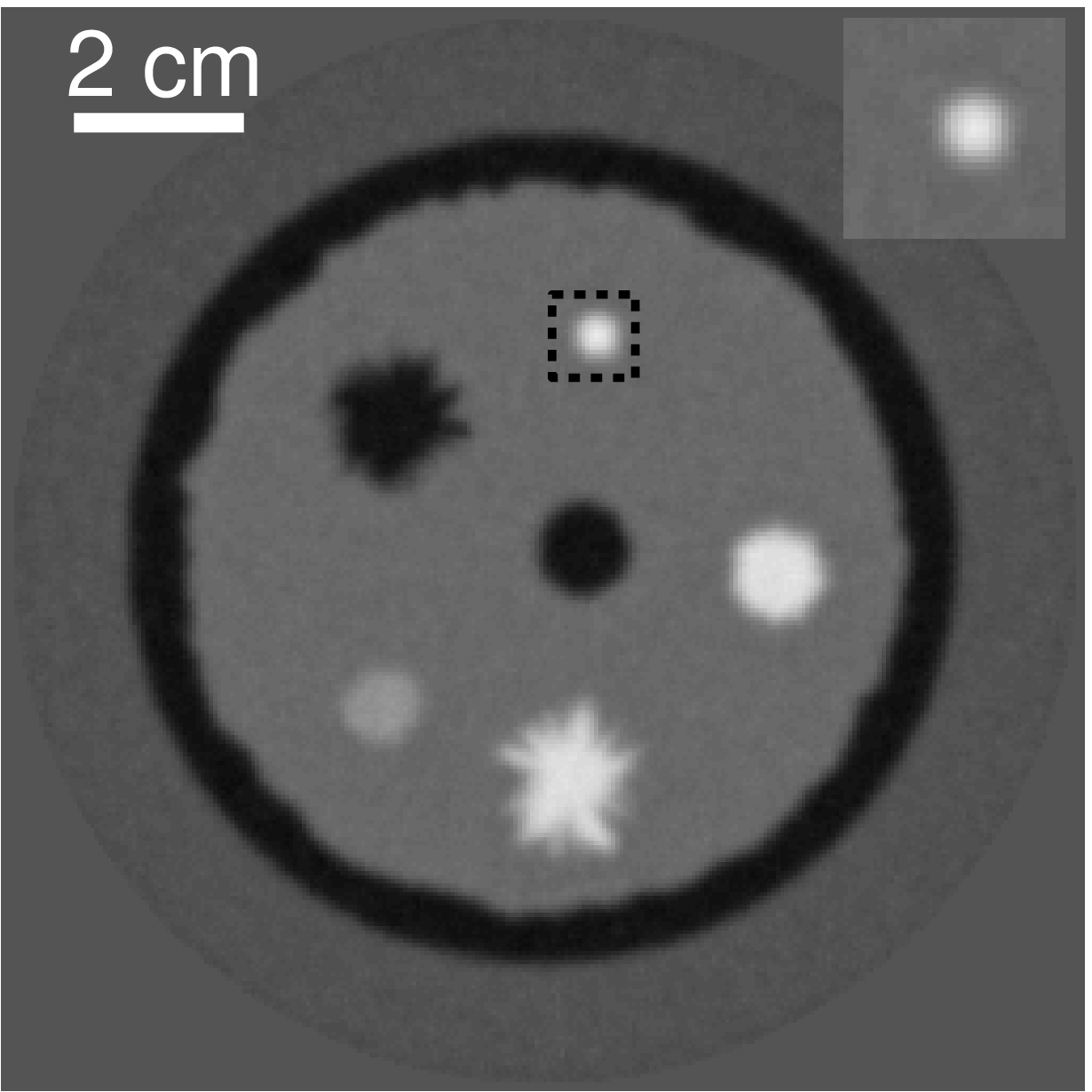}}
\subfloat[]{\includegraphics[width=5.6cm]{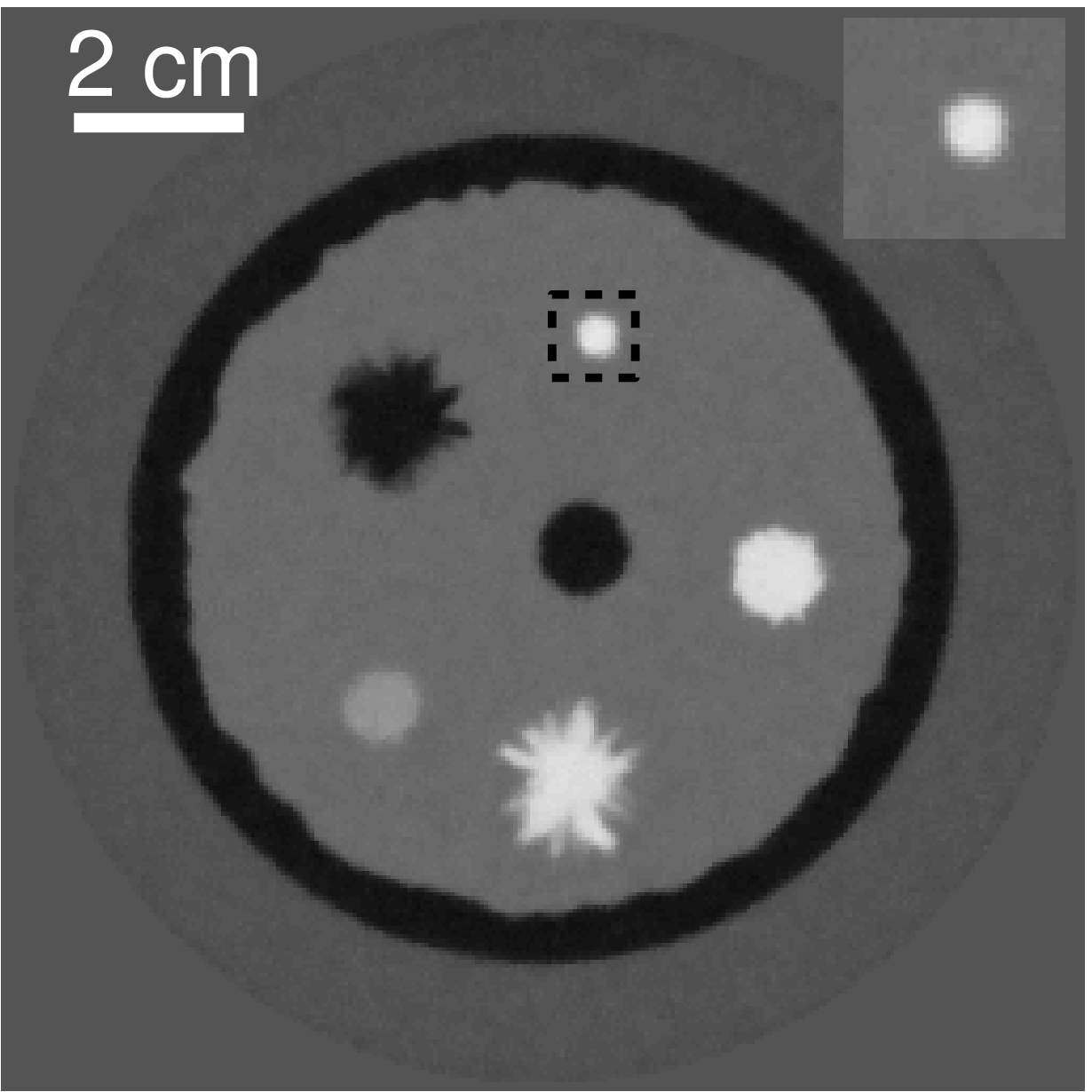}}
\caption{\label{fig:SimNoisy} 
Images reconstructed from non-attenuated data contaminated with Gaussian random noise. 
Images (a-c) were reconstructed by use of the WISE method with a quadratic penalty with 
$\beta^{\rm Q} = 1.0\times 10^{-3}$, $1.0\times 10^{-2}$, and $1.0\times 10^{-1}$, respectively. 
Image (d) was reconstructed by use of the WISE method with a TV penalty with $\beta^{\rm TV} = 5.0\times 10^{-4}$.
The insert in the up right corner of each image is the zoomed-in image of the dashed black box, which contains $35\times 35$ pixels ($17.5\times 17.5$ mm$^2$). 
The grayscale window is $[1.46,1.58]$ mm/$\mu$s.
}
\end{figure}

\begin{figure}[h]
\centering
\subfloat[]{\includegraphics[width=7cm]{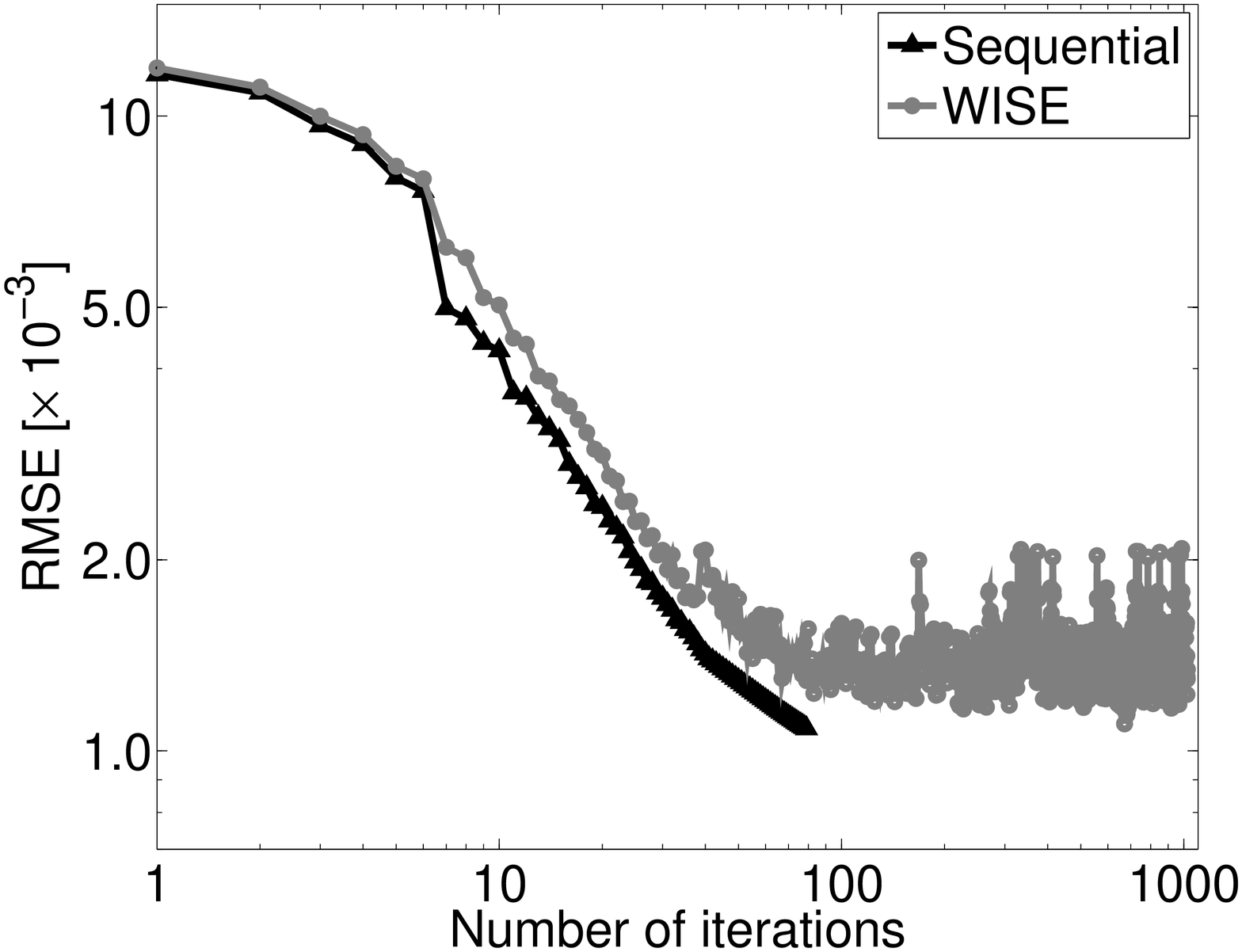}}
\subfloat[]{\includegraphics[width=7cm]{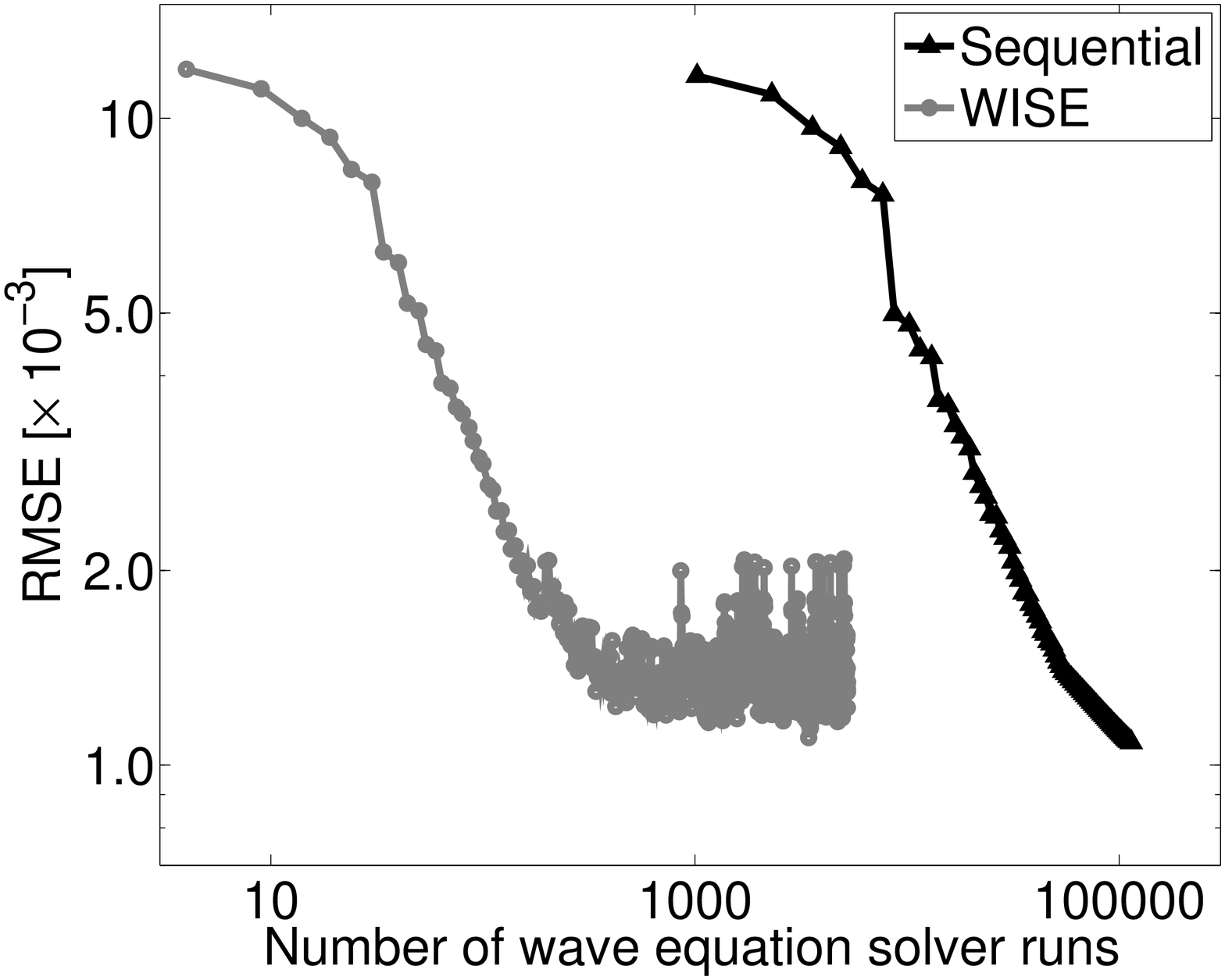}}\\
\subfloat[]{\includegraphics[width=7cm]{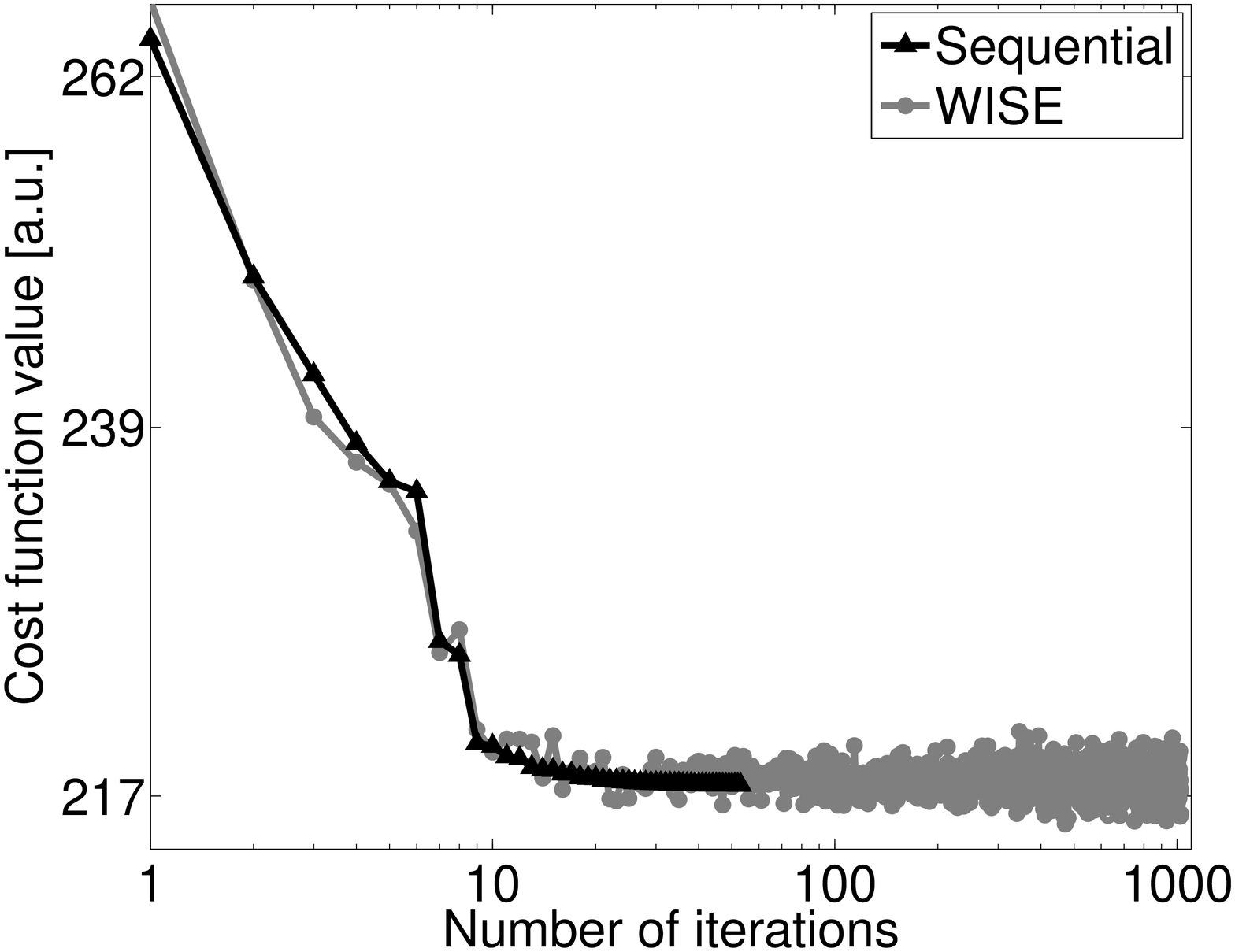}}
\caption{\label{fig:ConvergenceNoisy}
Plots of the root-mean-square errors (RMSEs) of the images reconstructed from the noisy data versus (a) the number of iterations and (b) the number of wave equation solver runs. 
(c) Plots of the cost function value versus the number of iterations. 
Both the WISE and the sequential waveform inversion methods employed a TV penalty with $\beta^{\rm TV}=5.0\times 10^{-4}$. }
\end{figure}

Images reconstructed by use of the WISE method with a quadratic penalty and  the WISE method with a TV penalty from noisy, non-attenuated, data are presented
in Fig.~\ref{fig:SimNoisy}.
All images were obtained after $1024$ algorithm iterations.
%In general, the WISE method is robust to noise as shown in Fig.~\ref{fig:SimNoisy}. 
The WISE method with a quadratic penalty effectively mitigates image noise as shown in Figs.~\ref{fig:SimNoisy}-(a-c),
 at the expense of image resolution, as expected.
%Also, as suggested in the literature \cite{XPanIP2009,Kun:PMB,LiBentRay09:TV},
% using a TV penalty can achieve a better balance between noise mitigation and resolution. 
Figure \ref{fig:SimNoisy}-(d) shows an  image reconstructed by use of the WISE method with a TV penalty. 
The image appears to possess a similar
 resolution but a lower noise level than the image  in Fig.~\ref{fig:SimNoisy}-(b)
that was reconstructed by use of the WISE method with a quadratic penalty.
% This observation is further confirmed by the profiles shown in Fig.~\ref{fig:ProfNoisy}.
%These results demonstrate that the WISE method is robust with respect to measurement noise.
We also compared the convergence rates of the WISE method and the sequential waveform inversion methods
when both utlize a TV penalty and the same regularization parameter.  As shown in Fig.~\ref{fig:ConvergenceNoisy}, 
the convergence properties of the penalized methods follow similar trends
as the un-penalized methods, which were discussed above and shown in Fig.~\ref{fig:Convergence}.
Even though it required a larger number of algorithm iterations, the WISE method
 reduced the computation time by approximately two-orders of magnitude as compared to the
sequential waveform inversion method.

\subsection{Images reconstructed from acoustically attenuated data}

\begin{figure}[h]
\centering
\subfloat[]{\includegraphics[height=5.6cm]{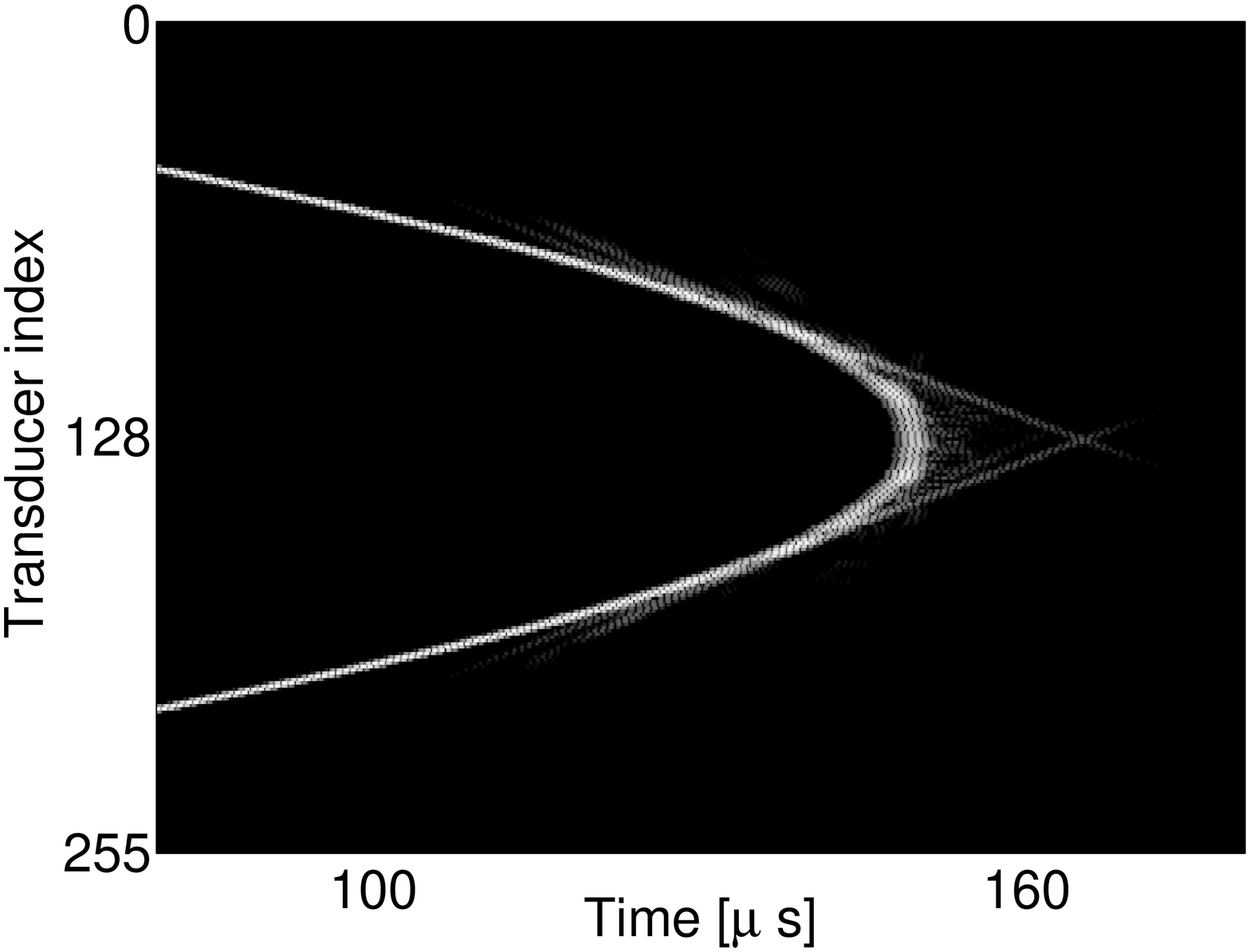}}
\subfloat[]{\includegraphics[height=5.6cm]{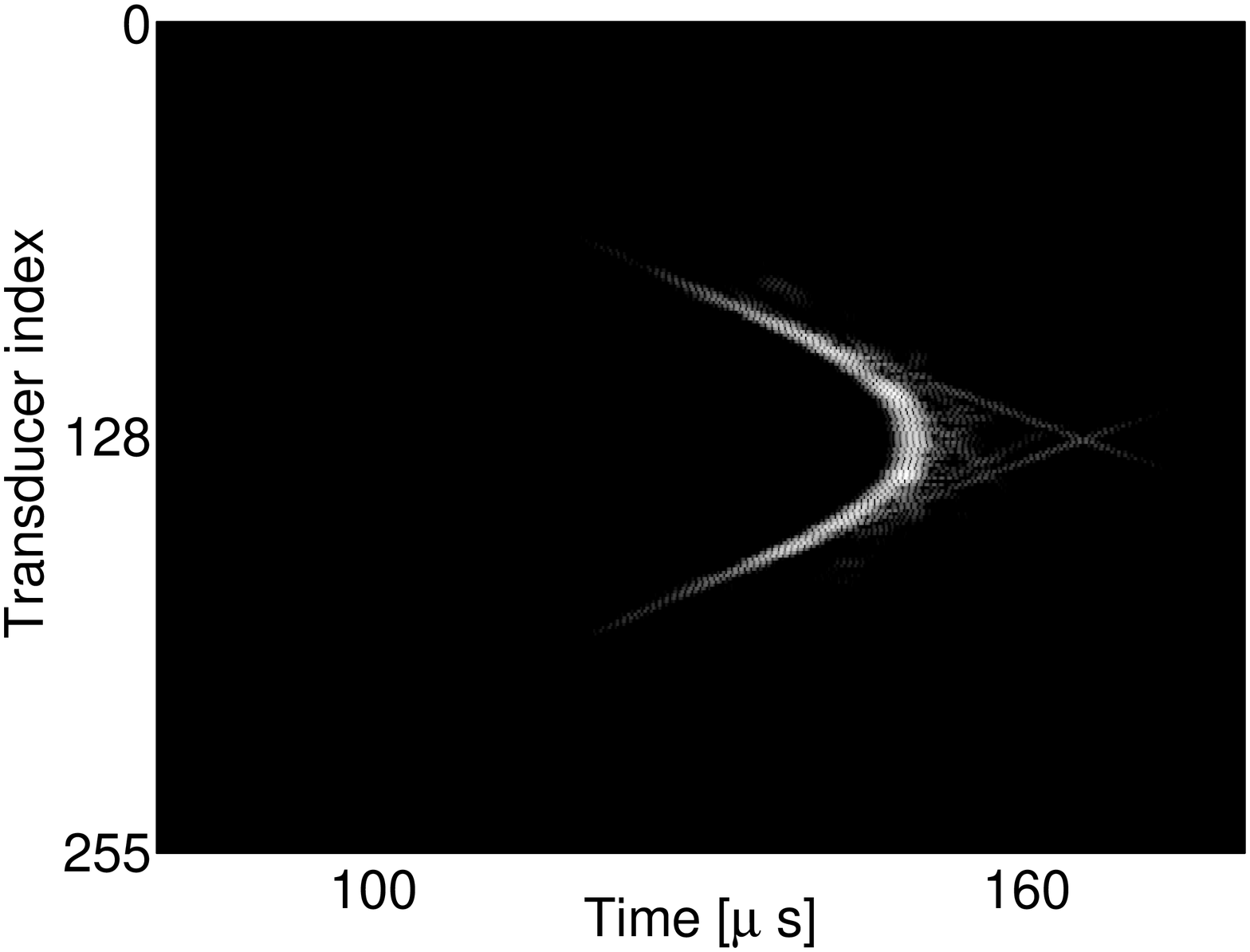}}\\
\subfloat[]{\includegraphics[height=5.4cm]{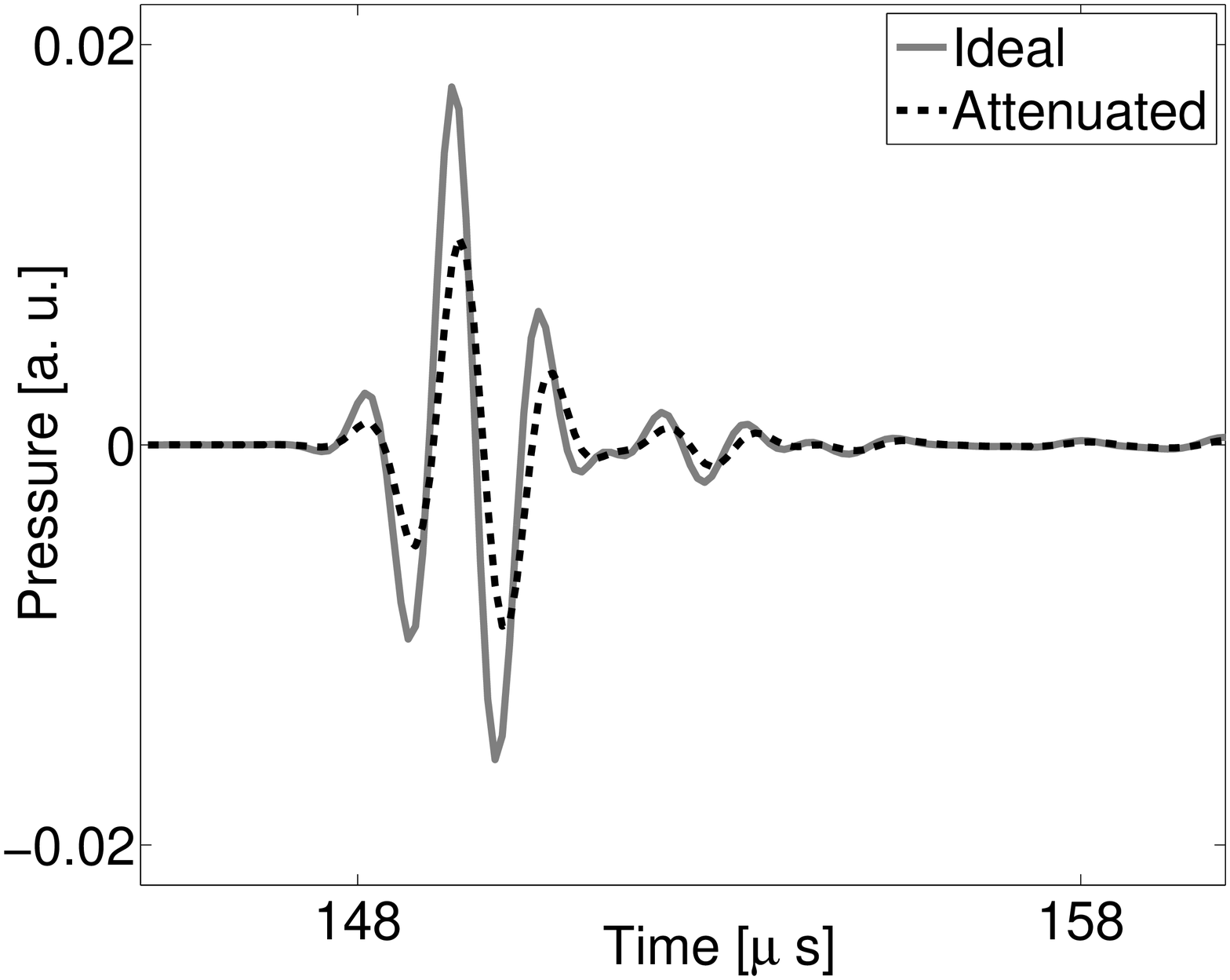}}
\subfloat[]{\includegraphics[height=5.4cm]{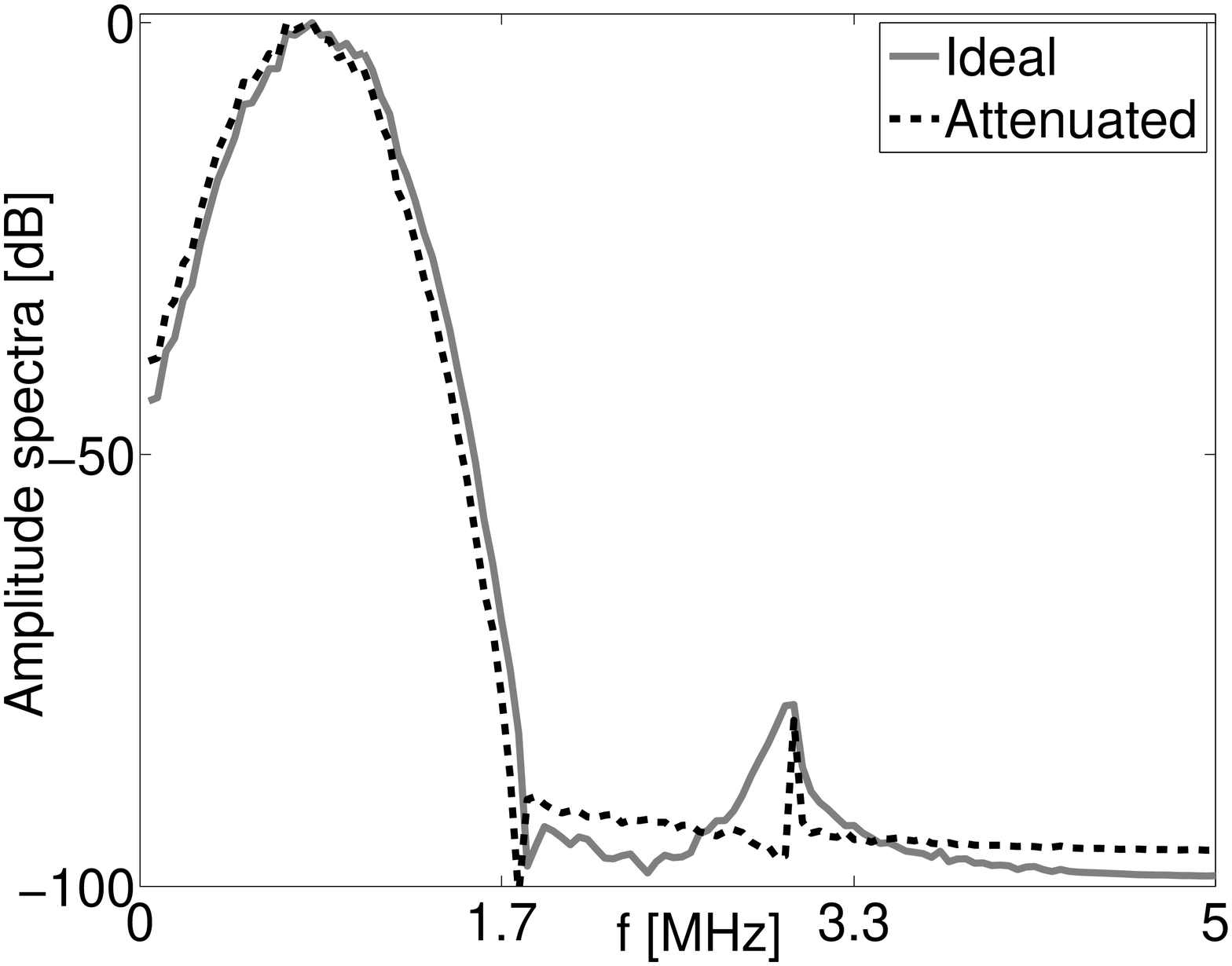}}
\caption{\label{fig:AttPre}
(a) Computer-simulated noise-free attenuated pressure of the $0$-th data acquisition. 
(b) The difference between the attenuated pressure data and the non-attenuated pressure data. 
(c) The temporal profiles and (d) the amplitude spectra of the pressure received by the $128$-th transducer.
The grayscale window for (a) and (b) is $[-45,0]$ dB. 
}
\end{figure}

Our current implementation of the WISE method assumes an absorption-free acoustic medium.
This assumption can be strongly violated in practice.  
In order to investigate the robustness of the the WISE method to model errors associated with ignoring medium acoustic absorption, we applied the algorithm to the acoustically attenuated data that were produced as described in Section \ref{Subsect:GenData}.
As shown in Fig.~\ref{fig:AttPre}, when acoustic absorption is considered, the amplitude of the measured pressure is attenuated by approximately a factor of $2$.
The wavefront (See Fig.~\ref{fig:AttPre}-(a)) remains very similar to that when medium absorption is ignored (See Fig.~\ref{fig:PreNoise}-(a)). 
Medium absorption has the largest impact on the pressure data received by transducers located opposite the emitter as shown in Fig.~\ref{fig:AttPre}-(b). 
The shape of the pulse profile remains very similar as shown in Fig.~\ref{fig:AttPre}-(c) and -(d), suggesting that waveform dispersion may be less critical than amplitude attenuation in image reconstruction for this phantom. 

\begin{figure}[h]
\centering
\subfloat[]{\includegraphics[width=5.6cm]{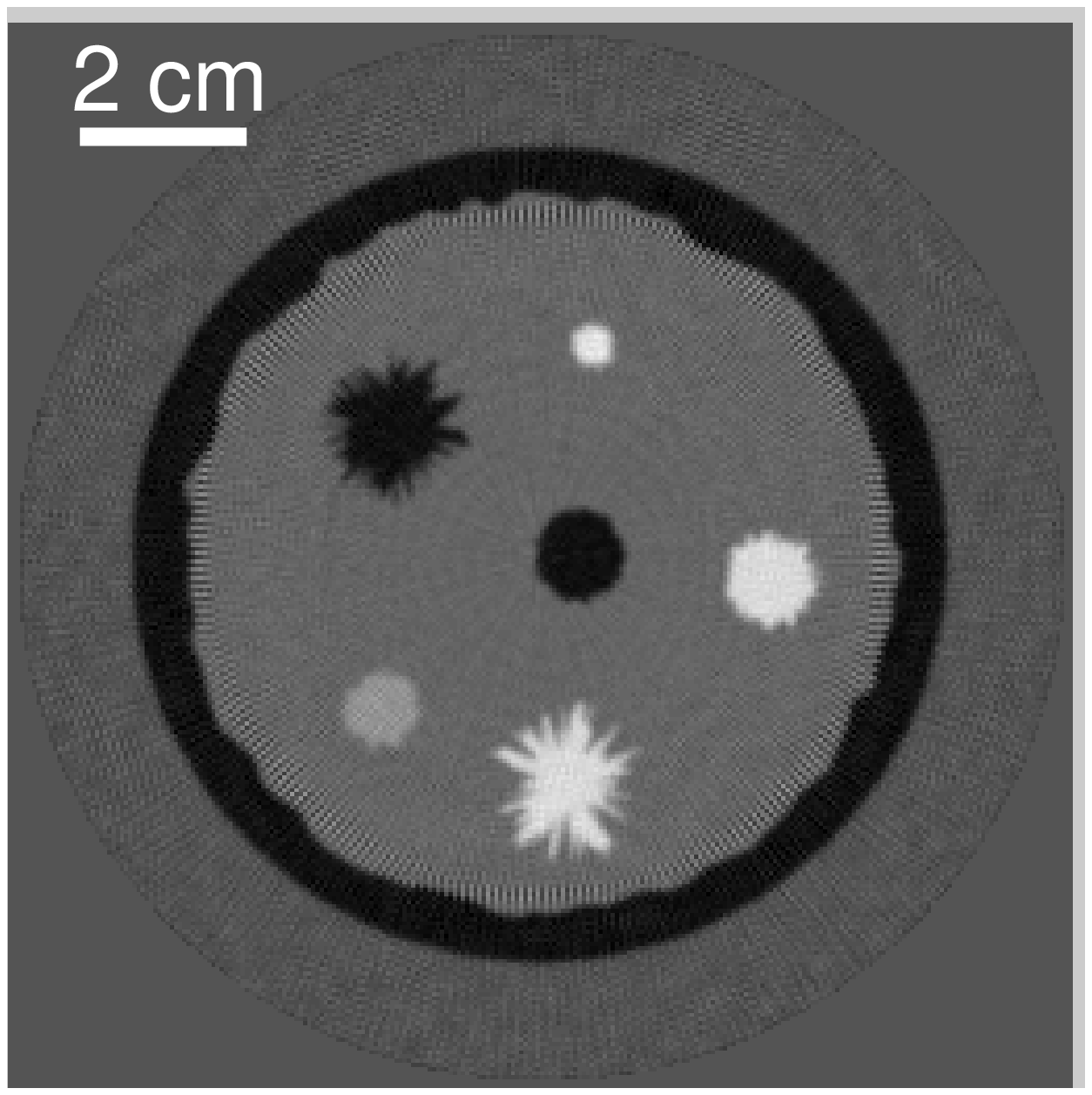}}
\subfloat[]{\includegraphics[width=5.6cm]{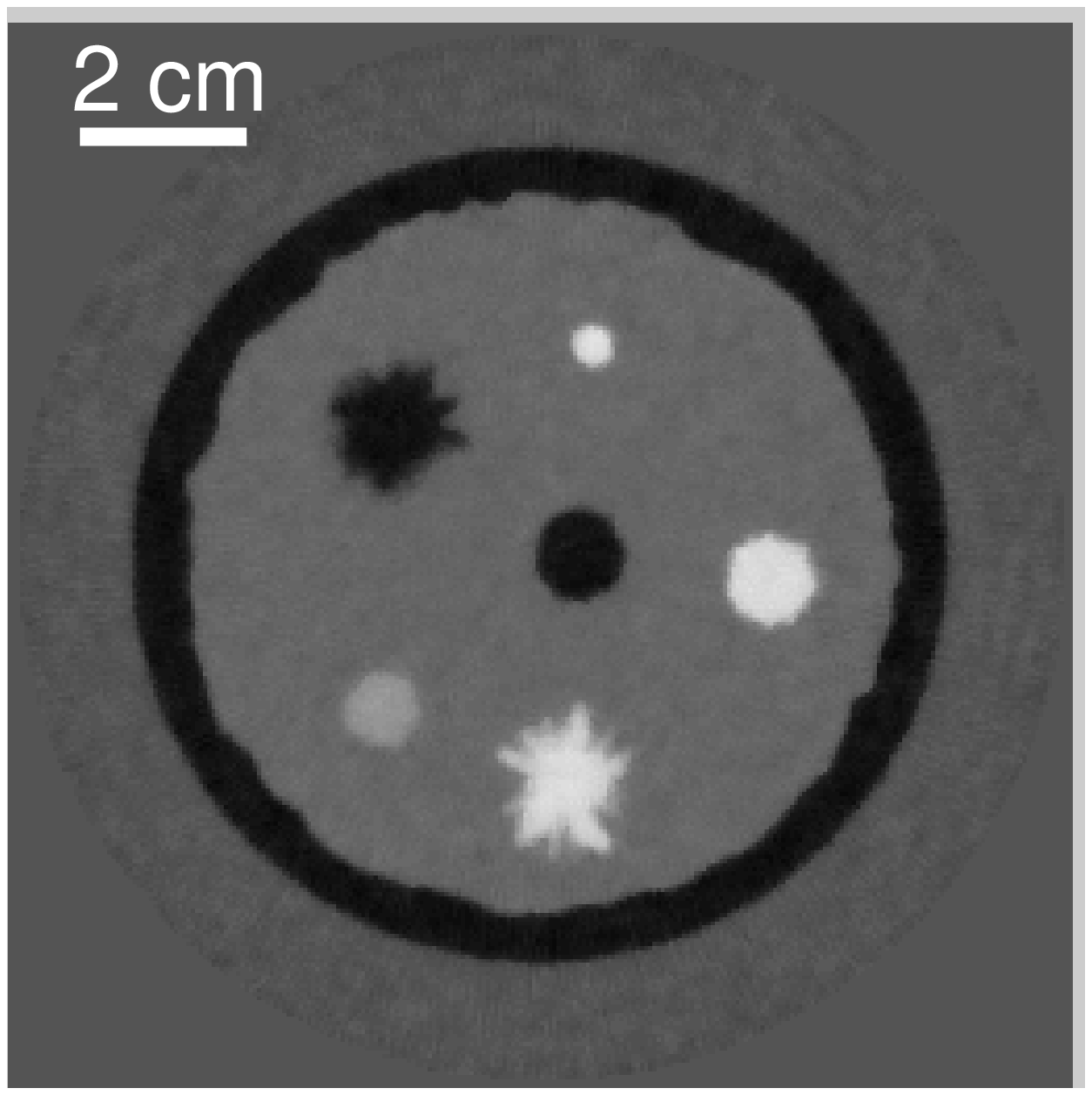}}\\
\subfloat[]{\includegraphics[width=7cm]{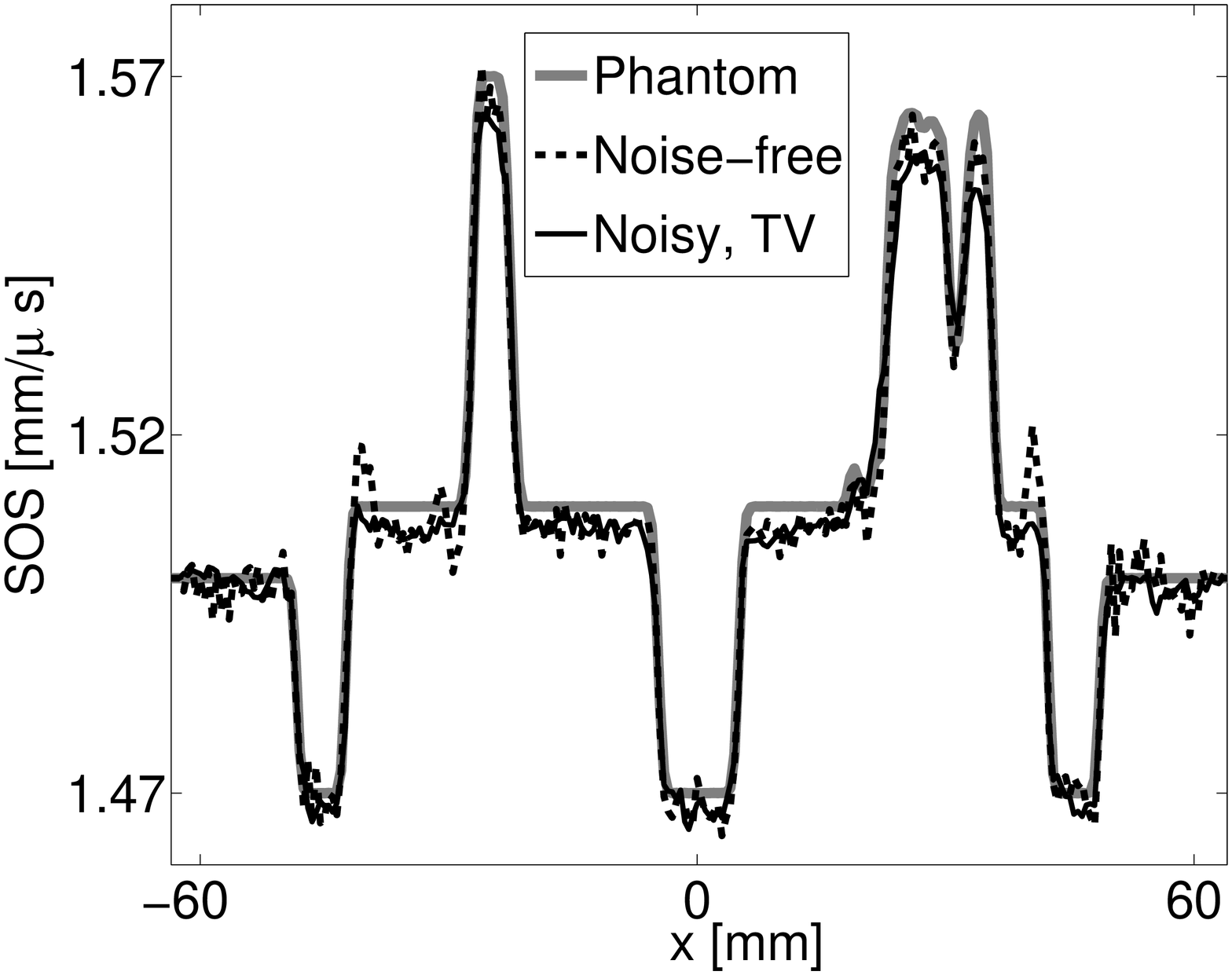}}
\caption{\label{fig:AttRecon}
(a) Image reconstructed by use of the WISE method from the noise-free attenuated data. 
(b) Image reconstructed by use of the WISE method with a TV penalty with $\beta^{\rm TV} = 5.0\times10^{-4}$, 
from the noisy attenuated data. 
The grayscale window is $[1.46,1.58]$ mm/$\mu$s.
(c) Profiles  at $y=6.5$ mm of the reconstructed images. 
}
\end{figure}

Images reconstructed by use of the WISE method with a TV penalty from noise-free and noisy attenuated
data are shown in  Figs.~\ref{fig:AttRecon}-(a) and (b). Image profiles are shown in  Fig.~\ref{fig:AttRecon}-(c).
Although these images contain certain artifacts that were not produced in the idealized data studies,
 most object structures remain readily identified. 
These results suggest that the WISE method with a TV penalty can tolerate data inconsistencies associated
with neglecting acoustic attenuation in the imaging model, at least to a certain level with regards
to feature detection tasks.
 %errors caused by ignoring the medium absorption, at least to a certain level. 

\subsection{Images reconstructed from idealized incomplete data} 

\begin{figure}[h]
\centering
\subfloat[]{\includegraphics[height=5.6cm]{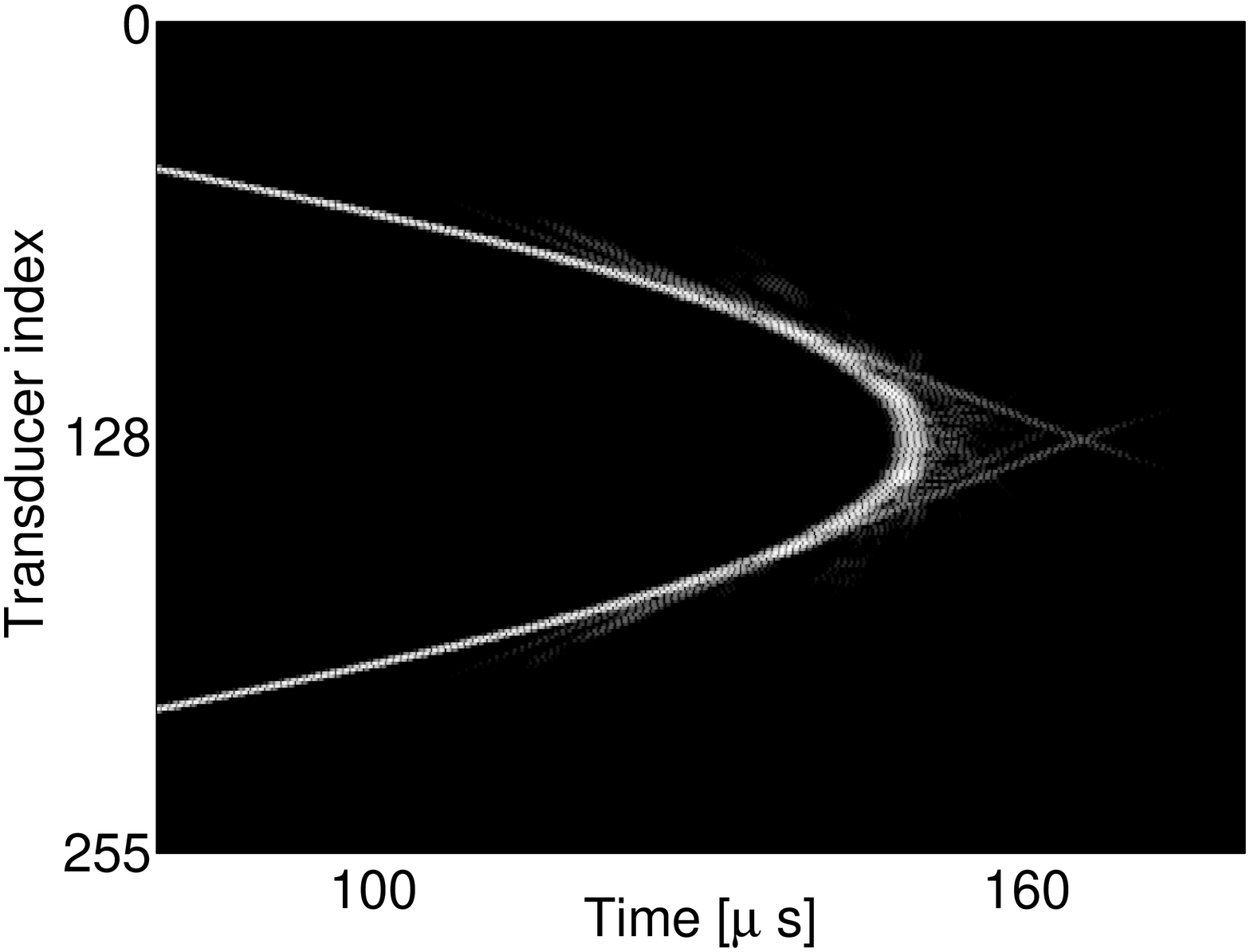}}
\subfloat[]{\includegraphics[height=5.6cm]{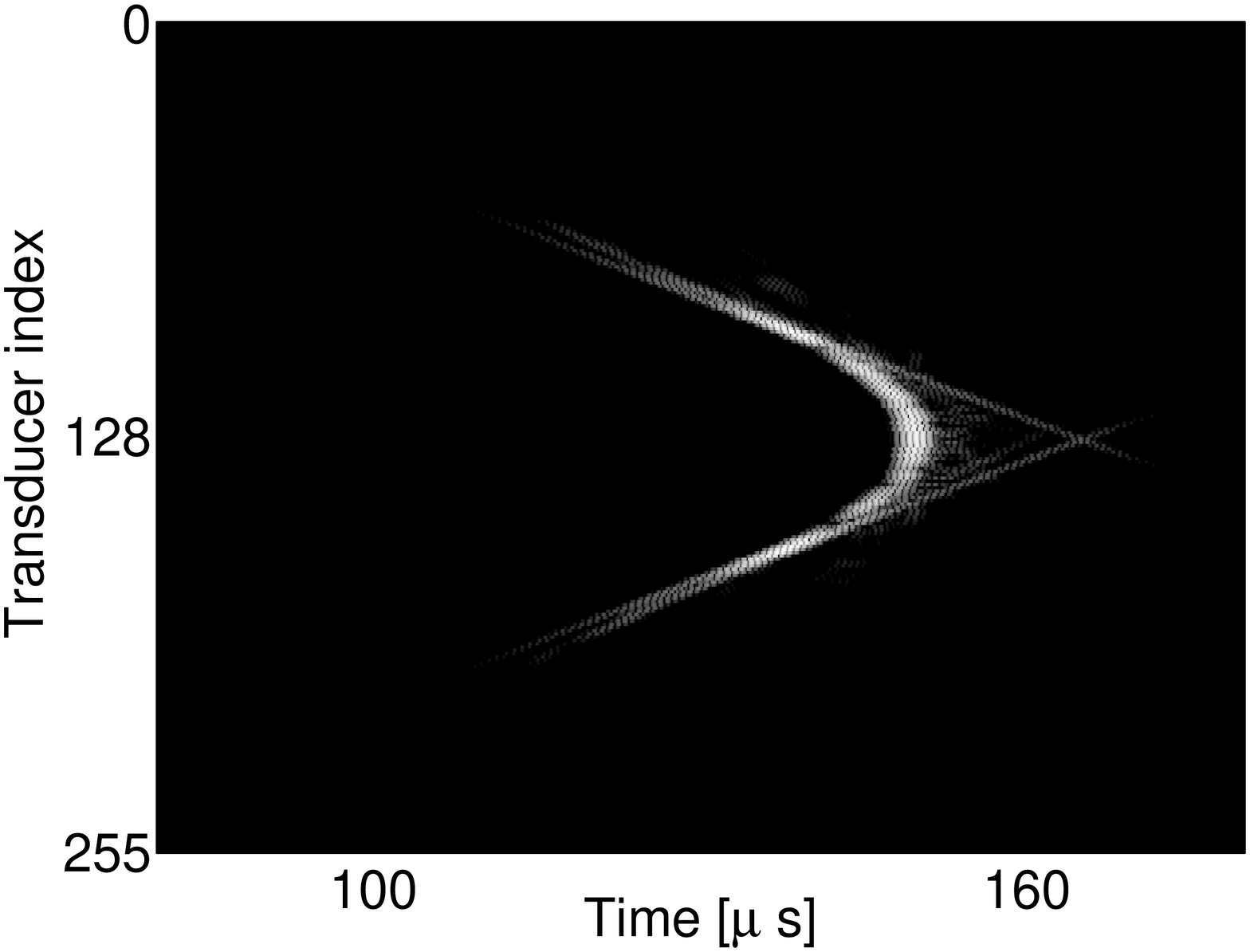}}\\
\subfloat[]{\includegraphics[height=5.6cm]{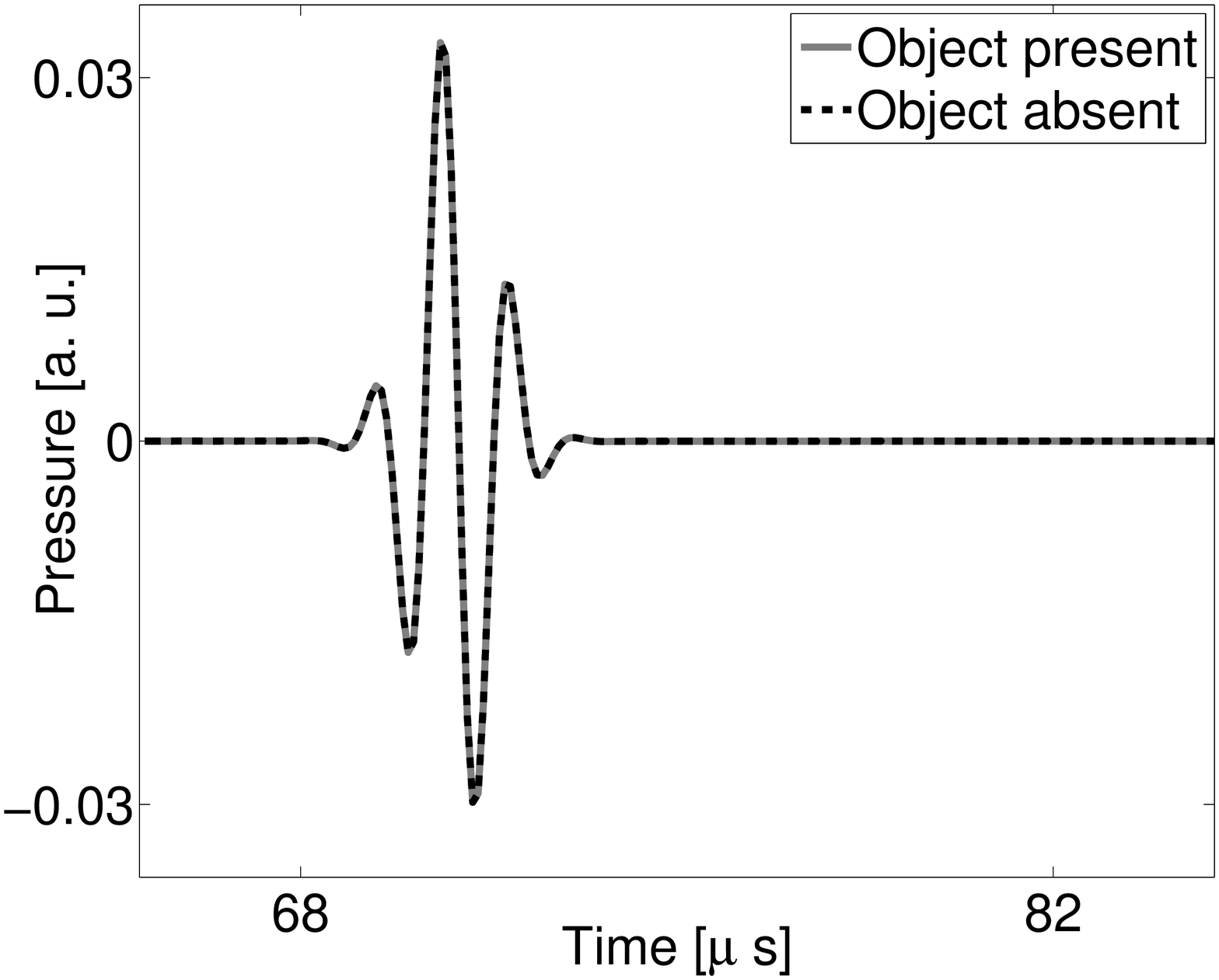}}
\caption{\label{fig:DataCompletionPre} 
(a) Computer-simulated noise-free non-attenuated pressure data when the object is absent. 
(b) The difference between the pressure data when object is present and the pressure data when the object is absent.
(c) Profiles of the pressure received by the $40$-th transducer. 
The grayscale window for (a) and (b) is $[-45,0]$ dB. 
}
\end{figure}

The wavefront of the noise- and attenuation-free pressure wavefield when the object is absent (Fig.~\ref{fig:DataCompletionPre}-(a)) appears to be very similar to that when the object is present (Fig.~\ref{fig:PreNoise}-(a)). 
As expected, the largest differences are seen in the signals received by the transducers located opposite of the emitter, as shown in Fig.~\ref{fig:DataCompletionPre}-(b). 
As seen in Fig.~\ref{fig:DataCompletionPre}-(c), the time traces received by the $40$-th transducer are nearly identical when object is present and absent. 
This is because the back-scattered wavefield is weak for breast imaging applications. 
These results establish the potential efficacy of the data completion strategy of filling the missing data with the pressure data corresponding to a water bath.

\begin{figure}[h]
\centering
\subfloat[]{\includegraphics[width=5.6cm]{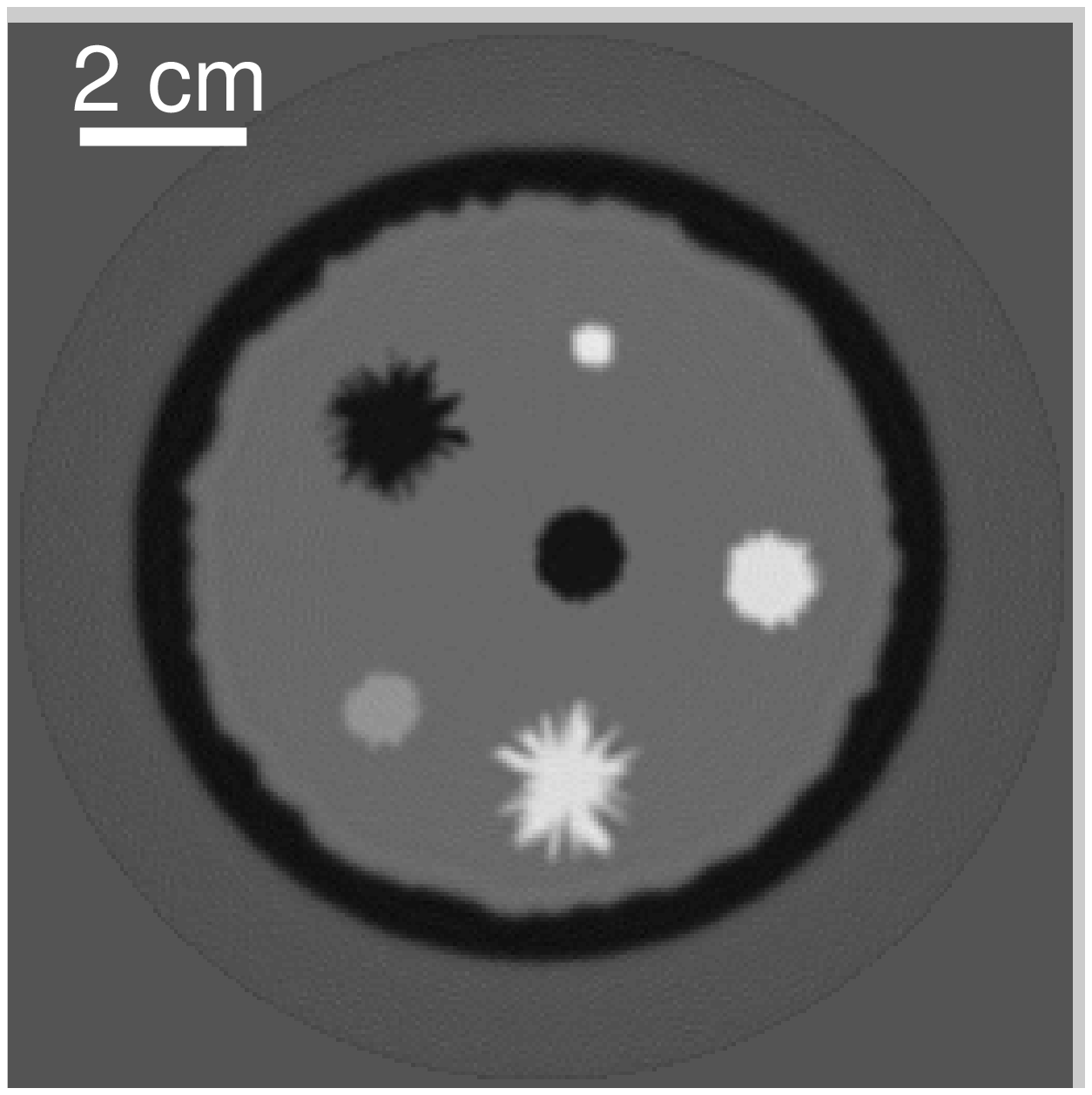}}
\subfloat[]{\includegraphics[width=5.6cm]{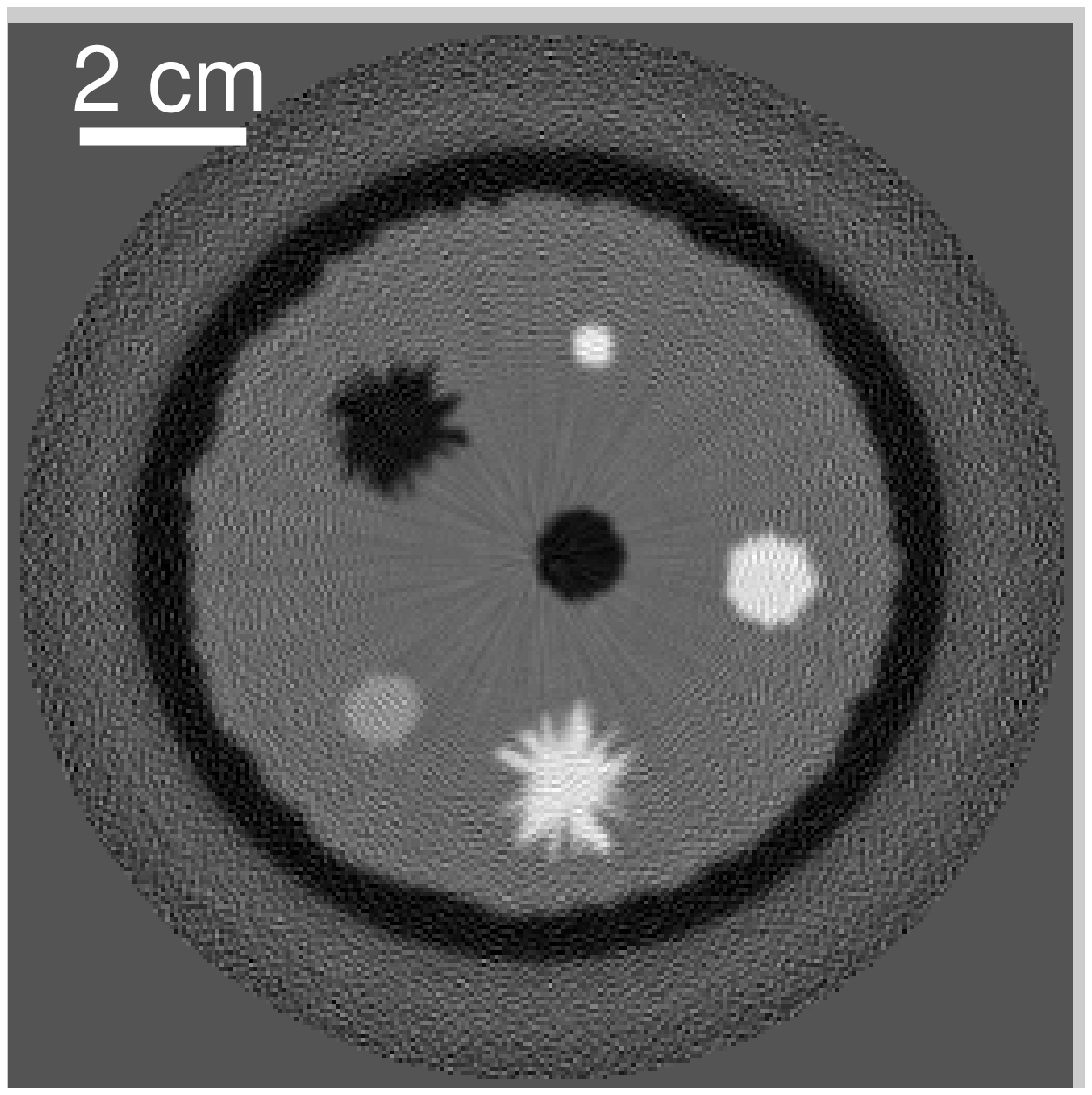}}\\
\subfloat[]{\includegraphics[width=7cm]{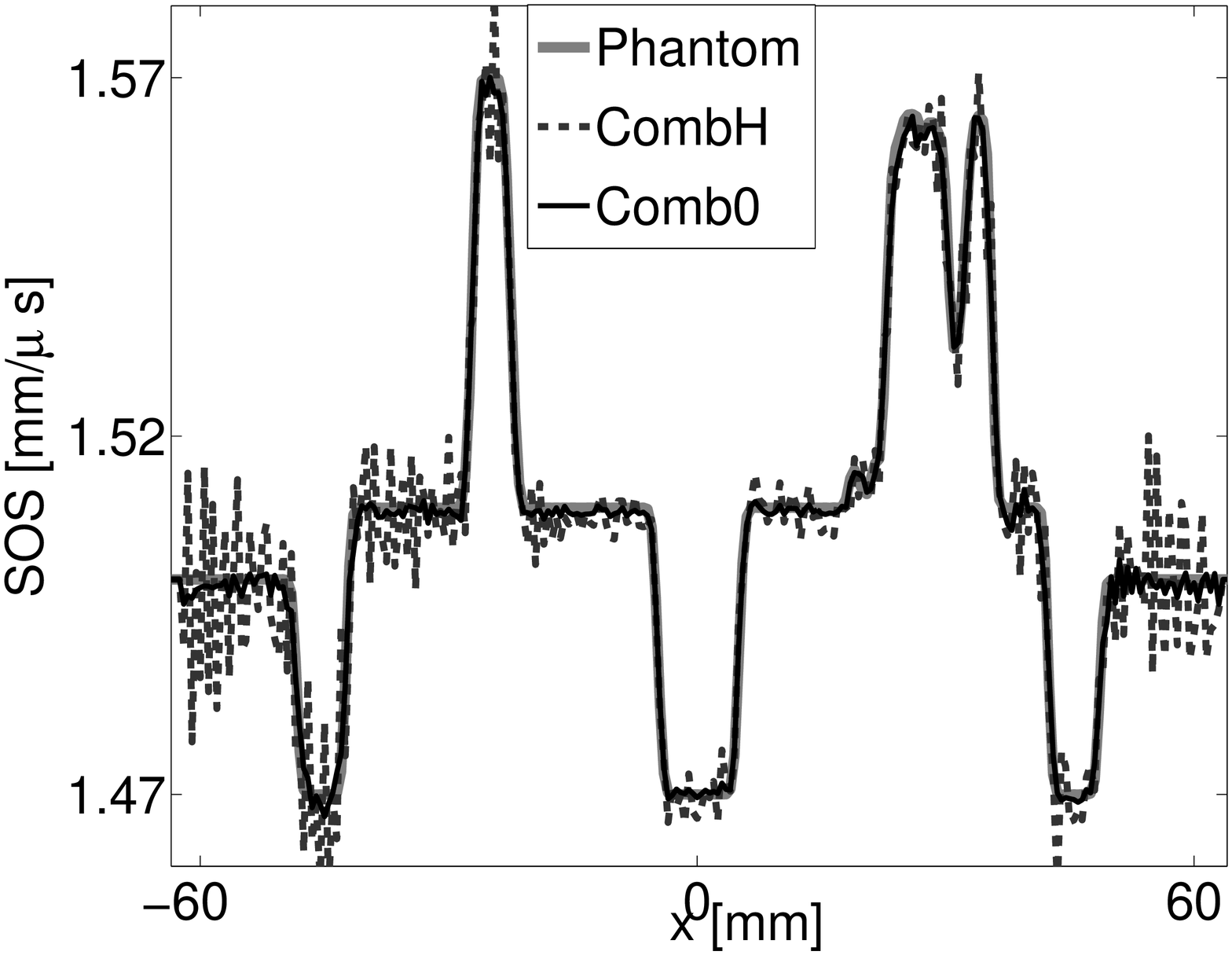}}
\caption{\label{fig:DataCompletionRecon} 
Images reconstructed by use of the WISE method from noise-free combined data that are completed
(a) with computer-simulated pressure corresponding to a homogeneous medium
% $\{\underline{\mathbf g^{\rm comb0}_m}\}$ 
and 
(b) with zeros. 
% from $\{\underline{\mathbf g^{\rm combH}_m}\}$. 
The grayscale window is $[1.46,1.58]$ mm/$\mu$s.
(c) Profiles at $y=6.5$ mm of the images reconstructed by use of the WISE method from the two 
combined data sets.
}
\end{figure}

The image reconstructed from the measurements completed with pressure data corresponding to a water bath
is shown in Fig.\ref{fig:DataCompletionRecon}-(a).
%method produces an accurate reconstruction from the data completed with pressure corresponding to a homogeneous medium. 
As revealed by the profile in Fig.\ref{fig:DataCompletionRecon}-(c), this image is highly accurate.
Alternatively, the image reconstructed from the the data completed with zeros contains strong artifacts as shown in Fig.~\ref{fig:DataCompletionRecon}-(b). 
%Profiles through the two reconstructed images are displayed in Fig.~\ref{fig:DataCompletionRecon}-(c).
These results suggest that the WISE method can be adapted to reconstruct images from incomplete data, which is particularly useful for emerging laser-induced USCT imaging systems \cite{manohar,AAO12:LUS,Jun13:USCT}.

\iffalse
\subsection{Images reconstructed from noisy reduced data}
As expected, the FBP algorithm is more sensitive to the reduced data as shown in Fig.~\ref{fig:SimNoisyV128}-(a). 
Meanwhile, the WISE method produces a significantly more accurate image. 
Except for the smallest structure, all other structures are accurate as shown in Fig.~\ref{fig:SimNV128Profile}. 
These results suggest that the WISE method will produce more accurate images when the transducer array only contains a limited number of elements. 
In order to improve the quantitative accuracy of the smallest structure, we employed the TV penalty. 
The image shown in Fig.~\ref{fig:SimNoisyV128}-(c) shows a significant improvement of the spatial resolution with little increase in noise level. 
The profiles in Fig.~\ref{fig:SimNV128Profile} also confirm the quantitative value of the smallest structure was estimated more accurately. 
\fi

\section{Experimental validation}
% Experimental phantom studies were conducted to validate the proposed WISE method. 

\subsection{Data acquisition}
Experimental data recorded by use of the SoftVue USCT scanner \cite{SoftVue:14}
was utilized to further validate the WISE method.
The scanner contained a ring-shaped  array of radius $110$ mm
that was populated with 2048 transducer elements.  
Each element had a center frequency of $2.75$ MHz, a pitch  of $0.34$ mm,  
and was elevationally focused to isolate a  $3$ mm thick slice of the to-be-imaged object.  
The transducer array was mounted in a water tank and could be translated with a motorized gantry in the vertical direction.  
Readers are referred to \cite{SoftVue:14} for additional details regarding the system.

\begin{figure}[h]
\centering
\includegraphics[height=5.6cm]{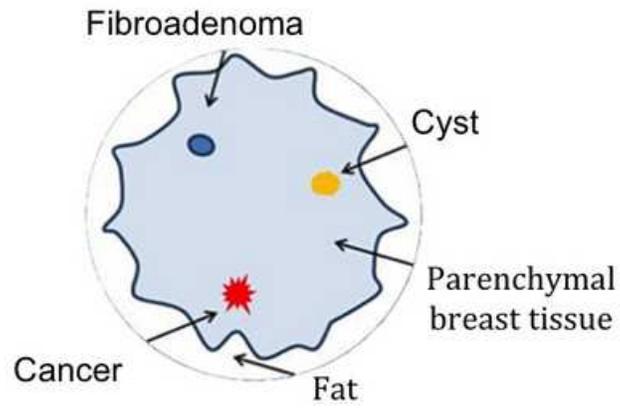}
\caption{\label{fig:SchExpPhantom}
Schematic of the breast phantom employed in the experimental study. 
}
\end{figure}

The breast phantom was built by Dr. Ernie Madsen from the University of Wisconsin and provides tissue-equivalent scanning characteristics of highly scattering, predominantly parenchymal breast tissue. 
The phantom mimics the presence of benign and cancerous masses embedded in glandular tissue, including a subcutaneous fat layer.  
Figure \ref{fig:SchExpPhantom} displays a schematic of one slice through the phantom. 
The diameter of the inclusions is approximately $12$ mm. 
Table \ref{tab:ExpPhantom} presents the known acoustic properties of the phantom.

During data acquisition, the breast phantom was placed near the center of the ring-shaped transducer array so that the distance between the phantom and each transducer was approximately the same.  
While scanning each slice, every other transducer element sequentially emits fan beam ultrasound signals towards the opposite side of the ring.
The forward scattered and backscattered ultrasound signals are subsequently recorded by the same transducer elements.
The received waveform  was sampled at a rate of 12 MHz.    
The 1024 data acquisitions required approximately 20 seconds in total. 
A calibration data set was also acquired in which the phantom object was absent.

\subsection{Data pre-processing}
48 bad channels were manually identified by visual inspection. 
After discarding these, the data set contained $M=976$ acquisitions.
Each acquisition contained $N^{\rm rec}=976$ time traces.
Each time trace contained $L=2112$ time samples.
The $976$ good channels were indexed from $0$ to $975$. 
The corresponding data acquisitions were indexed in the same way.
A Hann-window low-pass filter with a cutoff frequency of $4$ MHz was applied
 to every time trace in both the calibration and the measurement data.
This data filtering was implemented to mitigate numerical errors that could be introduced by our second-order wave equation solver. 

\subsection{Estimation of excitation pulse}
\label{SubSect:EstExct}
\begin{figure}[h]
\centering
\subfloat[]{\includegraphics[height=5.6cm]{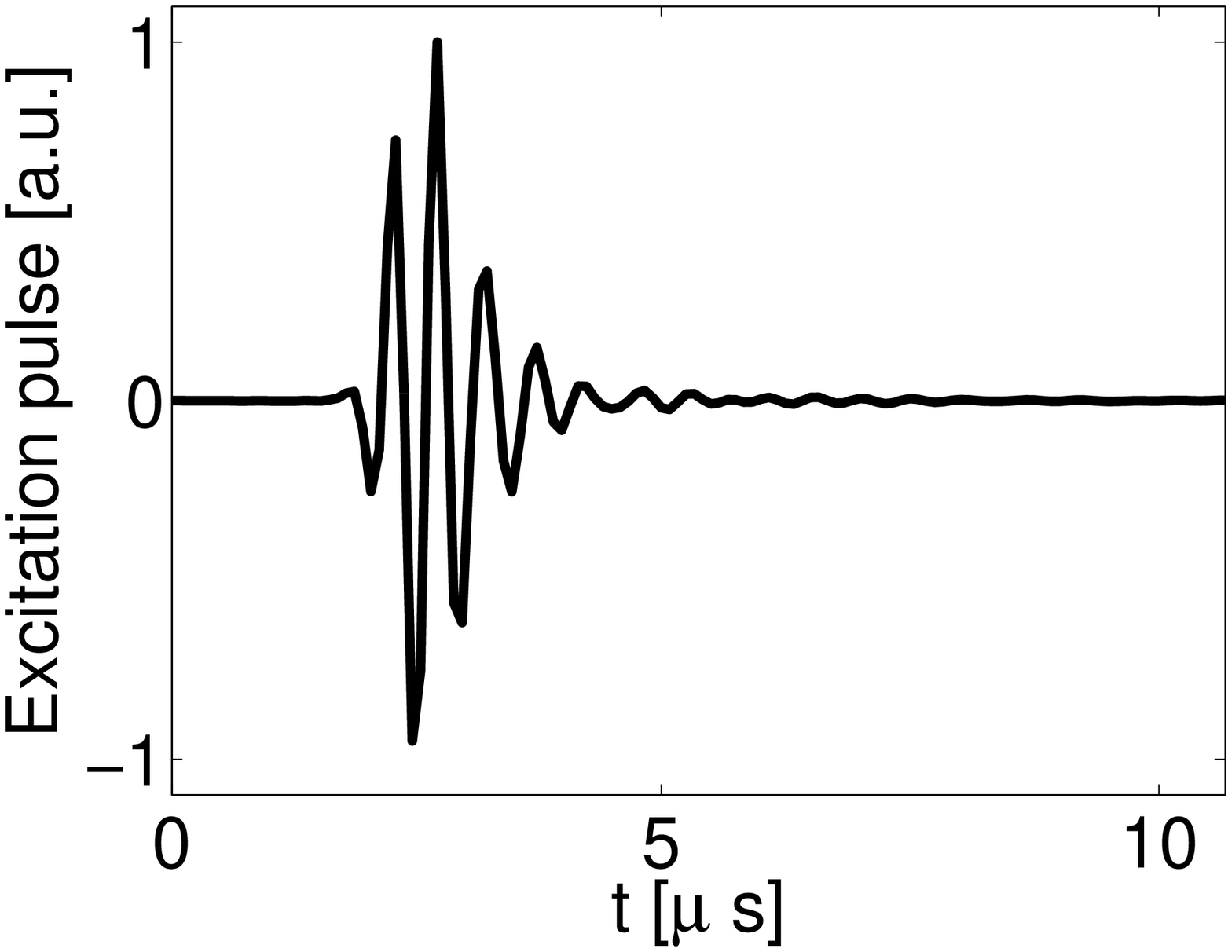}}
\hskip 0.2cm
\subfloat[]{\includegraphics[height=5.6cm]{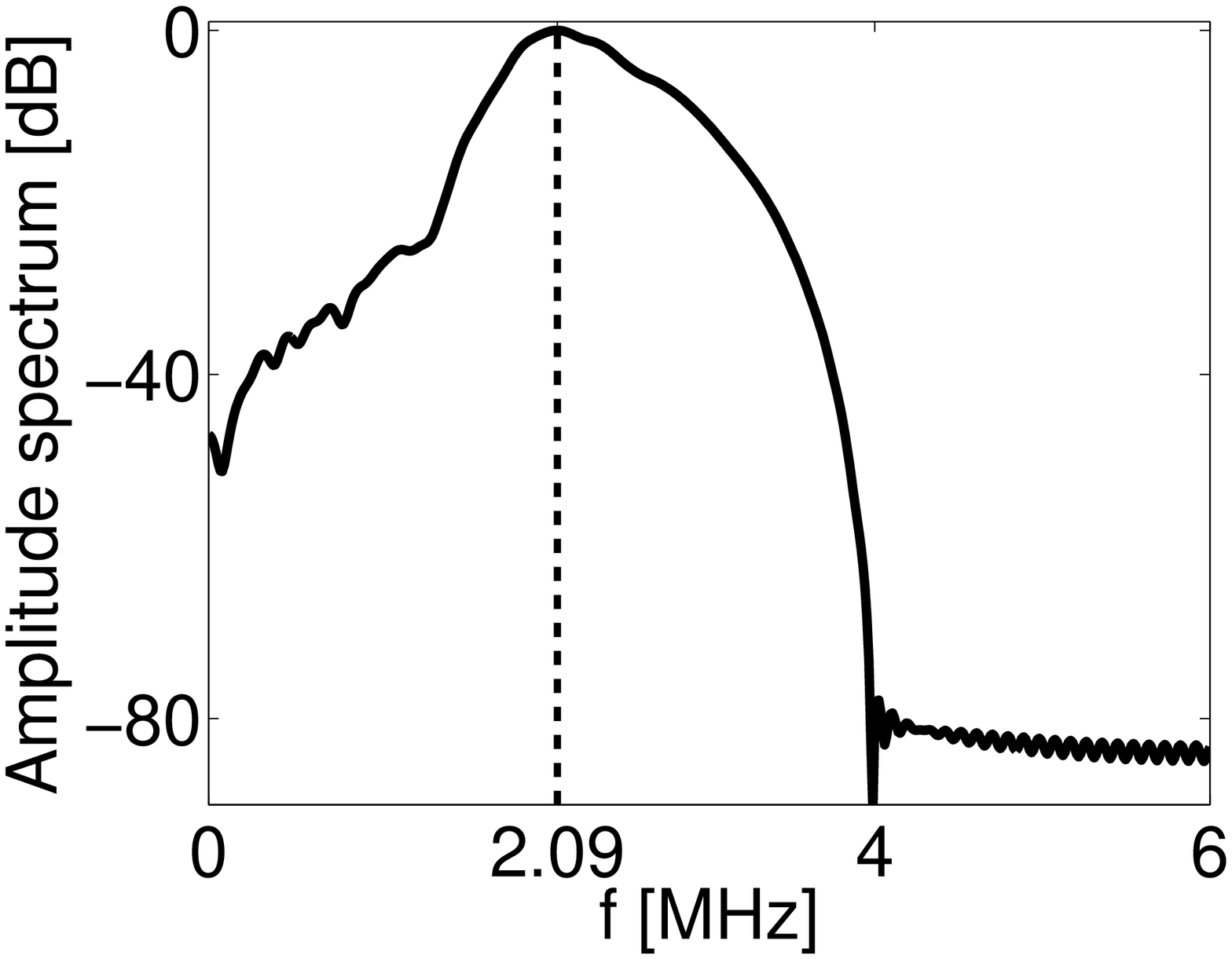}}
\caption{\label{fig:ExpExctPulse} 
(a) Normalized temporal profile and 
(b) amplitude spectrum of the excitation pulse employed in 
the experimental studies. 
The dashed line in (b) marks the center frequency of excitation pulse at $2.09$ MHz.  
}
\end{figure}

The shape of the excitation pulse was estimated as the time trace of the calibration data (after pre-processing) received by the $488$-th receiver at the $0$-th data acquisition. 
Note that the $488$-th receiver was approximated located on the axis of the $0$-th emitter, thus the received pulse was minimally affected by the finite aperture size effect of the transducers. 
Because our calibration data and measurement data were acquired using different electronic amplifier gains, the amplitude of the excitation pulse was estimated from the measurement data.  
More specifically, we simulated the $0$-th data acquisition using the second-order pseudospectral k-space method and compared the simulated time trace received by the $300$-th receiver with the corresponding measured time trace (after pre-processing). 
The ratio between the maximum values of these two traces was used to scale the excitation pulse shape. 
We selected the $300$-th receiver because it resided out of the fan-region indicated in Fig.~\ref{fig:Geo}; its received signals were unlikely to be strongly affected by the presence of the object.  
The estimated excitation pulse and its amplitude spectrum are displayed in Fig.~\ref{fig:ExpExctPulse}.
Note that the experimental excitation pulse contained higher frequency components than did the computer-simulated excitation pulse shown in Fig.~\ref{fig:SimExctPulse}. 

\subsection{Synthesis of combined data}
\begin{figure}[h]
\centering
\subfloat[]{\includegraphics[height=5.6cm]{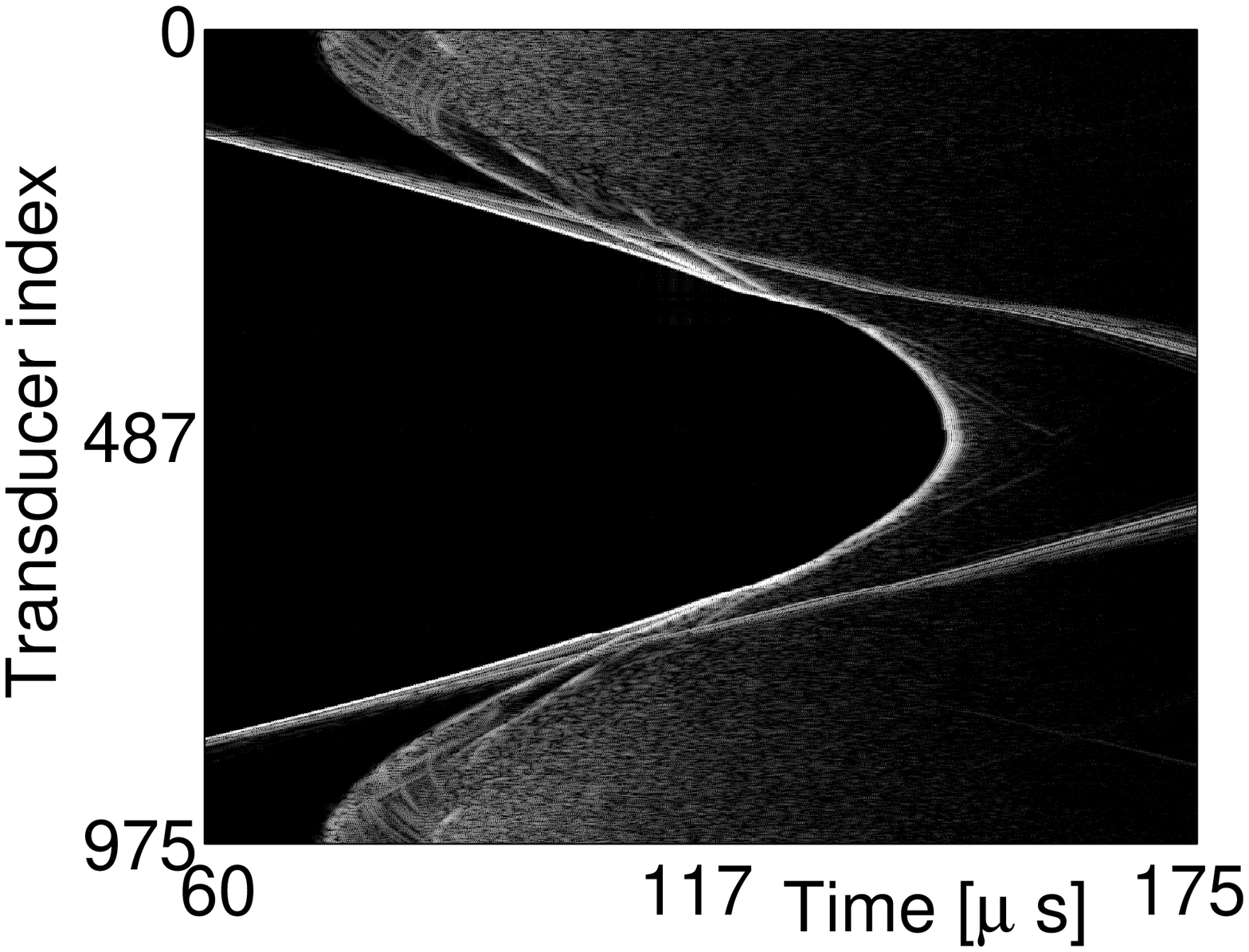}}
\subfloat[]{\includegraphics[height=5.6cm]{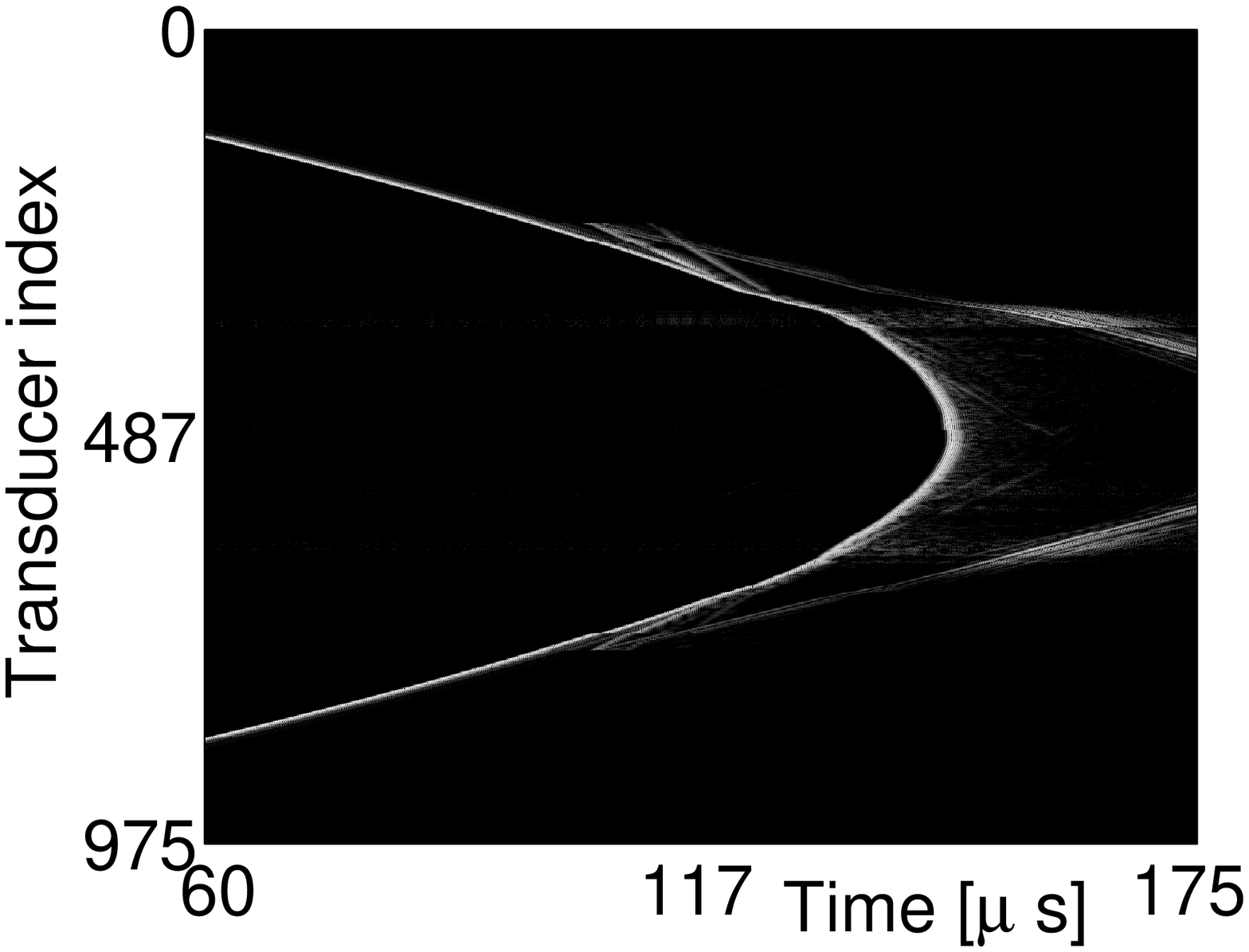}}\\
\subfloat[]{\includegraphics[height=5.6cm]{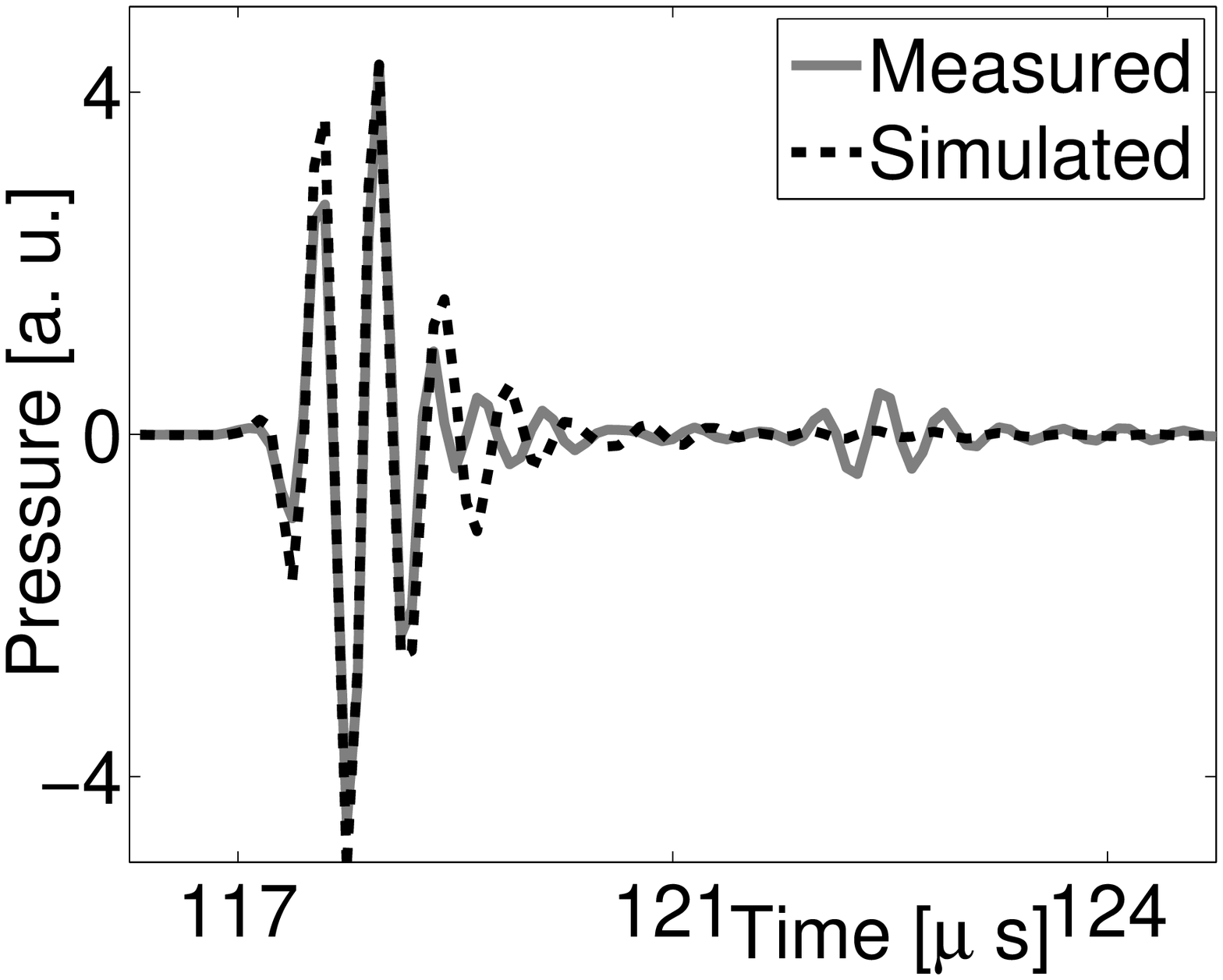}}
\subfloat[]{\includegraphics[height=5.6cm]{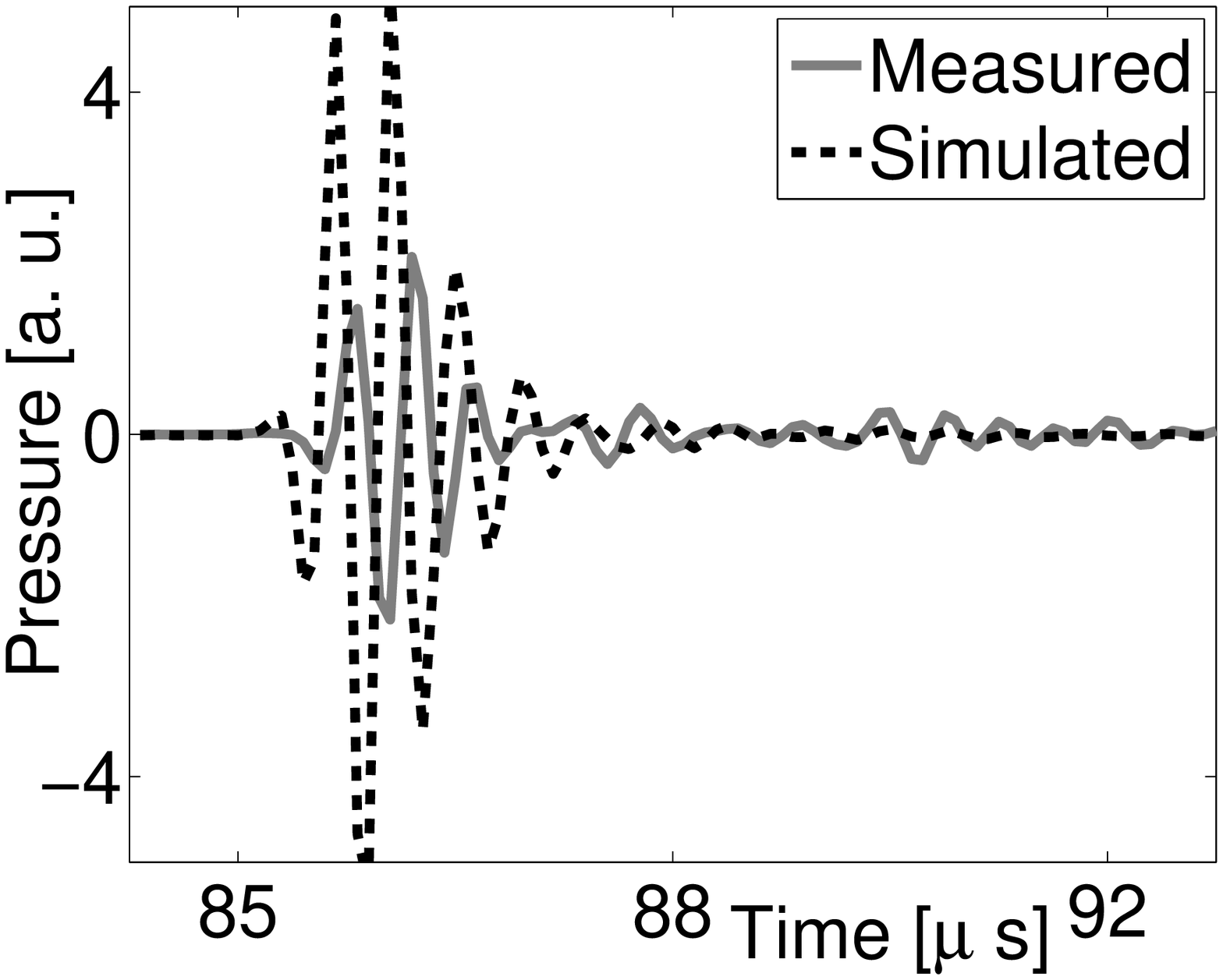}}
\caption{\label{fig:ExpPre} 
Zeroth acquisition of (a) the experimentally-measured raw data and 
(b) the combined data, respectively, and 
time traces at the $0$-th acquisition received by (c) the $300$-th receiver, 
and (d) the $200$-th receiver, respectively. 
The grayscale window for (a) and (b) is $[-45,0]$ dB. 
}
\end{figure}

As discussed in Section \ref{Sect:incomplete}, signals received by receivers located near the emitter can be unreliable \cite{Duric2007:SOSWave}. 
Our experimental data, as shown in Fig.~\ref{fig:ExpPre}-(a), contained noise-like measurements for the receivers indexed from $0$ to $200$, and from $955$ to $975$, in the case where the $0$-th transducer functioned as the emitter. 
Also, our point-like transducer assumption introduces larger model mismatches for the receivers located near the emitter.  
As shown in Figs.~\ref{fig:ExpPre}-(c) and -(d), even though the simulated time trace received by the $300$-th receiver matches accurately with the experimentally measured one, the simulated time trace received by the $200$-th receiver is substantially different compared with the experimentally measured one. 
%Additional discussions regarding model mismatch will be provided in Section \ref{Sect:Conclusions}. 
In order to minimize the effects of model mismatch, we replaced these unreliable
 measurements with computer-simulated water bath data, as described in Section \ref{Subsect:GenData}. 
We designated the time traces received by the $512$ receivers located on the opposite side of the emitter as the reliable measurements for each data acquisition. 
The $0$-th data acquisition of the combined data is displayed in Fig.~\ref{fig:ExpPre}-(b). 

\subsection{Estimation of initial guess}

The initial guess for the WISE method was obtained by use of the bent-ray reconstruction
 method described in Appendix \ref{Sect:BentRay}.
We first filtered each time trace of the raw data by a band-pass Butterworth filter ($0.5$MHz - $2.5$MHz). 
Subsequently, we extracted the TOF by use of the thresholding method with a thresholding value of $20\%$ of the peak value of each time trace. 
The bent-ray reconstruction algorithm was applied for image reconstruction with a measured background sound speed $1.513$ mm/$\mu$s. 
The resulting image is shown in Fig.~\ref{fig:WISEExp}-(a)
and has a pixel size of $1$ mm.      
Finally, the image was smoothed by convolving it with a $2$D Gaussian kernel with a standard deviation of $2$ mm. 
% How the initial guess affects the WISE method performance will be discussed in Section \ref{Sect:Conclusions}.

\subsection{Image reconstruction}
We applied the WISE method with a TV penalty to the combined data set. 
The second-order wave equation solver was employed with a calculation domain of dimensions $512.0\times512.0$ mm$^2$. 
The calculation domain was sampled on a $2560\times 2560$ Cartesian grid with a grid spacing of $0.2$ mm. 
On a platform consisting of dual quad-core CPUs with a $3.30$ GHz clock speed, $64$ GB RAM, and a single NVIDIA Tesla K20 GPU, each numerical solver run, took $40$ seconds to calculate the pressure data for $2112$ time samples.
Knowing the size of the phantom, we set the reconstruction region to be within a circle of diameter $128$ mm, i.e., only the sound speed values of pixels within the circle were updated during the iterative image reconstruction.  
% We stopped the iteration once the summed squared error between two consecutive estimates fell below $0.1\%$.    
We swept the value of $\beta^{\rm TV}$ over a wide range to investigate its impact on the reconstructed images.

\subsection{Images reconstructed from experimental data}

\begin{figure}[h]
\centering
\subfloat[]{\includegraphics[width=5cm]{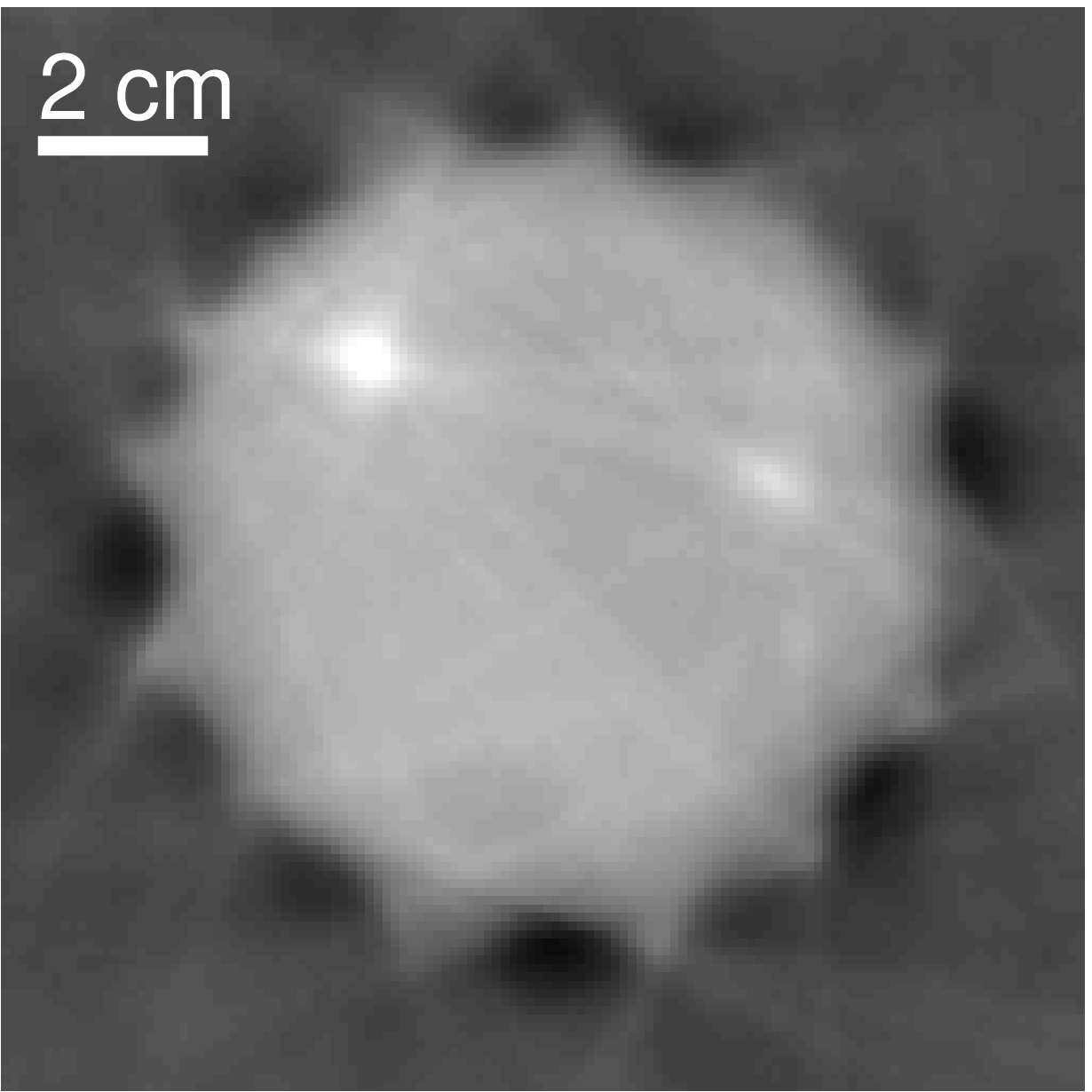}}
\subfloat[]{\includegraphics[width=5cm]{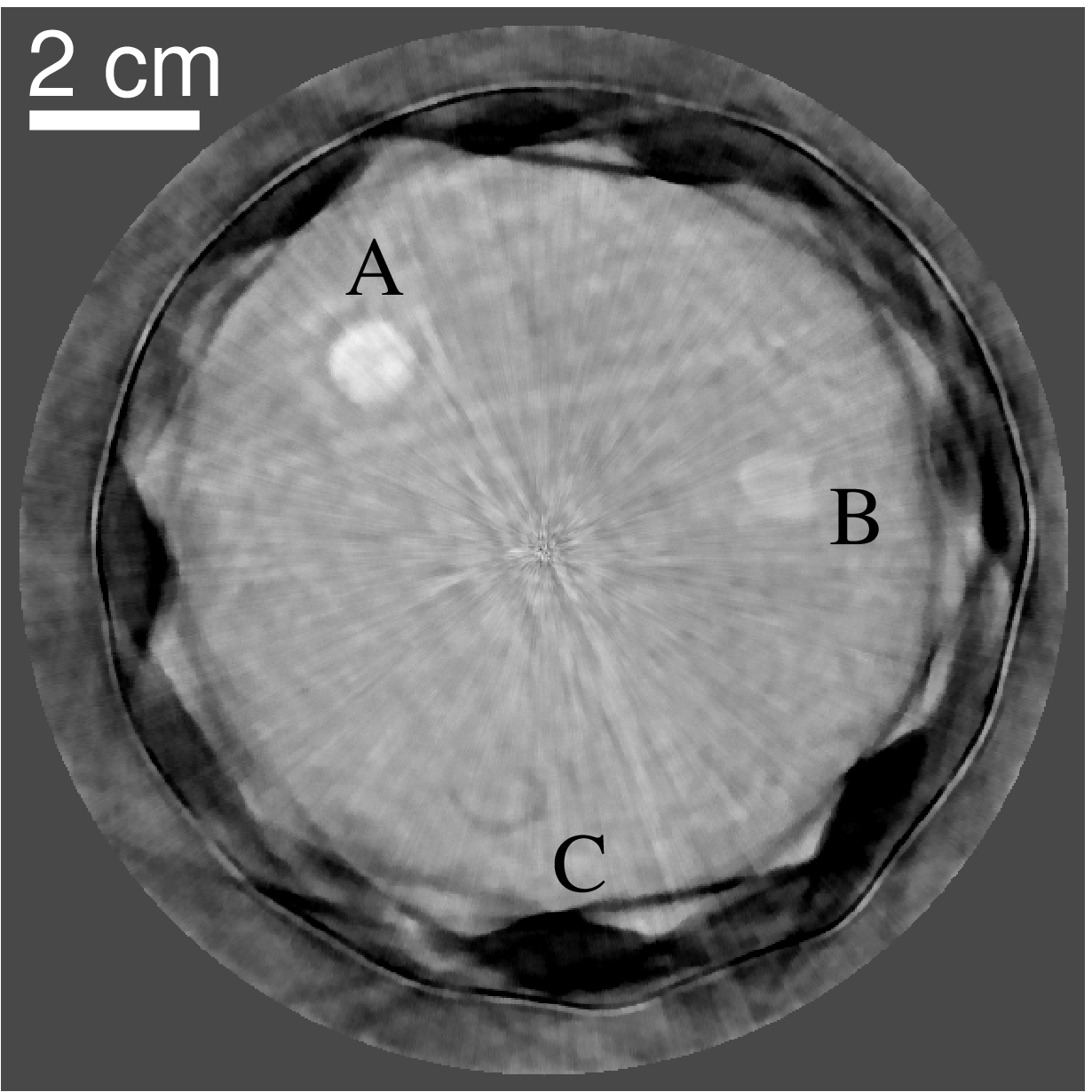}}
\caption{\label{fig:WISEExp} 
Images reconstructed from the experimentally measured phantom data 
by use of (a) the bent-ray model-based sound speed reconstruction method 
and (b) the WISE method with a TV penalty with ($\beta^{\rm TV} = 1.0\times 10^2$) after the $200$-th iteration. 
The grayscale window is $[1.49,1.57]$ mm/$\mu$s. 
}
\end{figure}

\begin{figure}[h]
\centering
\subfloat[]{\includegraphics[width=7cm]{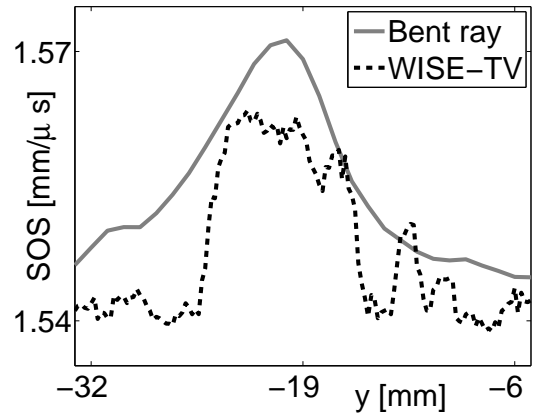}}\\
\vskip 0.5 cm
\subfloat[]{\includegraphics[width=7cm]{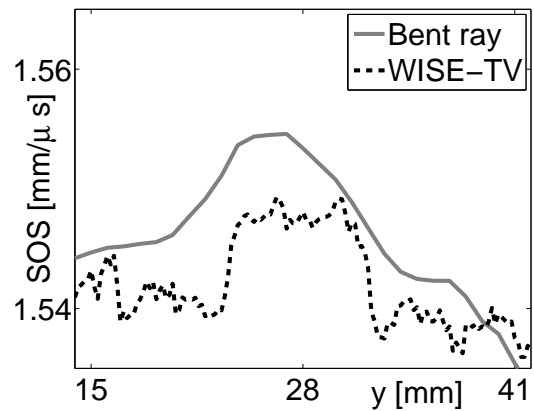}}
\caption{\label{fig:ExpProfile} 
Profiles at (a) $x = -24.0$ mm and (b) $x = 10.0$ mm 
of the reconstructed images by use of the bent-ray model-based sound speed reconstruction method (light solid) and 
the WISE method with a TV penalty with $\beta^{\rm TV} = 1.0\times 10^{2}$ (dark dashed) from experimentally measured data. 
}
\end{figure}

As shown in Fig.~\ref{fig:WISEExp}, the spatial resolution of the image reconstructed by use of the WISE method with a TV penalty is significantly higher than that reconstructed by use of the bent-ray model-based method.
In particular, the structures labeled `A' and `B' possess clearly-defined boundaries. 
This observation is further confirmed by the profiles of the two images shown in Fig.~\ref{fig:ExpProfile}.
In addition, the structure labeled `C' in Fig.~\ref{fig:WISEExp}-(b) is almost indistinguishable in the image reconstructed by use of the bent-ray model-based method (see Fig.~\ref{fig:WISEExp}-(a)).   
The improved spatial resolution is expected because the WISE method takes into account
  high-order acoustic diffraction, which is ignored
 by the bent-ray method  \cite{Duric2007:SOSWave}.  
Though not shown here, for the bent-ray method,
 we investigated multiple time-of-flight pickers \cite{Duric09:TOFPicker} and systematically tuned the regularization parameter. 
As such, it is likely that Fig.~\ref{fig:WISEExp}-(a) represents a nearly optimal bent-ray image in terms of the resolution.  
This resolution also appears to be similar to previous experimental results reported in the literature \cite{Roy10:BentRay}.
%Therefore, we believe that the outperformance of the WISE method over the BRSR method is, in general, valid as established in the literature  \cite{Duric2007:SOSWave}.  

\begin{figure}[h]
\centering
\subfloat[]{\includegraphics[width=5cm]{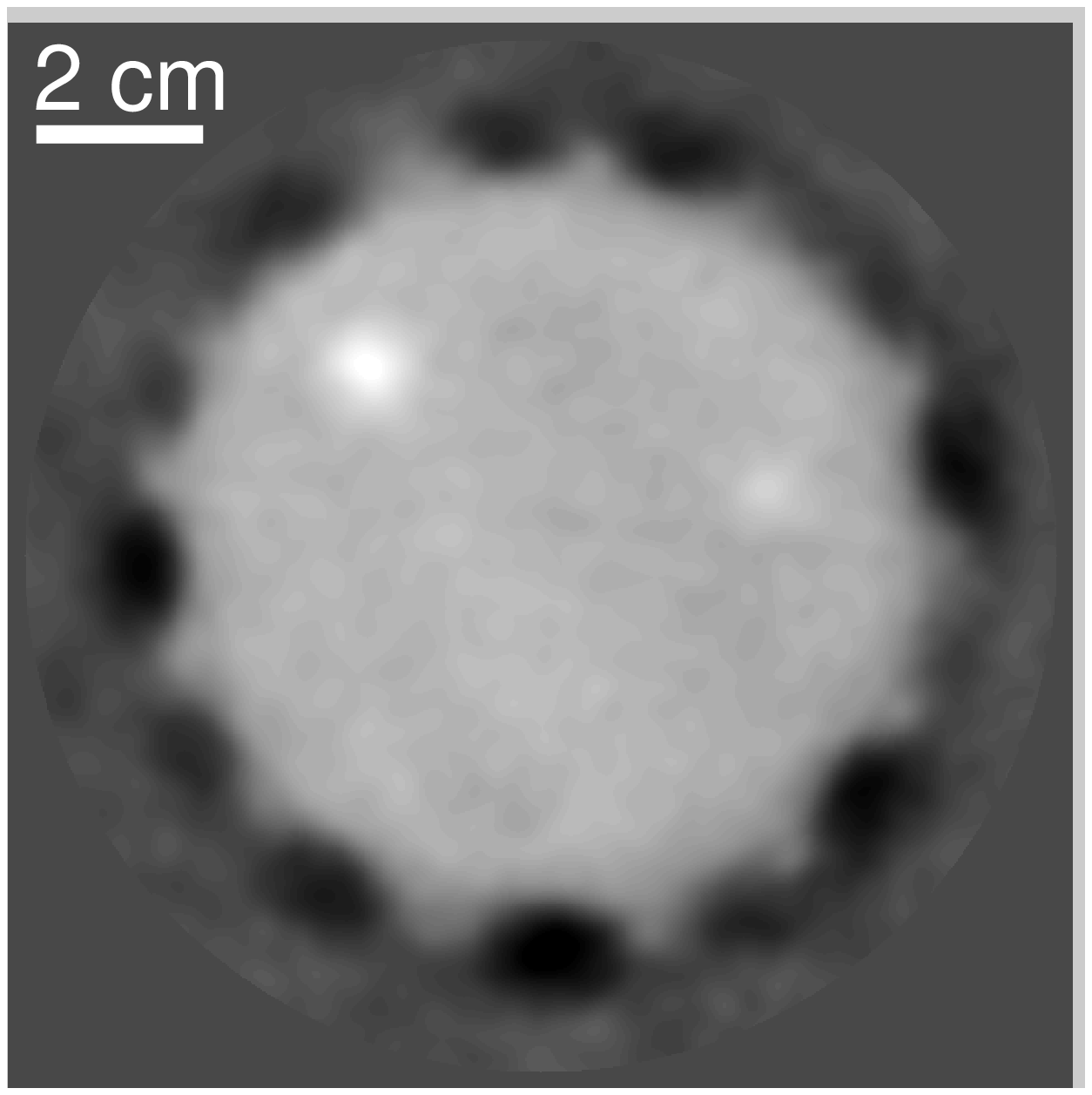}}
\subfloat[]{\includegraphics[width=5cm]{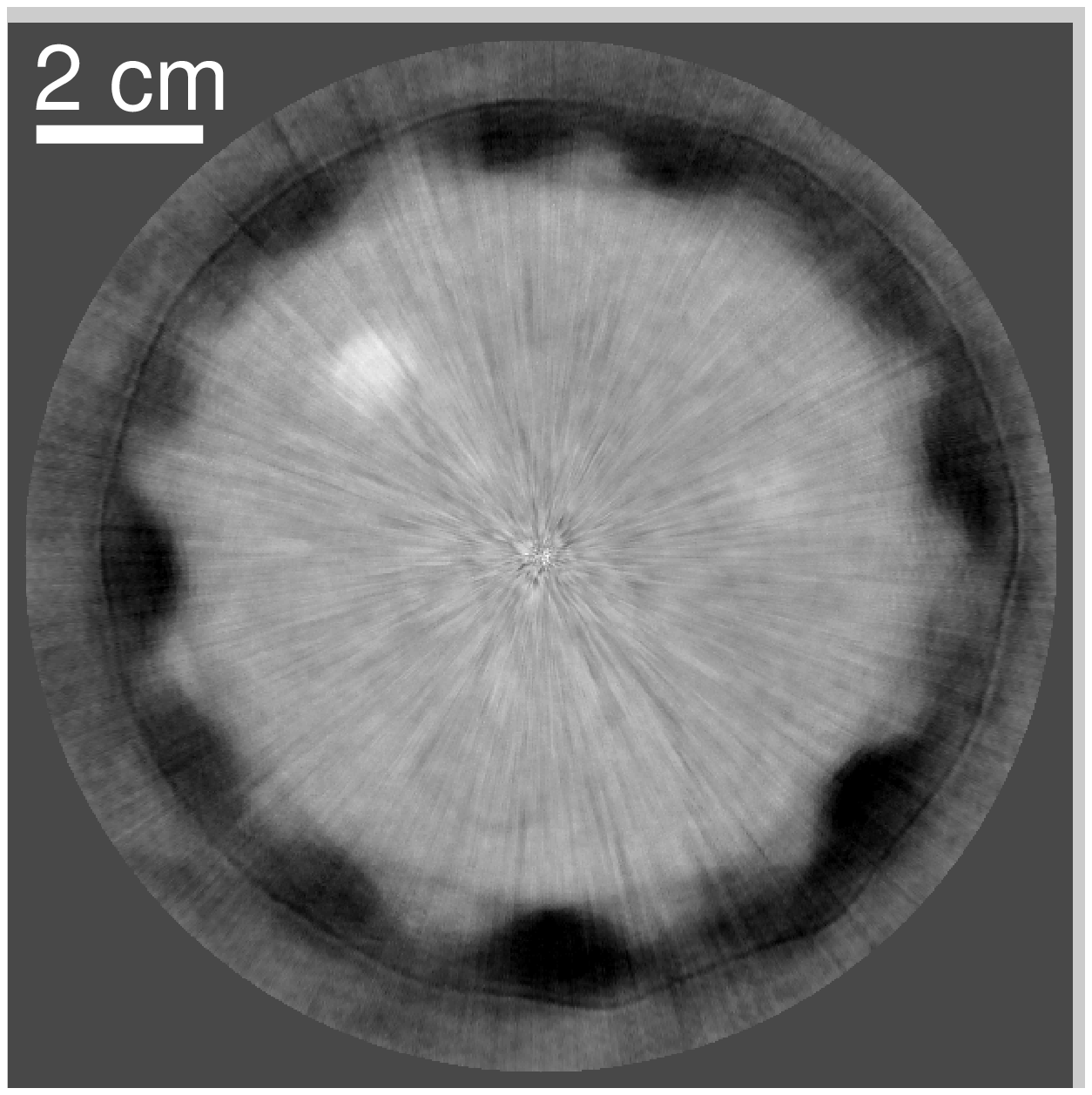}}\\
\subfloat[]{\includegraphics[width=5cm]{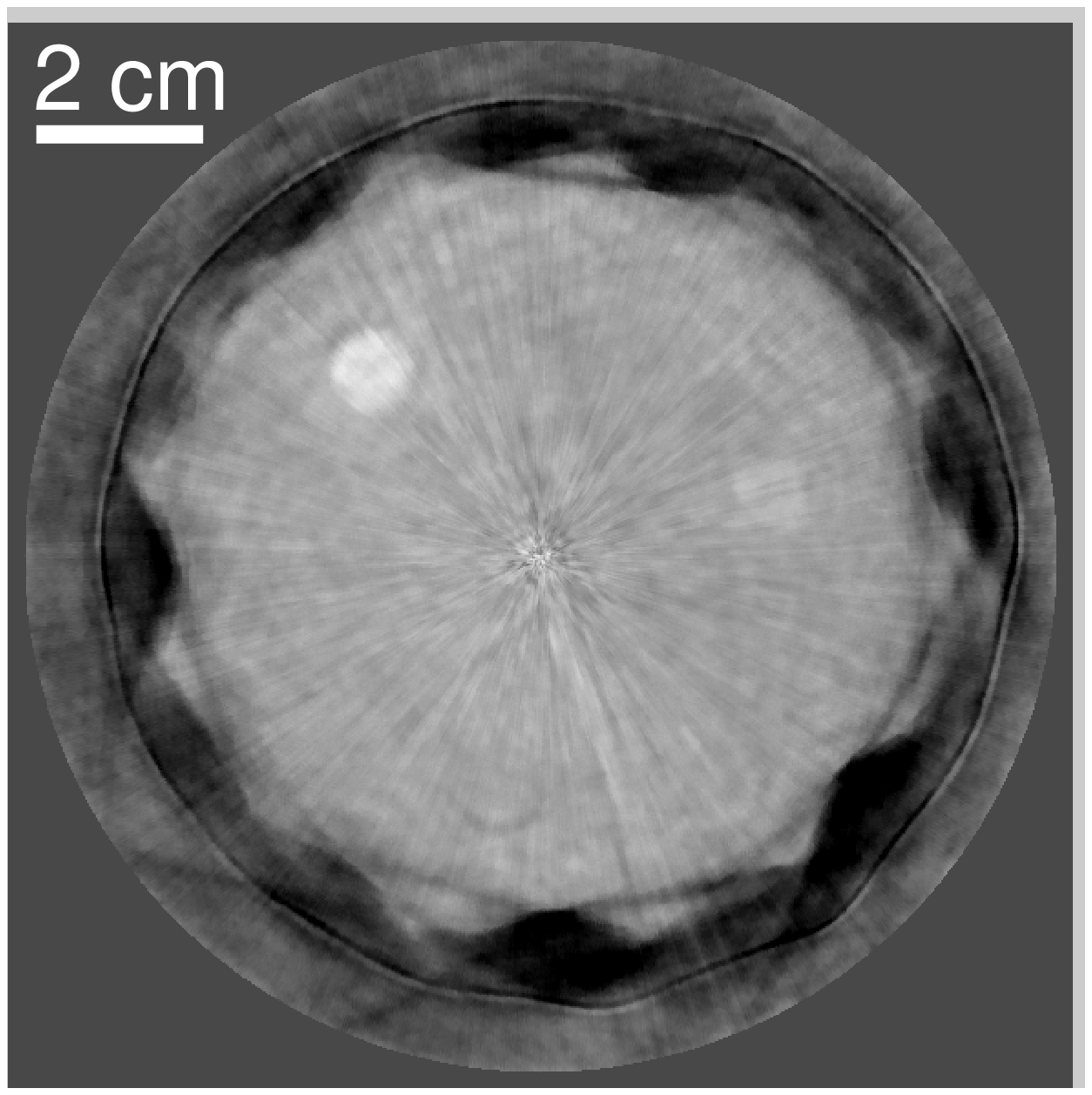}}
\subfloat[]{\includegraphics[width=5cm]{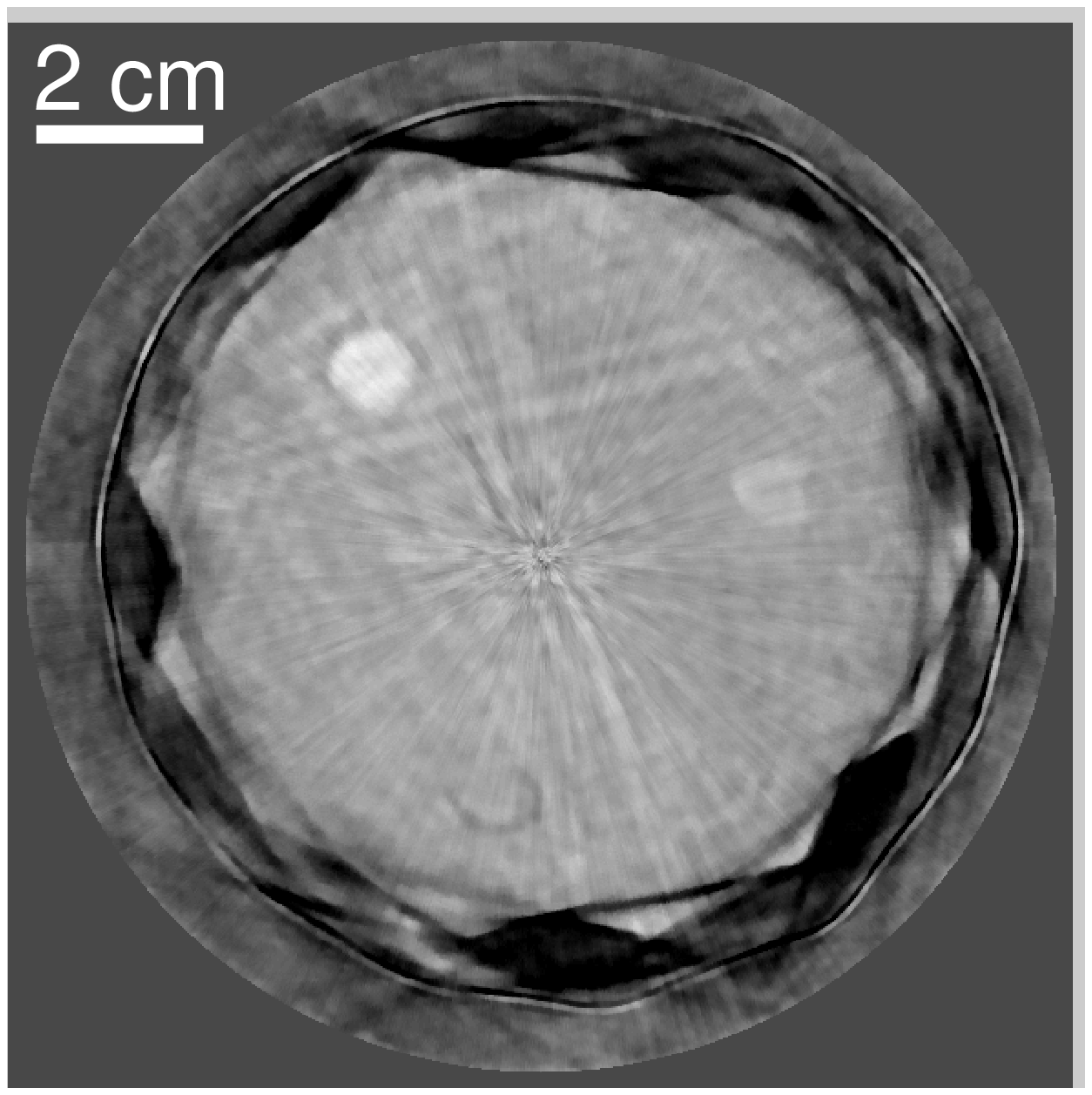}}
\caption{\label{fig:ExpConvergence} 
(a) The initial guess of the sound speed map 
and the images reconstructed by use of the WISE method with a TV penalty with ($\beta^{\rm TV} = 1.0\times 10^2$) 
after 
(b) the $10$-th, (b) the $50$-th
and (d) the $300$-th iteration, 
from the experimentally measured phantom data. 
The grayscale window is $[1.49,1.57]$ mm/$\mu$s. 
}
\end{figure}

The convergence properties
 of the WISE method with a TV penalty with experimental data were
 consistent with those observed in the computer-simulation studies.
Images reconstructed by use of 10, 50, and 300 algorithm iterations are
displayed in  Fig.~\ref{fig:ExpConvergence}. 
The image reconstructed by use of 10 iterations contains radial streak artifacts that are similar in nature
to those observed in the computer-simulation studies.
These artifacts were mitigated after more iterations. 
The image reconstructed after $300$ iterations (Fig.~\ref{fig:ExpConvergence}-(d)) appears to be similar to that after $200$ iterations (Fig.~\ref{fig:WISEExp}-(b)), suggesting that the WISE method with a TV penalty is close to convergence after about $200$ iterations. 
The time required to complete $200$ iterations was approximately $14$ hours. 
The estimated time it would take for the sequential waveform inversion method
to produce a comparable image is approximately one month, assuming the same number of iterations is required as in the computer-simulation studies (i.e., $40$).

\begin{figure}[h]
\centering
\subfloat[]{\includegraphics[width=5cm]{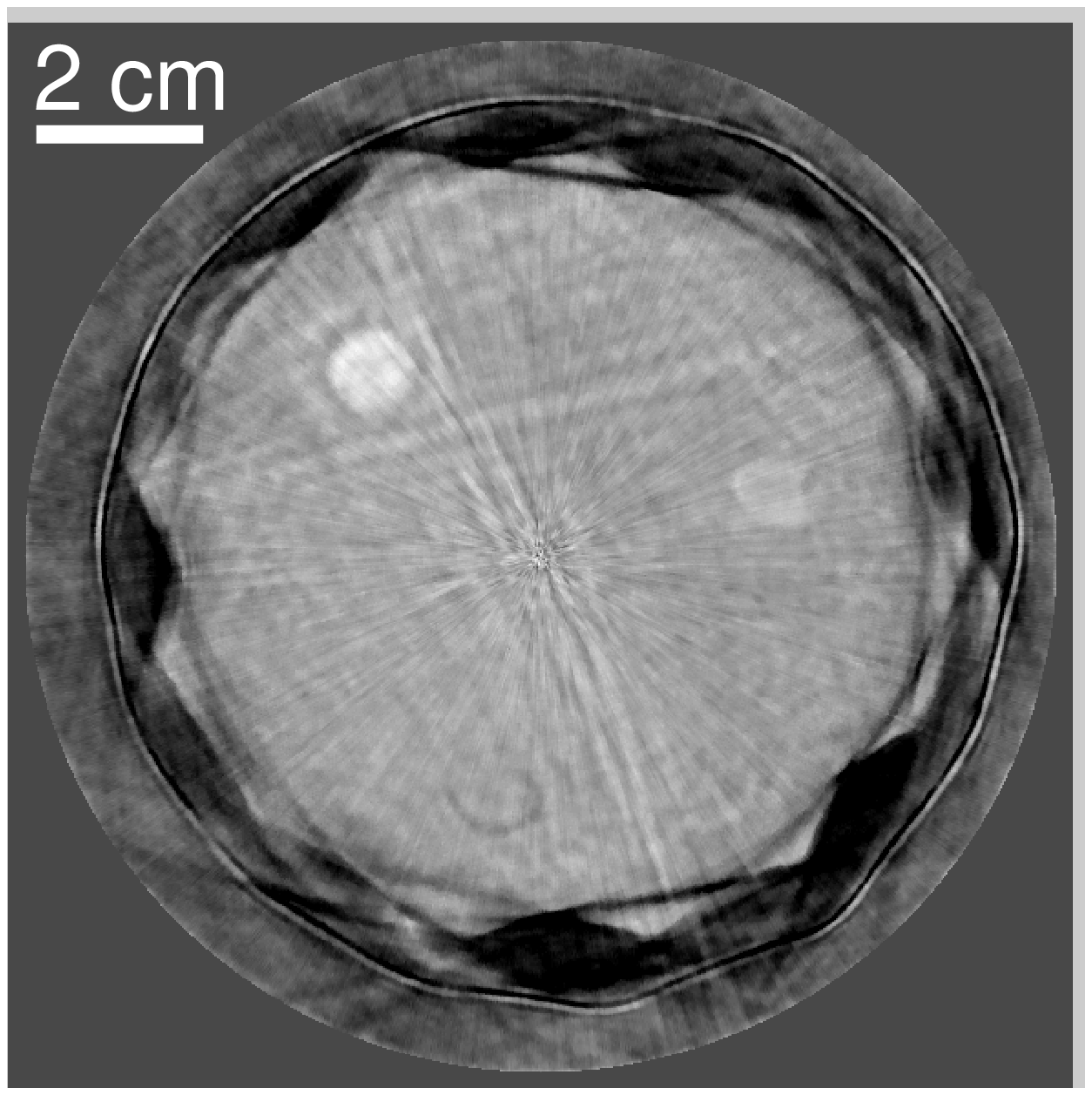}}
\subfloat[]{\includegraphics[width=5cm]{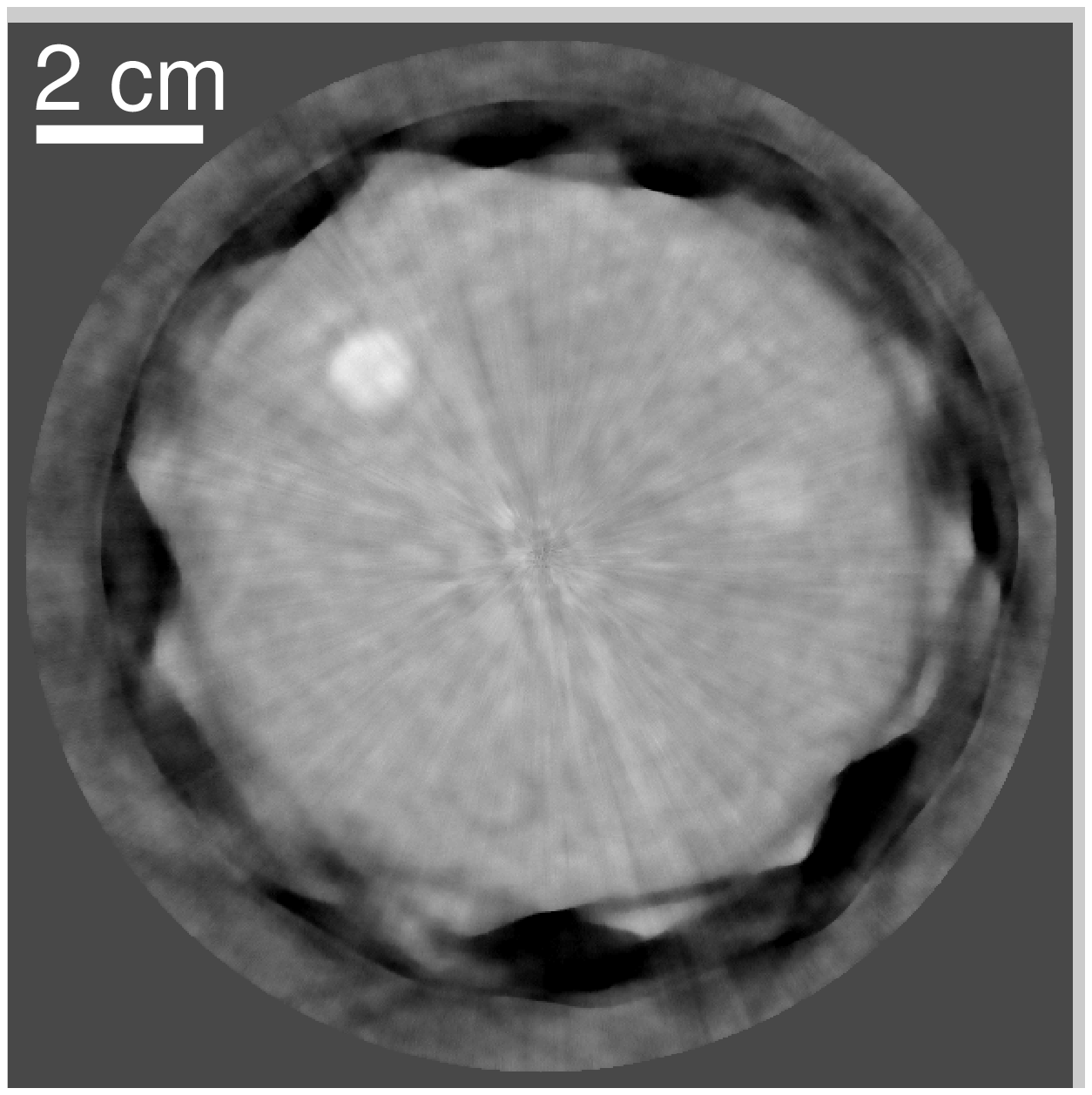}}
\caption{\label{fig:ExpReg} 
Images reconstructed by use of the WISE method with a TV penalty with 
(a) $\beta^{\rm TV} = 5.0\times 10^1$,
and 
(b) $\beta^{\rm TV} = 5.0\times 10^2$, 
from the experimentally measured phantom data. 
The grayscale window is $[1.49,1.57]$ mm/$\mu$s. 
}
\end{figure}

Despite the nonlinearity of the WISE method,  the impact of the TV penalty appears to be
 similar to that observed in other imaging applications \cite{Kun:PMB,Kun14TMI:Blob} (see Fig.~\ref{fig:ExpReg}). 
Though not shown here, the impact of the quadratic penalty is also similar. 
As expected, a larger value of $\beta$ reduced the noise level at the expense of spatial image resolution. 
These results suggest a predictable impact of the penalties on the images reconstructed by use of the WISE method.

\section{Summary}
\label{Sect:Conclusions}

 It is known that waveform inversion-based reconstruction methods
 can produce sound speed images that possess improved spatial resolution properties over those produced by ray-based methods.
However, waveform inversion methods are computationally demanding and have not
been applied widely in USCT breast imaging.
In this work, based on the time-domain wave equation and motivated by recent mathematical results in the geophysics literature,  the WISE method
was developed that circumvents the large computational burden of conventional waveform inversion methods.
This method encodes the measurement data using a random encoding vector and determines an estimate of the sound speed
 distribution by
 solving a stochastic optimization problem by use of a stochastic gradient descent algorithm.
With our current GPU-based implementation, the computation time was reduced from weeks to hours. 
The WISE method was systematically investigated in computer-simulation and experimental studies
involving a breast phantom.  The results suggest that the method holds value for USCT breast imaging
applications in a practical setting.
%may bring the power of waveform inversion  methods to bear on practical applications.
%In conjunction with practical considerations that are necessary when applying to experimental studies, we have demonstrated a robust and practical method that improves on the state of the art for USCT breast imaging, as has been demonstrated with laboratory phantom studies. 

Many opportunities remain to further improve the performance of the WISE method. 
As shown in Fig.~\ref{fig:WISEExp}, images reconstructed by use of the WISE
method can contain certain artifacts that are not present in the image reconstructed by use of
 the bent-ray method.  An example of such an artifact is the dark horizontal streak below the structure C.
Because of the nonlinearity of the image reconstruction problem, it is challenging to determine whether these artifacts are caused by imaging model errors or by the optimization algorithm, which might have arrived at a local minimum of the cost function. 
A more accurate imaging model can be developed to account for  out-of-plane scattering, the transducer finite aperture size effect, acoustic absorption, as well as other physical factors. 
Also, the stochastic gradient descent algorithm is one of the most basic stochastic optimization algorithms.
Numerous emerging optimization algorithms can be employed \cite{ArXiv:Haber14,ArXiv:Ascher14} to improve the convergence rate. 
In addition, there remains a great need to compare the WISE method with other existing sound speed reconstruction algorithms \cite{Chew10,Hesse2013:USCT}.

% Waveform inversion-based USCT tomography is, in general, computationally demanding. 
% We believe the computation time can be further reduced by the increasing computation capacity and through advanced optimization algorithms.

There remains a need to conduct additional investigations of the numerical properties of the WISE method.
Currently, a systematic comparison of the statistical properties of the WISE and the sequential waveform inversion method is prohibited by the excessively long computation times required by the latter method. 
This comparison will be interesting when a more efficient wave equation solver is available.
Given the fact that waveform inversion is nonlinear and sensitive to its initial guess, it becomes important to investigate how to obtain an accurate initial guess. 
We also observed that the performance of the WISE method is sensitive to how strong the medium heterogeneities are and the profile of the excitation pulse.
An investigation of the impact of the excitation pulse the numerical properties
of the image reconstruction may help optimize hardware design. 
In addition, quantifying the statistics of the reconstructed images will allow
application of task-based measures of image quality to be applied to guide 
system optimization studies.

% There remains a great space for the improvement of the computational efficiency. 
% First, the numerical wave equation solver can be implemented more efficiently. 
% Particularly, we can reduced the calculation domain by use of perfect matched layer (PML). 
% It can reduce the computation by at least to one quarter. 
% Second, more advanced optimization algorithm can be employed.
% n this study, we demonstrate the method by use of simple gradient descent algorithm. 
% The involved line search is computational demanding. 
% Developing advanced optimization algorithms can further improve the computational efficiency. 
% This study focuses on demonstration of the feasibility. 
% However, many numerical issues involved in the WISE method remains intersting for future studies. 
% For example, how the initial guess will affect the reconstructed images. 
% How the excitation pulse will affect the reconstructed images. 
% How the noise statistics will affect the reconstructed images. 

% or

\appendices

\section{Continuous-to-Discrete USCT Imaging Model}
\label{Sect:C-DModel}
In practice, each data function $g_m(\mathbf r, t)$ is spatially and temporally sampled
to form a data vector $\mathbf g_m \in\mathbb R^{N^{\rm rec}L}$,
where $N^{\rm rec}$ and $L$ denote the number of receivers and the number of time samples, respectively.
We will assume that $N^{\rm rec}$ and $L$  do not vary with excitation pulse.
Let $[\mathbf g_m]_{n^{\rm rec}L+l}$ denotes the $(n^{\rm rec}L+l)$-th element of $\mathbf g_m$.
When the receivers are point-like,  $\mathbf g_m$ is defined as
\begin{equation}
[\mathbf g_m]_{n^{\rm rec}L+l} = g_m(\mathbf r(m,n^{\rm rec}), l\Delta^{\rm t}),
\end{equation}
where the indices $n^{\rm rec}$ and $l$ specify the receiver location and temporal sample, respectively,
and $\Delta^{\rm t}$ is the temporal sampling interval.
The vector $\mathbf r(m,n^{\rm rec})\in\Omega_m$ denotes the location of the $n^{\rm rec}$-th receiver at the $m$-th data acquisition.

A C-D imaging model for USCT describes the mapping of $c(\mathbf r)$ to the data
vector $\mathbf g_m$ and can be expressed as
\begin{equation}\label{eqn:CDModel}
  [\mathbf g_m]_{n^{\rm rec}L+l}
= \mathcal M_m \mathcal H^{\rm c} s_m {(\mathbf r,t)}
   \big\vert_{\mathbf r=\mathbf r(m,n^{\rm rec}), t = l\Delta^{\rm t}} 
  \quad{\rm for}\quad
  \substack {  n^{\rm rec}=0,1,\cdots,N^{\rm rec}-1\\
     l=0,1,\cdots,L-1}.
\end{equation}
Note that the acousto-electrical impulse response \cite{TMI:transmodel} of the receivers can be incorporated into the C-D imaging model by temporally convolving $s_m(\mathbf r, t)$ in Eqn.~\eqref{eqn:WaveEqn} with the receivers' acousto-electrical impulse response if we assume all receiving transducers share an identical acousto-electrical impulse response.

\section{Fr\'echet derivative of data fidelity term}
\label{Sect:Frechet}

Consider the integrated squared-error data misfit function, \cite{Duric2007:SOSWave,Los12:SOSPE}
\begin{equation}\label{eqn:CDataMisfit}
  \mathcal F^{\rm CC}(c) = \frac{1}{2} \sum_{m=0}^{M-1}
    \int_{\Omega_m} d\mathbf r
    \int_0^T dt
    \big[\underline{g_m}(\mathbf r,t) - g_m(\mathbf r, t) \big]^2 ,
\end{equation}
where $\underline{g_m}(\mathbf r, t)$ and $g_m(\mathbf r,t)$ denote the measured
data function and the predicted data function computed by use of Eqn.\ (\ref{eqn:CImagingModel2}) with the current estimate
of $c(\mathbf r)$.

% Depending on the explicit form of Eqn.~\eqref{eqn:CCostFunc}, various optimization algorithms can be adopted accordingly.
Both the  sequential and WISE reconstruction method described in Section \ref{Sect:WISE} require knowledge of the Fr\'echet derivatives of $\mathcal F^{\rm CC}(c)$ and $\mathcal R^{\rm CC}(c)$ with respect to $c$, denoted by $\nabla_{\rm c} \mathcal F^{\rm CC}$ and $\nabla_{\rm c}\mathcal R^{\rm CC}$, respectively.
The calculation of $\nabla_{\rm c}\mathcal R^{\rm CC}$ can be readily accomplished for quadratic smoothness penalties \cite{Fessler:94,Kun:PMB}.
% Minimization of Eqn.~\eqref{eqn:CCostFunc} can be accomplished by use of a gradient-based optimization algorithm, which in general, requires the calculation of the Fr\'echet derivative of $\mathcal F^{\rm CC}(c)$ with respect to $c$.
For the integrated squared error data misfit function given in Eqn.~\eqref{eqn:CDataMisfit}, $\nabla_{\rm c} \mathcal F^{\rm CC}$ can be computed via an adjoint state method as \cite{Norton:Adjoint,AdjStaRev:06,Roy2010:SOSWave}
\begin{equation}\label{eqn:Frecht}
  \nabla_{\rm c} \mathcal F^{\rm CC} =
         \frac{1}{c^3(\mathbf r)}
\sum_{m=0}^{M-1}
      \int_{0}^T\!\! dt\, q_m(\mathbf r, T-t)
      \frac{\partial^2}{\partial^2 t} p_m(\mathbf r, t),
\end{equation}
where $q_m(\mathbf r, t) \in \mathbb L^2(\mathbb R^3\times[0,\infty))$ is the solution to the adjoint wave equation.
The adjoint wave equation is defined as
\begin{equation}\label{eqn:AdjWaveEqn}
   \nabla^2 q_m(\mathbf r, t) - \frac{1}{c^2(\mathbf r)} \frac{\partial^2}{\partial^2 t}q_m(\mathbf r, t) = - \tau_m(\mathbf r, t),
\end{equation}
where $\tau_m(\mathbf r, t) = g_m(\mathbf r, T-t) - \underline{g_m}(\mathbf r, T-t)$.
The adjoint wave equation is nearly identical in form to the wave equation in Eqn.~\eqref{eqn:WaveEqn} except for the different source term on the right-hand side, suggesting the same numerical approach can be employed to solve both equations.
Since one needs to solve Eqns.~\eqref{eqn:WaveEqn} and \eqref{eqn:AdjWaveEqn} $M$ times in order to calculate $\nabla_{\rm c}\mathcal F^{\rm CC}$, it is generally true that the sequential waveform inversion is computationally demanding even for a 2D geometry \cite{Operto09:FWIReview}.
% Calculation of the Fr\'echet derivative of $\mathcal R^{\rm CC}(c)$ with respect to $c$, denoted by $\nabla_{\rm c}\mathcal R^{\rm CC}$, depends on the choice of $\mathcal R^{\rm CC}$ \cite{Fessler:94,Kun:PMB}.

\section{Bent-ray model-based sound speed reconstruction}
\label{Sect:BentRay}
We developed an iterative image reconstruction algorithm based on a bent-ray imaging model.
The bent-ray imaging model assumes that an acoustic pulse travels along a ray path that connects the emitter and the receiver and accounts for the refraction of rays, also known as ray-bending, through an acoustically inhomogeneous medium. 
For each pair of receiver and emitter, the travel time, as well as the ray path, is determined by the medium's sound speed distribution. 
Given the travel times for a collection of emitter-and-receiver pairs distributed around the object, the medium sound speed distribution can be iteratively reconstructed.
This bent-ray model-based sound speed reconstruction (BRSR) method has been employed in the USCT literature \cite{Cuiping:SPIE:2009,Roy10:BentRay,jose:7262}. 

In order to perform the BRSR, we extracted a TOF data vector from the measured pressure data. 
Denoting the TOF data vector by $\underline{\mathbf T}\in\mathbb R^{MN^{\rm rec}}$, each element of $\underline{\mathbf T}$
represented the TOF from each emitter-and-receiver pair. 
% the $m$-th emitter to the $n^{\rm rec}$-th receiver.
% , denoted by $[\underline{\mathbf T}^{\rm d}]_{mN^{\rm rec}+n^{\rm rec}}$, for $m=0, 1, \cdots, M-1$, and $n^{\rm rec}=0,1,\cdots,N^{\rm rec}-1$, 
The extraction of the TOF was conducted in two steps. 
First, we estimated the difference between the TOF when the object was present and the TOF when the object was absent by use of a thresholding method \cite{Duric09:TOFPicker,Fatima:SPIE14}. 
In particular, $20\%$ of the peak value of each time trace was employed as the thresholding value. 
Second, a TOF offset was added to the estimated difference TOF for each emitter-and-receiver pair to obtain the absolute TOF, where the TOF offset was calculated according to the scanning geometry and the known background SOS. 

Having the TOF vector $\underline{\mathbf T}$, we reconstructed the sound speed by solving the following optimization problem: 
\begin{equation}\label{objectiveeq}
\hat{\bf s}=\arg \min\limits_{{\bf s}} \parallel {\underline{\bf T}}
- \mathbf K^{\rm s} \mathbf s \parallel^{2} + \beta \mathcal R({\bf s}),
\end{equation}
where $\bf s$ denotes the slowness (the reciprocal of the SOS) vector, 
and $\mathbf K^{\rm s}$ denotes the system matrix that maps the slowness distribution to the TOF data. 
The superscript `s' indicates the dependence of $\mathbf K^{\rm s}$ on the slowness map. 
At each iteration, using the current estimate of the SOS, a ray-tracing method \cite{Sethian20021996} was employed to construct the system matrix $\mathbf K^{\rm s}$. 
Explicitly storing the system matrix in the sparse representation, we utilized the limited BFGS method~\cite{lbfgsb:1995} to solve the optimization problem given in Eqn.~\eqref{objectiveeq}. 
The estimated slowness was then converted to the sound speed by taking the reciprocal of $\hat{\mathbf s}$ element-wisely. 
We refer the readers to \cite{born1999principles,Cuiping:SPIE:2009,Roy10:BentRay,jose:7262,Fatima:SPIE14} for more details about the BRSR method. 

% \section{}

\section*{Acknowledgment}
This work was supported in part by NIH awards EB010049, CA1744601, EB01696301 and DOD Award US ARMY W81XWH-13-1-0233.
\ifCLASSOPTIONcaptionsoff
  \newpage
\fi

\bibliographystyle{IEEEtran}
\bibliography{reflect}%,PAT-bib1,PAT-bib2,PAT-bib3}

\clearpage
\section*{Tables}

\begin{table}[h]
\caption{\label{tab:NumPhantom} Parameters of the numerical breast phantom 
\cite{Szabo04Book,Glide07:MedPhy,Li08:SPIE}}
\centering
\begin{tabular}[h]{c |c |c |c }
\hline\hline
         {Structure}
         &{Tissue type}
         &{Sound speed}
         &{Slope of attenuation}\\
index    & & [mm$\cdot\mu$s$^{-1}$] & [dB$\cdot($MHz$)^{-y}\cdot$cm$^{-1}$]\\[0.5ex]
\hline
$0$      &Adipose           & $1.47$  & $0.60$\\  
$1$      &Parenchyma     & $1.51$  & $0.75$\\ 
$2$      &Benign tumor   & $1.47$  & $0.60$\\
$3$      &Benign tumor   & $1.47$  & $0.60$\\
$4$      &Cyst           & $1.53$  & $0.00217$\\
$5$      &Malignant tumor& $1.565$ & $0.57$\\
$6$      &Malignant tumor& $1.565$ & $0.57$\\
$7$      &Malignant tumor& $1.57$  & $0.57$\\
\hline\hline
\end{tabular}
\end{table}
\clearpage

\begin{table}[h]
\caption{\label{tab:ExpPhantom} Parameters of the experimental breast phantom}
\centering
\begin{tabular}[h]{c |c |c }
\hline\hline 
 {Material} 
&{Sound speed}
&{Attenuation coefficient }\\
  & [mm$\cdot\mu$s$^{-1}$] & at $2.5$ MHz [dB/cm] \\
\hline
Fat & $1.467$ & $0.48$ \\
Parenchymal tissue  & $1.552$ & $0.89$ \\
Cancer & $1.563$ & $1.20$ \\
Fibroadenoma & $1.552$ & $0.52$ \\ 
Gelatin cyst & $1.585$ & $0.16$ \\
\hline \hline
\end{tabular}
\end{table}
\clearpage

\section*{Figures}

\clearpage

\clearpage

\clearpage

\clearpage

\clearpage

\clearpage

\clearpage

\clearpage

\iffalse
\begin{figure}[h]
\centering
\includegraphics[width=9cm]{Convergence}
\caption{\label{fig:Convergence}
Plots of the absolute (dark) and relative (gray) Euclidean distances versus the number of iterations. 
}
\end{figure}
\clearpage
\fi

\begin{figure}[h]
\centering
\subfloat[]{\includegraphics[width=7cm]{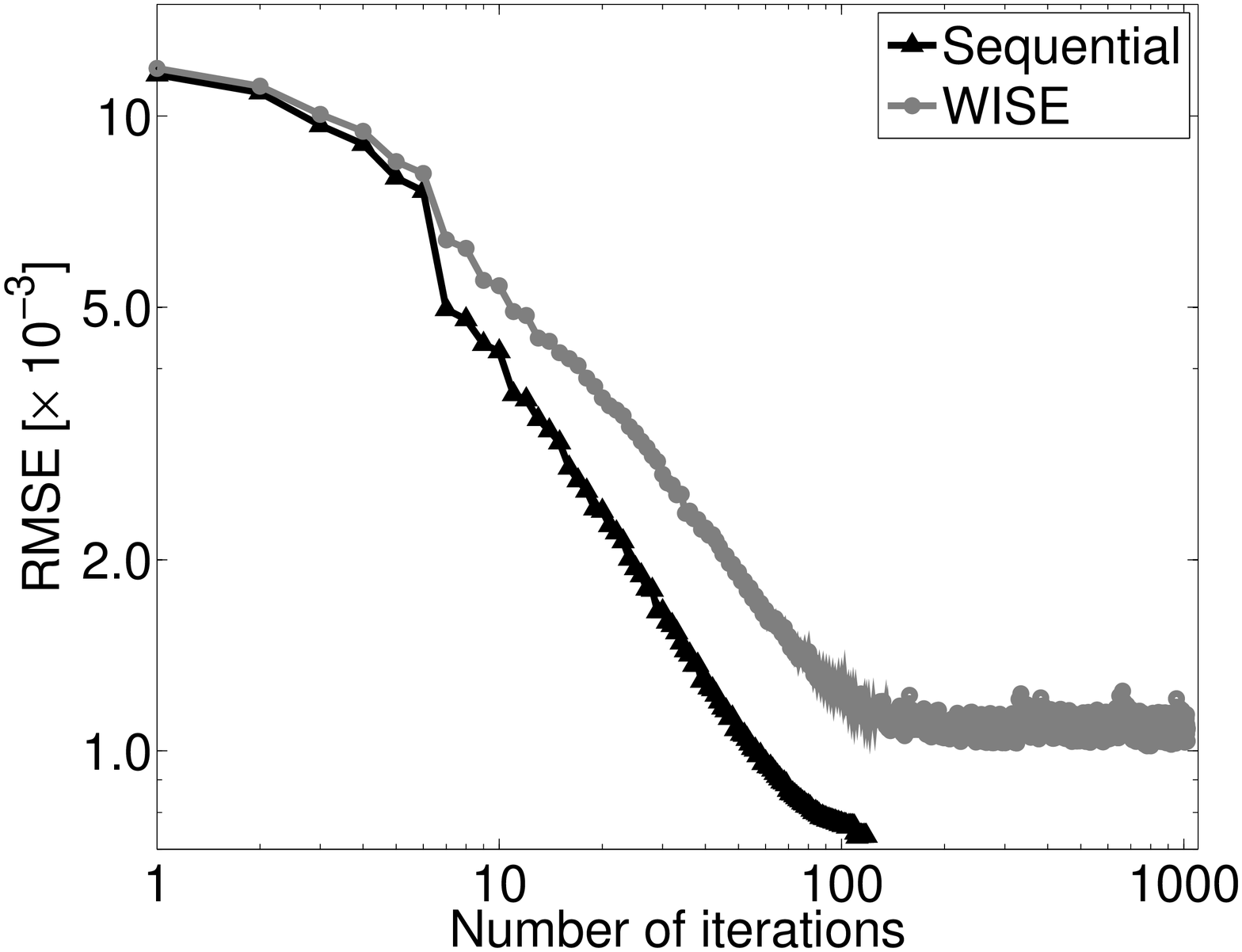}}
\subfloat[]{\includegraphics[width=7cm]{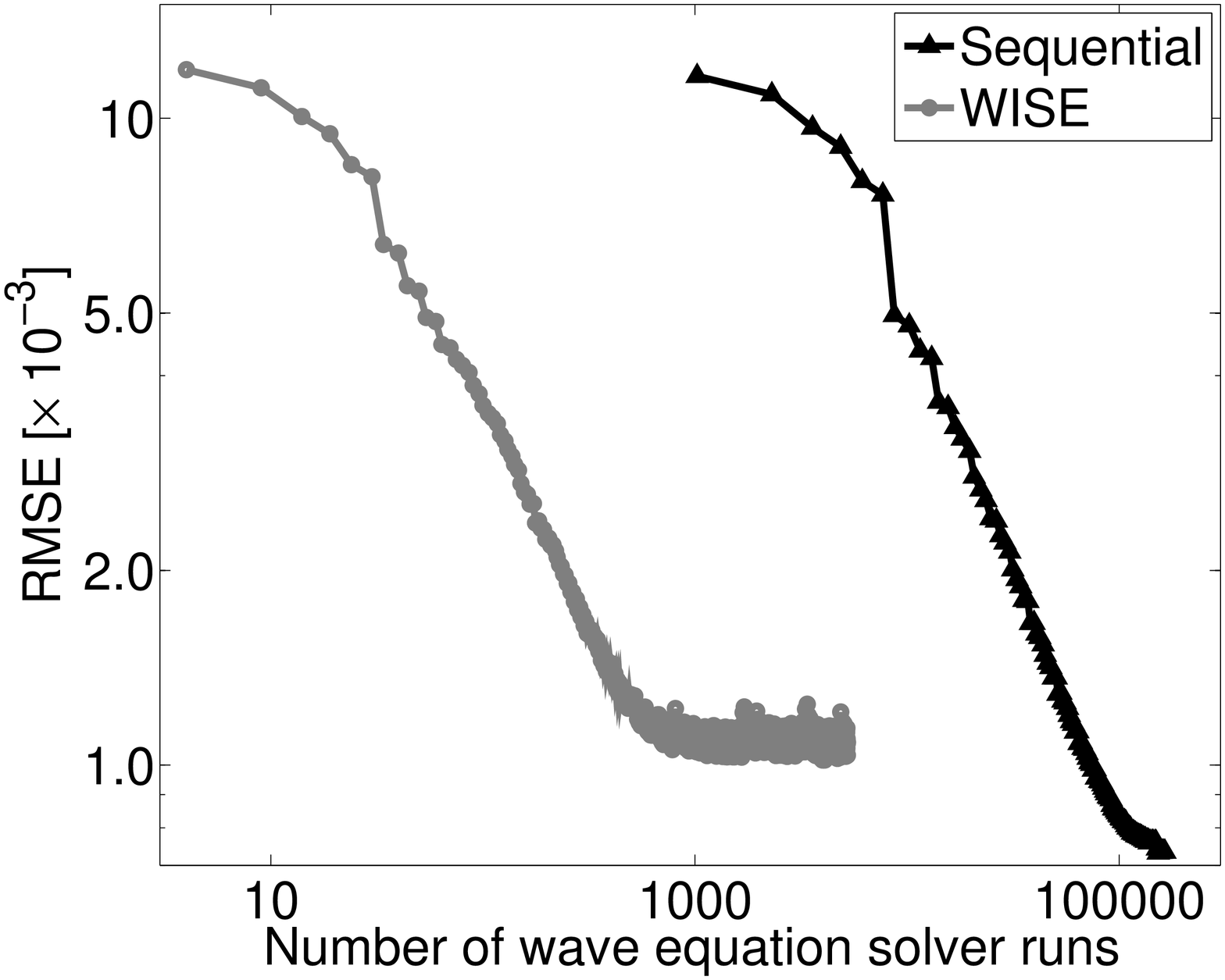}}\\
\subfloat[]{\includegraphics[width=7cm]{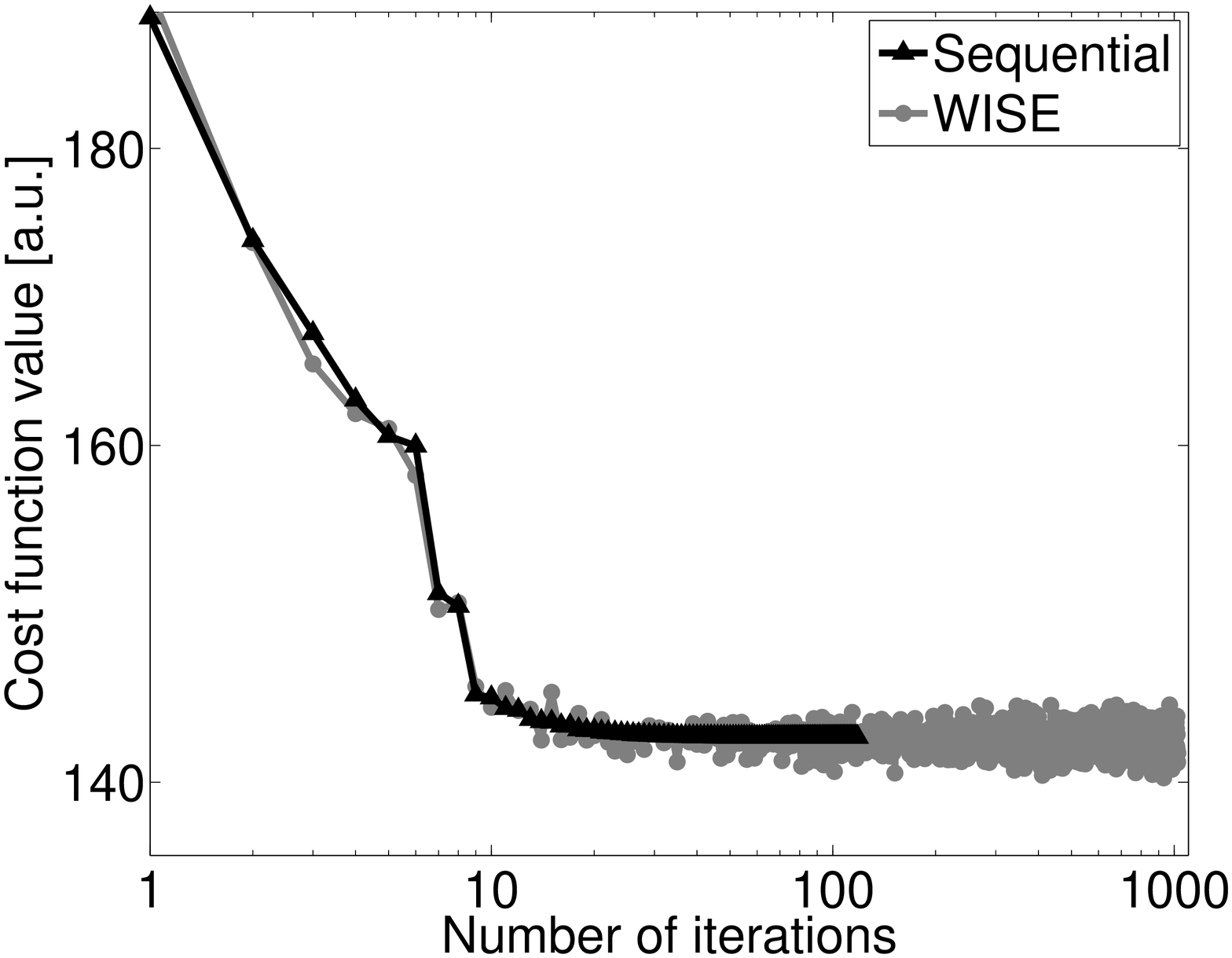}}
\caption{\label{fig:Convergence}
Plots of the root-mean-square errors (RMSEs) of the images reconstructed from the noise-free data versus (a) the number of iterations and (b) the number of wave equation solver runs. 
(c) Plots of the cost function value versus the number of iterations. }
\end{figure}
\clearpage

\clearpage

\clearpage

\clearpage

\clearpage

\clearpage

\clearpage

\clearpage

\clearpage

\clearpage

\clearpage

\clearpage

\clearpage

\clearpage

\end{document}